\documentclass[12pt,a4paper,notitlepage]{report}
\usepackage[american]{babel}
\usepackage{cmap} 
\usepackage{ifthen}
\usepackage[utf8]{inputenc}
\usepackage[T1]{fontenc}
\usepackage{tikz}
\usetikzlibrary{automata, positioning, arrows.meta}
\usepackage{algorithm}
\usepackage{algpseudocode}
\usepackage{amsthm}
\usepackage{amssymb}
\usepackage{stmaryrd}
\usepackage{makeidx}
\ifthenelse{\isundefined{\hypersetup}}{
  \usepackage[colorlinks=true,linkcolor=blue,urlcolor=magenta,citecolor=red,plainpages=false,pdfpagelabels]{hyperref}
  \usepackage{bookmark}
}{}
\usepackage{csquotes}
\usepackage[style=verbose, style=numeric, backend=biber]{biblatex}
\makeindex

\algrenewcommand{\algorithmiccomment}[1]{{\hfill\textit{\footnotesize //#1}}}

\theoremstyle{plain}
\newtheorem{theorem}{Theorem}

\newtheorem{corollary}{Corollary}

\theoremstyle{definition}
\newtheorem{definition}{Definition}
\newtheorem{xmpl}{Example}
  
\newenvironment{example}[1][]{\ifthenelse{\equal{#1}{}}%
{\begin{xmpl}}%
{\begin{xmpl}[#1]}%
\pushQED{\qed}%
}%
{\popQED\end{xmpl}}

\theoremstyle{remark}
\newtheorem{proposition}{Proposition}

\newcommand{\refchp}[1]{\ref{#1}}
\newcommand{\refsec}[1]{\ref{#1}}
\newcommand{\refdef}[1]{Definition \ref{#1}}
\newcommand{\refthm}[1]{Theorem \ref{#1}}

\newcommand{\refcor}[1]{Corollary \ref{#1}}
\newcommand{\refeqn}[1]{(\ref{#1})}
\newcommand{\refxmp}[1]{Example \ref{#1}}
\newcommand{\refalg}[1]{Algorithm \ref{#1}}
\newcommand{\reffig}[1]{Figure \ref{#1}}

\newcommand{\POSIX}{\textsc{posix}}
\newcommand{\submatchsym}{\$}
\newcommand{\submatch}[1]{\ensuremath{\submatchsym_{#1}}}

\newcommand{\emptystring}{\ensuremath{\varepsilon}}
\newcommand{\N}{\mathbb{N}}
\newcommand{\define}[1]{\emph{#1}}
\newcommand{\length}[1]{|#1|}

\newcommand{\set}[1]{\ensuremath{\hspace{-0.2ex}\left\{#1\right\}}}
\newcommand{\inv}[1]{\ensuremath{#1'}}
\newcommand{\eq}[1]{\ensuremath{\equiv_{#1}}}
\newcommand{\union}{+}
\newcommand{\Union}{\sum}
\newcommand{\intersection}{\times}

\newcommand{\anchorsym}{}
\newcommand{\banchor}[1]{\ensuremath{{\anchorsym}_{#1}}}
\newcommand{\eanchor}[1]{\ensuremath{{_{#1}}\anchorsym}}
\renewcommand{\bot}{\banchor{\dashv}}
\newcommand{\bol}{\banchor{\prec}}
\newcommand{\bow}{\banchor{\langle}}

\newcommand{\eow}{\eanchor{\rangle}}
\newcommand{\eol}{\eanchor{\succ}}
\newcommand{\eot}{\eanchor{\vdash}}

\newcommand{\tagsym}{}
\newcommand{\etag}[1]{\ensuremath{{\tagsym}_{\lfloor}^{\mathnormal{#1}}}}
\newcommand{\ltag}[1]{\ensuremath{{\tagsym}_{\rfloor}^{\mathnormal{#1}}}}
\newcommand{\bank}[1]{\ensuremath{\mu_{\mathnormal{#1}}}}
\newcommand{\slot}[2]{\ensuremath{\mu^{\mathnormal{#1}}_{\mathnormal{[#2]}}}}
\newcommand{\slotvalue}[2]{\ensuremath{\mu^{\mathnormal{#1}}_{\mathnormal{#2}}}}

\newcommand{\tevalsym}{\tau}
\newcommand{\teval}[1]{\ensuremath{\tevalsym{#1}}}

\DeclareRobustCommand{\n}[1]{%
   \ifthenelse{ \equal{#1}{} }%
      {\ensuremath{\nu}}%
      {\ensuremath{\ensuremath{\nu(#1)}}}%
}

\newcommand{\optsubscript}[3][]{%
   \ifthenelse{ \equal{#1}{} }
      {\ifthenelse{ \equal{#3}{} }
        {\ensuremath{#2}}
        {\ensuremath{#2^{#3}}}}%
      {\ifthenelse{ \equal{#3}{} }
        {\ensuremath{#2_{#1}}}
        {\ensuremath{#2^{#3}_{#1}}}}%
}

\newcommand{\A}[2][]{\optsubscript[#1]{\Sigma}{#2}}
\newcommand{\B}[2][]{\optsubscript[#1]{\Lambda}{#2}}
\newcommand{\T}[2][]{\optsubscript[#1]{\Gamma}{#2}}

\newcommand{\D}[2]{\ensuremath{\partial_{\mathnormal{#1}}#2}}
\newcommand{\V}[2]{\ensuremath{\nabla_{\mathnormal{#1}}#2}}

\newcommand{\optsupscript}[3][]{%
   \ifthenelse{ \equal{#1}{} }
      {\ifthenelse{ \equal{#3}{} }
        {\ensuremath{#2}}
        {\ensuremath{#2_{#3}}}}%
      {\ifthenelse{ \equal{#3}{} }
        {\ensuremath{#2^{#1}}}
        {\ensuremath{#2^{#1}_{#3}}}}%
}

\renewcommand{\L}[2][]{\optsupscript[#1]{L}{#2}}
\newcommand{\R}[2][]{\optsupscript[#1]{R}{#2}}

\DeclareRobustCommand{\d}[2][]{%
  \ifthenelse{ \equal{#1}{} }
    {\ifthenelse{ \equal{#2}{} }
      {\ensuremath{\delta}}
      {\ensuremath{\delta({#2})}}%
    }%
    {\ifthenelse{ \equal{#2}{} }
      {\ensuremath{\delta^{#1}}}
      {\ensuremath{\delta^{#1}({#2})}}%
    }%
}

\newcommand{\rx}[1]{\ensuremath{\mathsf{#1}}}
\newcommand{\comp}[1]{\ensuremath{\neg#1}}
\newcommand{\rxlang}[1]{\llbracket\rx{#1}\rrbracket}
\newcommand{\nt}[1]{\emph{#1}}
\DeclareRobustCommand{\dca}[1][]{%
  \ifthenelse{ \equal{#1}{} }
    {\ensuremath{C}}
    {\ensuremath{C({#1})}}%
}

\newcommand{\itemlabel}[1]{\textnormal{\textit{#1}}}
\newcommand{\rxstrut}{\rule[-2mm]{0mm}{7mm}}

\tikzset{every picture/.style={%
    line width=0.90pt,%
    >=Latex,%
    -{Latex},%
    node distance=2.3cm,%
    on grid,%
    auto,%
    initial text=}}

\algloopdefx{Return}[1]{\textbf{return} #1}

\begin{document}

\title{\vspace{-2cm}\sffamily\bfseries Regular languages, derivatives and finite automata}
\author{\sffamily Ola Wingbrant}
\date{\sffamily\today}


\maketitle
\thispagestyle{empty}

\begin{abstract}
  This report is mostly written for educational purposes. It is meant
  as a self contained introduction to regular languages, regular
  expressions, and regular expression matching by using Brzozowski
  derivatives. As such it is mostly based on the work by
  Brzozowski\autocite{brz} and Owens et al.\autocite{owens} The
  language basics material have been inspired by
  books\autocite{aho1986compilers} and web material\autocite{cs390}.

  Chapter \refchp{chp-fundamentals} introduces the fundamental
  concepts of formal languages, as well as the idea of string
  derivatives. In chapter \refchp{chp-regular-languages} we define the
  class of regular languages, and further develops the theory of
  derivatives for that class. We use derivatives to prove the
  Myhill-Nerod theorem, the Pumping lemma, and the closure of regular
  languages under all Boolean connectives. In chapter
  \refchp{chp-regular-expressions} we introduce regular expressions
  and regular expression matching. Chapter
  \refchp{chp-finite-automata} connects the theory of regular
  languages and derivatives with that of finite automata. Chapter
  \refchp{chp-anchors} looks at the concept of anchors, and how this
  can be incorporated into a matcher based on derivatives. Chapter
  \refchp{chp-submatching} discusses submatching using derivatives
  with an approach inspired by Laurikari and his work on tagged
  transitions\autocite{laurikari-2}. This is the part we consider as
  our main contribution to the field. In the last chapter, chapter
  \refchp{chp-putting-it-together}, we summarize by giving a regular
  expression matching algorithm using the previously discussed
  techniques. We also discuss related work by others.
\end{abstract}

\tableofcontents

\chapter{Language fundamentals}\label{chp-fundamentals}

In this chapter we define the basic elements that we are going to work
with, alphabet, strings, and languages, as well as proving a set of
properties for these elements. These properties will come handy in the
following chapters when we discuss a more specialized class of
languages.

\section{Alphabets and strings}

\begin{definition}[Alphabet] \index{alphabet}
  An \define{alphabet} is any, possibly infinite, set of symbols. We
  will use \A{} to denote an alphabet with a non-empty, finite set of
  symbols.
\end{definition}

We are interested in studying sequences of symbols drawn from some
alphabet \A{}. We therefore make the following definitions.

\begin{definition}[String]\index{string}
  A \define{string} $s$ over some alphabet \A{} is a, possibly
  infinite, sequence of symbols $s=a_1a_2\dots a_i\dots$, with
  $a_i\in\A{}$. We note the special case of a string with no symbols,
  called the \define{empty string}, and denote it by \emptystring.
\end{definition}

\begin{definition}[Exponentiation] \index{language!exponentiation}
  For an alphabet \A{} we define the \define{exponentiation}, or
  powers of \A{} by
  \begin{enumerate}
    \item $\A{0} = \set{\emptystring}$
    \item $\A{n+1} = \A{n}\A{} = \set{sa : s\in\A{n}, a\in\A{}}$
      $n\in\N$
  \end{enumerate}
  Hence, the elements $s\in\A{n}$ for some $n\in\N$ are sequences
  of symbols drawn from \A{}. We may also note that formally
  \A{} and \A{1} are distinct. \A{} is a set of
  symbols, whereas \A{1} is a set of sequences where each
  sequence is one symbol.
\end{definition}

Although strings may be infinite, for our purposes we will only
concern ourselves with finite strings. For any finite string $s$, it
is clear that $s \in \A{n}$ for some $n$.

\begin{definition}[String length]\index{string!length}
  Let $s\in\A{n}$ be a string. We say that the \define{length} of
  $s$ is $n$, written $\length{s} = n$, and hence the length is the
  number of consecutive symbols. As a special case we have
  $\length{\emptystring} = 0$.
\end{definition}

\begin{definition}[Kleene closure]\index{Kleene closure}
  \index{operator!closure}\index{*}
  Let \A{} be an alphabet, then we denote the set of all finite
  strings over \A{} by \A{*}.
\end{definition}

\begin{theorem}\label{thm-A-kleene-eq-A-inf}
  For any finite alphabet \A{} the following holds. 
  \begin{equation}
    \A{*} = \Union_{n\in\N}{\A{n}}
  \end{equation}
\end{theorem}

\begin{proof}
  Let $s\in\A{*}$ be a string. By definition $s$ is
  finite. Suppose $\length{s} = m$, then $s\in\A{m} \subseteq
  \Union_{n\in\N}{\A{n}}$. Since $s$ is arbitrary it follows
  that
  \begin{equation}
    \A{*} \subseteq \Union_{n\in\N}{\A{n}}.
  \end{equation}
  Conversely, suppose that $s\in\Union_{n\in\N}{\A{n}}$. Then there
  is an $m\in\N$ such that $s\in\A{m}$. Thus we have $\length{s} = m$,
  and $s$ is finite and therefore $s\in\A{*}$. Again, since $s$ is
  arbitrary it follows that
  \begin{equation}
    \Union_{n\in\N}{\A{n}} \subseteq \A{*} 
  \end{equation}
  and the theorem follows.
\end{proof}

\begin{definition}[Concatenation]\index{string!concatenation}
  Suppose that $s\in\A{m}$ and $t\in\A{n}$ are strings over some
  alphabet. The \define{concatenation} of $s$ and $t$ written $s \cdot
  t$ or $st$, is the string formed by letting the sequence of symbols
  in $s$ be followed by the sequence of symbols in $t$, i.e.

  \begin{equation}
    s \cdot t =
    a_1a_2...a_m \cdot b_1b_2...b_n =
    a_1a_2...a_mb_1b_2...b_n = st \in\A{m+n}
  \end{equation}
\end{definition}
We note that $s\cdot\emptystring=\emptystring\cdot s=s$. Hence
\emptystring\ is the identity element under concatenation.  We also
note that if $s\in\A{*}$ and $t\in\A{*}$ then $st\in\A{*}$, and hence
\A{*} is closed under string concatenation.

\begin{theorem}[Associative law]
  Let $s\in\A{k}$, $t\in\A{m}$ and $u\in\A{n}$, then
  \begin{equation}
    s \cdot (t \cdot u) = (s \cdot t) \cdot u
  \end{equation}
\end{theorem}

\begin{proof}
  It follows directly from the definition that
  \begin{eqnarray}
    && s \cdot (t \cdot u) \nonumber\\
    && \quad = a_1a_2...a_k \cdot (b_1b_2...b_m \cdot c_1c_2...c_n) \nonumber\\
    && \quad = a_1a_2...a_kb_1b_2...b_mc_1c_2...c_n \nonumber\\
    && \quad = (a_1a_2...a_k \cdot b_1b_2...b_m) \cdot c_1c_2...c_n \nonumber\\
    && \quad = (s \cdot t) \cdot u 
  \end{eqnarray}
\end{proof}

\begin{definition}(Substring)
  Suppose that $s,t,u,v$ are strings such that $s = tuv$, then $u$ is
  called a \define{substring} of $s$. Further, if at least one of $t$
  and $v$ is not \emptystring\ then $u$ is called a \define{proper
    substring} of $s$.
\end{definition}

\begin{definition}(Prefix)
  Suppose that $s,t,u$ are strings such that $s = tu$, then $t$ is
  called a \define{prefix} of $s$. Further, $t$ is called a
  \define{proper prefix} of $s$ if $u\neq\emptystring$,
 
\end{definition}

\begin{definition}(Suffix)
  Suppose that $s,t,u$ are strings such that $s = tu$, then $u$ is
  called a \define{suffix} of $s$. Further, $u$ is called a
  \define{proper suffix} of $s$ if $t\neq\emptystring$
\end{definition}

\section{Languages}

\begin{definition}[Language]
  A language \L{} over some alphabet \A{} is a subset of \A{*},
  i.e. $\L{} \subseteq \A{*}$.
\end{definition}
Since languages are sets of strings, all set operations can be applied
to languages, and we will still have a language. So we can use
e.g. union, intersection and complement to create new languages. The
complement of a language over \A{} is with respect to \A{*},
i.e. $\inv{\L{}} = \A{*}\setminus\L{}$.

Some useful properties of sets that we leave without
proofs \footnote{Proofs can be found in any textbook on basic set
  theory, abstract algebra, logic, or discrete mathematics.} are as
follows.

\begin{proposition}\label{prp-set-fundamentals}
  Let $P$, $Q$ and $R$ be sets.
  \begin{eqnarray}
    P \union P = P & & P \intersection P = P \\
    P \union \emptyset = P & & P \intersection \emptyset = \emptyset \\
    P \union Q = Q \union P & & P \intersection Q = Q \intersection P \\
    P \union (Q \union R) &=& (P \union Q) \union R \\
    P \intersection (Q \intersection R) &=& (P \intersection Q) \intersection R  \\
    P \union (Q \intersection R) &=& (P \union Q) \intersection (P \union R) \\
    P \intersection (Q \union R) &=& (P \intersection Q) \union (P \intersection R)
  \end{eqnarray}
\end{proposition}

\begin{proposition}\label{prp-de-morgans-law}
  We also state the well known De Morgan's Laws.
  \begin{eqnarray}
    \inv{(P \union Q)} &=& \inv{P} \intersection \inv{Q} \\
    \inv{(P \intersection Q)} &=& \inv{P} \union \inv{Q} 
  \end{eqnarray}
\end{proposition}

\begin{definition}[Concatenation]
  Let \L{1} and \L{2} be languages. The \define{concatenation} of
  \L{1} and \L{2}, written $\L{1}\cdot\L{2}$, or \L{1}\L{2} is defined by

  \begin{equation}
    \L{1}\L{2} = \set{s \cdot t=st : s \in \L{1}, t \in \L{2}}.
  \end{equation}
\end{definition}
Hence the concatenation of two languages is formed by concatenating
all strings in the first language with all strings in the second.

\begin{theorem}[Laws of concatenation]
  \begin{eqnarray}
    \L{1}\cdot\emptyset = \emptyset\cdot\L{1}
      & = & \emptyset \quad \mbox{nullify law} \\
    \L{1}(\L{2}\L{3}) & = & (\L{1}\L{2})\L{3} \quad \mbox{associative law} \\
    \L{1}(\L{2}\star\L{3}) & = & \L{1}\L{2}\star\L{1}\L{3}
    \quad \mbox{distributive law}
  \end{eqnarray}
  where $\star$ denotes either union or intersection.
\end{theorem}

\begin{proof}
  The nullify law follows directly from the definition.
  \begin{eqnarray}
    \L{1}\cdot\emptyset &=& \set{s \cdot t : s\in\L{1}, t \in \emptyset} = \emptyset\\
    \emptyset\cdot\L{1} &=& \set{s \cdot t : s\in\emptyset, t \in \L{1}} = \emptyset
  \end{eqnarray}
  The associative law follows from the definition and the associative
  law for strings,
  \begin{eqnarray}
    && \L{1}(\L{2}\L{3})
    = \set{s \cdot (t \cdot u) : s \in \L{1}, t \in \L{2}, u \in \L{3}} \nonumber\\
    && \quad =
    \set{(s \cdot t) \cdot u : s \in \L{1}, t \in \L{2}, u \in \L{3}}
    = (\L{1}\L{2})\L{3}.
  \end{eqnarray}
  Likewise, for the distributive law we have,
  \begin{eqnarray}
    && \L{1}(\L{2}\star\L{3})
    = \{st : s\in\L{1}, t \in \L{2}\star\L{3}\} \nonumber\\
    && \quad = \set{st : st \in \L{1}\L{2}\star\L{1}\L{3}}
    = \L{1}\L{2}\star\L{1}\L{3}
  \end{eqnarray}
\end{proof}

Similar to operations on alphabets, we also define exponentiation, or
powers of a language, and the Kleene closure of a language.

\begin{definition}[Exponentiation]
  Let \L{} be a language. The \define{exponentiation} or powers of
  \L{} is defined by
  \begin{enumerate}
    \item $\L[0]{} = \set{\emptystring}$
    \item $\L[n+1]{} = \L[n]{}\L{} \quad n\in\N$
  \end{enumerate}
\end{definition}

\begin{definition}[Kleene closure]
  Let \L{} be a language. \L[*]{} is defined by
  \begin{enumerate}
    \item $\emptystring \in \L[*]{}$
    \item For any $s \in \L[*]{}$ and $t \in \L{}$, $st \in \L[*]{}$ 
    \item Nothing else is in \L[*]{}
  \end{enumerate}  
\end{definition}
As will be clear from \refthm{thm-lstar-extraction}, \L[*]{} is the
concatenation of zero or more strings from \L{}. We also note that
\L[*]{} is closed under string concatenation, hence the name.

\section{Properties of the Kleene closure}

Note that these properties are valid for all languages over a finite
alphabet \A{}, specifically they are valid for \A{1}.
\begin{theorem}\label{thm-l-subeq-lstar}
  For any $n\in\N$
  \begin{equation}\label{eqn-l-subeq-lstar}
    \L[n]{} \subseteq \L[*]{}
  \end{equation}  
\end{theorem}

\begin{proof}
  The proof is by induction over n. Since $\L[0]{}=\set{\emptystring}$,
  by definition, and $\emptystring\in\L[*]{}$ then
  $\L[0]{}\subseteq\L[*]{}$. Assume that \refeqn{eqn-l-subeq-lstar}
  holds for $n=k$, and let $s\in\L[k+1]{}$. Then by definition there
  exist $t\in\L[k]{}\subseteq\L[*]{}, u\in\L{}$ such that $s=tu$. But
  then $s\in\L[*]{}$ by the second step in the definition of
  \L[*]{}. Thus we have that $\L[k+1]{}\subseteq\L[*]{}$ and the
  theorem follows.
\end{proof}

\begin{theorem}\label{thm-lstar-eq-lunion}
  \begin{equation}
    \L[*]{} = \Union_{n\in\N} \L[n]{}
  \end{equation}
\end{theorem}

\begin{proof}
  Let $s\in\Union_{n\in\N}\L[n]{}$. Then $s\in\L[k]{}$ for some
  $k\in\N$. By \refthm{thm-l-subeq-lstar} we have that
  $\L[k]{}\subseteq\L[*]{}$. Since $s$ is arbitrary, it follows that
  \begin{equation}
    \Union_{n\in\N}\L[n]{}\subseteq\L[*]{}
  \end{equation}
  The proof of the reverse relation is by induction. By definition
  $\emptystring\in\L[*]{}$ and $\emptystring\in\L[0]{}$. Assume that for some
  $s\in\L[*]{}$ it also holds that $s\in\Union_{n\in\N}\L[n]{}$. Then
  $s\in\L[k]{}$ for some $k\in\N$. Let $t\in\L{}$, then $st\in\L[*]{}$
  by definition. But then also $st\in\L[k+1]{}$ by definition of
  \L[n]{}. Therefore $st\in\Union_{n\in\N}\L[n]{}$. It follows that
  \begin{equation}
    \L[*]{}\subseteq\Union_{n\in\N}\L[n]{}
  \end{equation}
  and the theorem follows.
\end{proof}

\begin{theorem}[Extraction]\label{thm-lstar-extraction}
  \begin{equation}
    \L[*]{} = \set{\emptystring}\union\L{}\L[*]{} \\
  \end{equation}
\end{theorem}

\begin{proof}
  Let $s\in\L[*]{}$. By the proof of \refthm{thm-lstar-eq-lunion} it
  is clear that $s\in\L[k]{}$ for some $k\in\N$. If $k=0$ then
  $s=\emptystring$ and $s\in\set{\emptystring}\union\L{}\L[*]{}$. Now,
  assume $k>0$, then $s\in\L{}\L[k-1]{}\subseteq\L{}\L[*]{}$ again by
  \refthm{thm-lstar-eq-lunion}. Since $s$ is arbitrary it follows that
  \begin{equation}
    \L[*]{} \subseteq\set{\emptystring}\union\L{}\L[*]{}
  \end{equation}
  Conversely, let $s\in\set{\emptystring}\union\L{}\L[*]{}$. If
  $s=\emptystring$, then $s\in\L[*]{}$. Assume
  $s\in\L{}\L[*]{}$. Again by the proof of
  \refthm{thm-lstar-eq-lunion} it is clear that $s\in\L{}\L[k-1]{} =
  \L[k]{}$ for some $k\in\N$. By \refthm{thm-l-subeq-lstar}
  $\L[k]{}\subseteq\L[*]{}$. Hence, since $s$ is arbitrary we have
  that
  \begin{equation}
    \set{\emptystring}\union\L{}\L[*]{} \subseteq \L[*]{}
  \end{equation}
  and the theorem follows.
\end{proof}

\begin{corollary}
  If $\emptystring\in\L{}$ then $\L[*]{} = \L{}\L[*]{}$.
\end{corollary}

\begin{theorem}[Jumping star]
  \begin{equation}
    \L{}\L[*]{} = \L[*]{}\L{}
  \end{equation}
\end{theorem}

\begin{proof}
  By \refthm{thm-lstar-eq-lunion} we have
  \begin{equation}
    \L{}\L[*]{} = \L{}\Union_{n\in\N} \L[n]{} = \Union_{n\in\N} \L[n+1]{} =
    \Union_{n\in\N} \L[n]{}\L{} = \L[*]{}\L{}.
  \end{equation}
\end{proof}

\begin{theorem}[Idempotence]
  \begin{equation}
    \L[*]{} = (\L[*]{})^*
  \end{equation}
\end{theorem}

\begin{proof}
  Let $K = \L[*]{}$, then by \refthm{thm-l-subeq-lstar} we have that
  \begin{equation}
    \L[*]{} = K \subseteq K^* = (\L[*]{})^*
  \end{equation}
  We prove the converse relation $(\L[*]{})^* \subseteq \L[*]{}$ as
  follows:

  Let $s \in K^* = (\L[*]{})^*$. By \refthm{thm-lstar-eq-lunion}
  $K^* = \Union_{n\in\N}K^n$, and $s \in K^m$ for some $m\in\N$. Thus
  $s=t_1t_2...t_m$ with $t_i \in K, 1\le i \le m$. However, since $K =
  \L[*]{}$ it means that $t_i\in\L[*]{}$, and therefore for each $i$,
  $t_i\in\L[j_i]{}$ for some $j_i\in\N$. Hence
  $s\in\L[j_1]{}\L[j_2]{}...\L[j_m]{}\subseteq\L[*]{}$. Since
  $s\in(\L[*]{})^*$ is arbitrary, $(\L[*]{})^*\subseteq\L[*]{}$, and
  the theorem follows.
\end{proof}

\begin{corollary}[Absorbtion]\label{cor-star-absorbtion}
  $\L[*]{} = \L[*]{}\L[*]{}$ since $\L[*]{} = (\L[*]{})^* =
  \set{\emptystring}\union\L[*]{}(\L[*]{})^* = \L[*]{}(\L[*]{})^* =
  \L[*]{}\L[*]{}$
\end{corollary}

The following theorem\footnote{This is an adaption. In his original
  paper Arden proved the statement for $X = XL_1 \union L_2$. The
  proof given here is an adaption of the proof given by
  Verma \autocite{verma}.}, known as Arden's Rule or Arden's
Lemma\autocite{arden}, gives us a way to exchange a recurrence for the
closure operation.

\begin{theorem}[Arden's Rule]\label{thm-ardens-rule}
  Let $L_1\subseteq\A{*}$, $L_2\subseteq\A{*}$ and
  $X\subseteq\A{*}$ be languages. The equation
  \begin{equation}
    X = L_1X \union L_2
  \end{equation}
  has a solution
  \begin{equation}
    X = L_1^*L_2.
  \end{equation}
  Furthermore, if $\emptystring \notin L_1$ then this solution is unique.
\end{theorem}

\begin{proof}
  It is easy to verify that $X=L_1^*L_2$ satisfies the equation. For
  the left-hand side we trivially have $X=L_1^*L_2$. For the
  right-hand side we have
  \begin{equation}
    L_1X \union L_2 = L_1L_1^*L_2 \union L_2 =
    (L_1L_1^* \union \emptystring)L_2 = L_1^*L_2.
  \end{equation}
  In order to establish uniqueness we expand the right-hand side by
  recursively putting in the value of $X$ over and over again.
  \begin{eqnarray*}
    X & = & L_1X \union L_2 \\  
    & = & L_1(L_1X \union L_2) \union L_2 \\
    & = & L_1(L_1(L_1X \union L_2) \union L_2) \union L_2 \\  
    & \vdots & \\
    & = & L_1^{n+1}X \union L_1^{n}L_2 \union L_1^{n-1}L_2 \union \cdots
          \union L_1L_2 \union L_2 \\
    & = & L_1^{n+1}X \union (L_1^{n} \union
          L_1^{n-1} \union \cdots \union L_1 \union \emptystring)L_2 \\
    & = & L_1^{n+1}X \union \left(\Union_{k=0}^{n} L_1^{k}\right)L_2
  \end{eqnarray*}
  where $n$ is arbitrary. Now, let $s\in X$, and we can assume that
  $\length{s} = m$. Substituting $m$ for $n$ in the above equation we get
  \begin{equation}
    X = L_1^{m+1}X \union \left(\Union_{k=0}^{m} L_1^{k}\right)L_2.
  \end{equation}
  Given that $\emptystring \notin L_1$ we have $s\notin L_1^{m+1}X$
  since the length of the shortest string contained in $L_1^{m+1}X$
  will be at least $m+1$. But $s\in X$ by choice, therefore $s\in
  \left(\Union_{k=0}^{m} L_1^{k}\right)L_2.$

  Conversely, let $s\in L_1^*L_2$, then there must be an $m$ such
  that $s\in \left(\Union_{k=0}^{m} L_1^{k}\right)L_2$, but then
  $s\in L_1^{m+1}X \union \left(\Union_{k=0}^{m} L_1^{k}\right)L_2 = X$.
\end{proof}

\begin{example}
  If the condition that $\emptystring \notin L_1$ does not hold in
  Arden's Rule, i.e. $\emptystring \in L_1$, it is easy to find other
  solutions, e.g. \A{*} being one.
  \begin{displaymath}
    L_1\A{*} \union L_2 = \left[s\in L_1 \Rightarrow L_1\A{*}
    = \A{*}\right] = \A{*} \union L_2 = \A{*}
  \end{displaymath}
\end{example}

\section{Nullable languages and the nullify function}

\begin{definition}[Nullable]
  A language \L{} is said to be \define{nullable} if
  $\emptystring\in\L{}$, and we define the \define{nullify} function \n{} by
  \begin{equation}
    \n{\L{}} = \left\{ \begin{array}{cl}
      \set{\emptystring} & \mbox{if } \emptystring\in\L{} \\
      \emptyset & \mbox{if } \emptystring\notin\L{} 
    \end{array}\right.
  \end{equation}
\end{definition}

\noindent It is clear from the definition that
\begin{eqnarray}
  \n{\emptyset} &=& \emptyset \\
  \n{\set{\emptystring}} &=& \set{\emptystring} \\
  \n{\L[*]{}} &=& \set{\emptystring}
\end{eqnarray}

\begin{theorem}
  Let \L{}, \L{1} and \L{2} be languages. Then
  \begin{eqnarray}
    \n{\L{1}\union\L{2}} &=& \n{\L{1}}\union\n{\L{2}} \\
    \n{\L{1}\intersection\L{2}} &=& \n{\L{1}}\intersection\n{\L{2}} \\
    \n{\L{1}\L{2}} &=& \n{\L{1}}\n{\L{2}} \\
    \n{\inv{\L{}}} &=& \left\{ \begin{array}{cl}
      \set{\emptystring} & \mbox{if } \n{\L{}} = \emptyset \\
      \emptyset & \mbox{if } \n{\L{}} = \set{\emptystring}
    \end{array}\right.
  \end{eqnarray}
\end{theorem}

\begin{proof}
  They all follow easily from the definition and noting the different
  cases when the languages contain or do not contain \emptystring.
\end{proof}

\section{Derivatives}

The main point in this section will be to derive a criterion to decide
whether a string is contained in a language or not.

\begin{definition}[String derivative]\label{def-derivative}
  The \define{derivative} of a language $\L{}\subseteq\A{*}$ with
  respect to a string $s\in\A{*}$ is defined to be
  \begin{equation}
    \D{s}{\L{}} = \set{t : s \cdot t \in\L{}}
  \end{equation}
\end{definition}

The following theorem gives us a way to calculate the
derivative recursively.\autocite[Theorem 3.2]{brz}

\begin{theorem}\label{thm-rec-derivative}
  \begin{eqnarray}
    \D{\emptystring}{\L{}} & = & \L{} \\
    \D{sa}{\L{}} & = & \D{a}{(\D{s}{\L{}})} \quad s\in\A{*}, a\in\A{1} 
  \end{eqnarray}
\end{theorem}

\begin{proof}
  Both properties are immediate from the definition.
\end{proof}

\begin{theorem}\label{thm-derivative}
  Rules \autocite[Theorem 3.1]{brz} for computing the derivative with
  respect to a string $a\in\A{1}$.
  \begin{eqnarray} 
    \D{a}{\emptyset} &=& \emptyset \label{eqn-Da0} \\
    \D{a}{\set{\emptystring}} &=& \emptyset \label{eqn-Dae} \\
    \D{a}{\set{a}} &=& \set{\emptystring} \label{eqn-Daa} \\
    \D{a}{\set{b}} &=& \emptyset \;\: \mbox{for} \;\: a \neq b \in \A{1}
      \label{eqn-Dab} \\
    \D{a}{(\L{1}\union\L{2})} &=& \D{a}{\L{1}}\union\D{a}{\L{2}} \label{eqn-DaOr} \\
    \D{a}{(\L{1}\intersection\L{2})} &=& \D{a}{\L{1}}\intersection\D{a}{\L{2}}
      \label{eqn-DaAnd} \\
    \D{a}{(\L{1}\cdot\L{2})} &=&
      \D{a}{\L{1}}\cdot\L{2}\union\n{\L{1}}\cdot\D{a}{\L{2}} \label{eqn-DaCat} \\
    \D{a}{(\L[*]{})} &=& \D{a}{\L{}}\cdot\L[*]{} \label{eqn-DaStar} \\
    \D{a}{(\inv{\L{}})} &=& \inv{(\D{a}{\L{}})} \label{eqn-DaInv}
  \end{eqnarray}
\end{theorem}

\begin{proof}
  We proof each equation in turn.
  \begin{itemize}
  \item[\refeqn{eqn-Da0}]
    $\D{a}{\emptyset} = \{s:a \cdot s\in\emptyset\} = \emptyset$
  \item[\refeqn{eqn-Dae}]
    $\D{a}\{\emptystring\} = \{s:a \cdot s\in\{\emptystring\}\} = \emptyset$
  \item[\refeqn{eqn-Daa}]
    $\D{a}\{a\} = \{s:a \cdot s\in\{a\}\} = \{\emptystring\}$
  \item[\refeqn{eqn-Dab}]
    $\D{a}\{b\} = \{s:a \cdot s\in\{b\}\} = \emptyset$ since $a \neq b$.
  \item[\refeqn{eqn-DaOr}]
    \begin{eqnarray*}
       \D{a}{(\L{1}\union\L{2})} &=& \{s:a \cdot s\in(\L{1}\union\L{2})\} \\
       &=& \{s:a \cdot s\in\L{1}\}\union\{t:a \cdot t\in\L{2})\}
       = \D{a}{\L{1}}\union\D{a}{\L{2}}
    \end{eqnarray*}
  \item[\refeqn{eqn-DaAnd}]
    \begin{eqnarray*}
       \D{a}{(\L{1}\intersection\L{2})} &=& \{s:a \cdot t\in(\L{1}\intersection\L{2})\} \\
       &=& \{s:a \cdot s\in\L{1}\}\intersection\{t:a \cdot t\in\L{2}\}
       = \D{a}{\L{1}}\intersection\D{a}{\L{2}}
    \end{eqnarray*}
  \item[\refeqn{eqn-DaCat}] Let $\L{1} = \n{\L{1}}\union\L{0}$ where
    $\n{\L{0}} = \emptyset$. Then
    \begin{eqnarray*}
      \D{a}{(\L{1}\cdot\L{2})} &=& \{s:a \cdot
        s\in(\n{\L{1}}\union\L{0})\cdot\L{2}\} \\
      &=& \{s:a \cdot s\in \n{\L{1}}\cdot\L{2}\} \union
          \{t:a \cdot t\in \L{0}\cdot\L{2}\} \\
      &=& \n{\L{1}}\cdot\D{a}\L{2} \union
          \{t_0 \cdot t_2: a \cdot t_0\in\L{0}, t_2\in\L{2}\} \\
      &=& \n{\L{1}}\cdot\D{a}\L{2} \union
          \{t_0 : a \cdot t_0\in\L{0}\}\cdot\L{2} \\
      &=& \n{\L{1}}\cdot\D{a}\L{2} \union \D{a}{\L{0}}\cdot\L{2}
    \end{eqnarray*}
    But $\D{a}{\L{1}} = \D{a}{(\n{\L{1}}\union\L{0})} =
    \D{a}{\n{\L{1}}}\union\D{a}{\L{0}} = \D{a}{\L{0}}$. Hence, we have
    \begin{displaymath}
      \D{a}{(\L{1}\cdot\L{2})} = \n{\L{1}}\cdot\D{a}\L{2} \union
      \D{a}{\L{0}}\cdot\L{2} = \n{\L{1}}\cdot\D{a}\L{2} \union
      \D{a}{\L{1}}\cdot\L{2},
    \end{displaymath} which is equation \refeqn{eqn-DaCat}
  \item[\refeqn{eqn-DaStar}]
    Assume that
    \begin{equation}\label{eqn-DaCat-k}
      \D{a}{(\L[k]{})} = \D{a}{(\L{})}\L[k-1]{} \mbox{ for } k \ge 1.
    \end{equation}
    Then
    \begin{eqnarray}
      \D{a}{(\L[*]{})} &=& \D{a}{\Union_{k\in\N}\L[k]{}} =
      \D{a}{(\L[0]{})} \union \Union_{k\in\N \atop k \neq 0}\D{a}{(\L[k]{})} \nonumber\\
      &=& \emptyset \union \Union_{k\in\N \atop k \neq 0}\D{a}{(\L{})}\L[k-1]{} =
      \D{a}{(\L{})}\Union_{k\in\N}\L[k]{} = \D{a}{(\L{})}\L[*]{}
    \end{eqnarray}
    Which is \refeqn{eqn-DaStar}. Now in order to prove \refeqn{eqn-DaCat-k}
    we note that it is trivially true for $k=1$. Assume that it holds
    for $k=m$, then
    \begin{displaymath}
      \D{a}{(\L[m+1]{})} = \D{a}{(\L{})}\L[m]{} \union \n{\L{}}\D{a}{(\L[m]{})} =
      \D{a}{(\L{})}\L[m]{} \union \n{\L{}}\D{a}{(\L{})}\L[m-1]{}
    \end{displaymath}
    Thus, there are two cases.
    \begin{eqnarray*}
      \n{\L{}} = \emptyset &\Rightarrow& \D{a}{(\L[m+1]{})} = \D{a}{(\L{})}\L[m]{} \\
      \n{\L{}} = \{\emptystring\} &\Rightarrow& \D{a}{(\L[m+1]{})} =
      \D{a}{(\L{})}\L[m]{} \union \D{a}{(\L{})}\L[m-1]{} = \D{a}{(\L{})}\L[m]{}
    \end{eqnarray*}
    The last equality follows from the fact that $\L[m-1]{}\subseteq
    \L[m]{}$ when $\emptystring \in\L{}$. Hence, \refeqn{eqn-DaCat-k}
    holds and we are done.
  \item[\refeqn{eqn-DaInv}]
    We note that
    \begin{equation}
      \D{a}{(\L{})}\union\D{a}{(\inv{\L{}})} = \D{a}{(\L{}\union\inv{\L{}})} =
      \D{a}{\A{*}} = \D{a}{(\A{1})}\A{*} = \A{*}
    \end{equation}
    and that
    \begin{equation}
      \D{a}{(\L{})}\intersection\D{a}{(\inv{\L{}})} =
      \D{a}{(\L{}\intersection\inv{\L{}})} =
      \D{a}{\emptyset} = \emptyset
    \end{equation}
    These two equations taken together gives
    \begin{equation}
      \D{a}{(\inv{\L{}})} = \inv{(\D{a}{\L{}})}
    \end{equation}
  \end{itemize}
\end{proof}

\begin{theorem}\label{thm-derivative-decide}
  Let \L{} be a language, then \autocite[Theorem 4.2]{brz}
  \begin{equation}
    s\in\L{} \Leftrightarrow
    \emptystring\in\D{s}{\L{}} \label{eqn-derivative-decide}
  \end{equation}
\end{theorem}

\begin{proof}
  If $\emptystring\in\D{s}{\L{}}$ then, by \refdef{def-derivative},
  $s\cdot\emptystring\in\L{}$. Conversely, if $s\in\L{}$, then
  $s\cdot\emptystring\in\L{}$ and $\emptystring\in\D{s}{\L{}}$, again
  by \refdef{def-derivative}.
\end{proof}

What \refthm{thm-derivative-decide} says, is that deciding if
$s\in\L{}$ is equivalent to deciding if
$\emptystring\in\D{s}{\L{}}$. In section \refsec{sec-re-matching} the
following corollary will prove useful in order to build efficient
recognizers for a particular class of languages.

\begin{corollary}\label{cor-nL-eq-inL}
  $\n{\D{s}{\L{}}} = \{\emptystring\} \Leftrightarrow s\in\L{}$
\end{corollary}

\begin{theorem}\label{thm-derivative-expansion}
  Any language \L{} can be written in the form
  \begin{equation}\label{eqn-derivative-expansion}
    \L{} = \n{\L{}} \union  \Union_{a\in\A{1}}\set{a}\D{a}{\L{}}
  \end{equation}
  where the terms are disjoint.\autocite[Theorem 4.4]{brz}
\end{theorem}

\begin{proof}
  First \L{} may or may not contain \emptystring. This is taken care
  of by $\n{\L{}}$. If \L{} contains a string $s$, that string must
  begin with a prefix $a\in\A{1}$. In view of the definition of
  derivative, the set $\set{a}\D{a}{\L{}}$ is exactly the set of strings in
  \L{} with prefix $a$. The terms in the sum are obviously disjoint,
  for a string in one term begin with a prefix in \A{1} different from
  those in another term.
\end{proof}

\section{Derivative classes}\label{sec-derivative-classes}

\begin{definition}[Distingushing extension]
  Let $\L{}\subseteq\A{*}$ be a language, and $s,t\in\A{*}$ strings. A
  \define{distinguishing extension} is a string $u\in\A{*}$ such that
  either $su\in\L{}$ or $tu\in\L{}$, but not both.
\end{definition}

\begin{definition}\label{def-indistinguishable-strings}
  Define the relation \eq{\L{}}, ``\define{\L{}-equivalent}'', or
  ``\define{equivalent with respect to \L{}}'', on strings by the rule
  \begin{equation}
    s \eq{\L{}} t \Leftrightarrow \set{u: su\in\L{}} = \set{u: tu\in\L{}},
  \end{equation}
  i.e. $s \eq{\L{}} t$ if there is no distinguishing extension for $s$
  and $t$. For future reference we note that by \refdef{def-derivative}
  \begin{equation}\label{eqn-indistinguishable-strings-eq-derivatives}
    s \eq{\L{}} t \Leftrightarrow \D{s}{\L{}} = \D{t}{\L{}}
  \end{equation}  
\end{definition}

\begin{theorem}
  \eq{\L{}} is an equivalence relation. 
\end{theorem}

\begin{proof}
  We need to show that \eq{\L{}} has the three properties of an
  equivalence relation.
  \begin{description}
  \item[\itemlabel{Reflexive:}] \eq{\L{}} is reflexive, since 
    \begin{equation}
      \D{s}{\L{}} = \D{s}{\L{}} \Leftrightarrow s \eq{\L{}} s
    \end{equation}
  \item[\itemlabel{Symmetric:}] It is also symmetric since
    \begin{equation}
      s \eq{\L{}} t
      \Leftrightarrow \D{s}{\L{}} = \D{t}{\L{}}
      \Leftrightarrow \D{t}{\L{}} = \D{s}{\L{}}
      \Leftrightarrow t \eq{\L{}} s
    \end{equation}
  \item[\itemlabel{Transitive:}] Suppose $r \eq{\L{}} s$ and $s
    \eq{\L{}} t$, then $r \eq{\L{}} t$ holds, since
    \begin{equation}
      r \eq{\L{}} s
      \Leftrightarrow \D{r}{\L{}} = \D{s}{\L{}} = \D{t}{\L{}}
      \Leftrightarrow r \eq{\L{}} t
    \end{equation}
  \end{description}
  We will denote the equivalence class of a string $s$ by
  $[s]_{\eq{\L{}}} = \set{t: s \eq{\L{}} t}$ or simply $[s]$ when it is
  clear from the context which relation we are talking about.
\end{proof}

\begin{corollary}
  \begin{equation}\label{eqn-eqclass-eq-derivative}
    [s]_{\eq{\L{}}} = \set{t: \D{s}{\L{}} = \D{t}{L}}
  \end{equation}
  In the special case when $\length{s} = 1$, i.e. $s = a
  \in \A{1}$ we have
  \begin{equation}
    [a]_{\eq{\L{}}} = \set{b: \D{a}{\L{}} = \D{b}{L}, b \in \A{1}}
  \end{equation}
  and we can write equation \refeqn{eqn-derivative-expansion} as
  \begin{equation}
    \L{} = \n{\L{}} \union  \Union_{a_k}[a_k]\D{a_k}{\L{}}
  \end{equation}
  where all \D{a_k}{\L{}} are disjoint and $[a_k]$ is the partitioning
  of \A{1} by the \eq{\L{}} relation.
\end{corollary}

\begin{definition}[Derivative class]
  We call $[s]_{\eq{\L{}}}$ the \define{derivative class} of \L{} with
  respect to $s$.
\end{definition}

Derivative classes can be useful. If we somehow know
$[s]_{\eq{\L{}}}$, equation \refeqn{eqn-eqclass-eq-derivative} tells
us that by calculating \D{s}{\L{}} we know the derivative \D{t}{\L{}}
for all $t \in [s]_{\eq{\L{}}}$. This property will come handy in
section \refsec{sec-sets-of-symbols} where we try to optimize the
construction of recognizers for a particular class of languages.

We end this section with a theorem regarding equivalence and composed
languages.\autocite[Lemma 4.1]{owens}

\begin{theorem}\label{thm-derivative-class-composition}
  Let \L{1} and \L{2} be languages over some alphabet \A{}, and let
  $s$ and $t$ be strings such that $s \eq{\L{1}} t$ and $s \eq{\L{2}}
  t$ then the following equations hold:
  \begin{eqnarray}
    && s \eq{\L{1}\L{2}} t \label{eqn-derivative-class-intersection}\\
    && s \eq{\L{1} \union \L{2}} t \\
    && s \eq{\L{1} \intersection \L{2}} t \\
    && s \eq{\L[*]{1}} t \\
    && s \eq{\inv{\L{1}}} t
  \end{eqnarray}
\end{theorem}

\begin{proof}
  We prove equation \refeqn{eqn-derivative-class-intersection}, the
  others follow by similar arguments. By equation
  \refeqn{eqn-indistinguishable-strings-eq-derivatives} we have $s
  \eq{\L{1}} t \Leftrightarrow \D{s}{\L{1}} = \D{t}{\L{1}}$ and $s
  \eq{\L{2}} t \Leftrightarrow \D{s}{\L{2}} = \D{t}{\L{2}}$. Hence
  \begin{eqnarray}
    \D{s}{(\L{1}\L{2})}
    &=& \D{s}{(\L{1})}\L{2} \union \n{\L{1}}\D{s}{(\L{2})} \nonumber\\
    &=& \D{t}{(\L{1})}\L{2} \union \n{\L{1}}\D{t}{(\L{2})} \nonumber\\
    &=& \D{t}{(\L{1}\L{2})}
  \end{eqnarray}
  Thus
  \begin{equation}
    \D{s}{(\L{1}\L{2})} = \D{t}{(\L{1}\L{2})}
      \Leftrightarrow s \eq{\L{1}\L{2}} t
  \end{equation}
\end{proof}

\chapter{Regular Languages}\label{chp-regular-languages}

We will now turn our interest to a special class of languages, called
regular languages. Regular languages are important in many
applications as they form the building blocks of more advanced
languages.

\begin{definition}[Regular language]\label{def-regular-language}
  Let \A{} be an alphabet. A \define{regular language} over \A{} is
  defined recursively by
  \begin{enumerate}
  \item $\emptyset$ and $\set{\emptystring}$ are regular languages.
  \item\label{eqn-regular-base} For each $a\in\A{1}$, \set{a} is a regular language.
  \item If \R{} is a regular language, then \R[*]{} is a regular
    language.
  \item If \R{1} and \R{2} are regular languages, then so is
    $\R{1}\R{2}$ and $\R{1}\union\R{2}$.
  \item No other languages over \A{} are regular.
  \end{enumerate}  
\end{definition}

\begin{theorem}
  If \R{} is a regular language over \A{}, then so is \D{s}{\R{}} with
  respect to any string\autocite[Theorem 4.1]{brz} $s\in\A{*}$.
\end{theorem}

\begin{proof}
  It is clear from \refthm{thm-rec-derivative} that
  \D{\emptystring}{\R{}} is regular, and that \D{s}{\R{}} is regular
  if \D{a}{\R{}} is regular for $a\in\A{1}$. The proof of
  \refthm{thm-derivative} shows that \D{a}{R} is regular, since only a
  finite number of regular operations are needed to calculate the
  derivative.
\end{proof}

\section{Characteristic equations}

In this section we will derive a set of equations, called the
characteristic equations, for a regular language \R{}. This in turn
will help us in the next section, where we extend the set of operators
in regular languages, and prove that regular languages are closed
under complement, and therefore under intersection and other Boolean
set operators as well.

\begin{theorem}\label{thm-regular-finite-derivative}
  Every regular language \R{} has a finite number $d_{\R{}}$ of
  different derivatives.\autocite[Theorem 4.3 a)]{brz}
\end{theorem}

\begin{proof}
  The proof is by induction over the number $n$ of regular
  operators. The theorem is true for \R{} with $n = 0$ since, with
  $a\in\A{1}$,
  \begin{center}
    \setlength{\tabcolsep}{0.3em}
    \renewcommand{\arraystretch}{1.25}
    \begin{tabular}{rcl}
      \D{s}{\emptyset} &=& $\emptyset$ for all $s\in\A{*}$ \\
      \D{\emptystring}{\set{\emptystring}} &=& $\set{\emptystring}$ and
        $\D{s}{\set{\emptystring}} = \emptyset$ for all $s\in\A{*},
        s\neq\emptystring$ \\
      \D{\emptystring}{\set{a}} &=& $\set{a}$.
        $\D{a}{\set{a}} = \set{\emptystring}$.
        $\D{s}{\set{a}} = \emptyset$ for all $s\in\A{*}, s\neq\emptystring, a$
    \end{tabular}
  \end{center}
  Hence we have $d_\emptyset = 1$, $d_\emptystring = 2$, and $d_a = 3$.

  Assume that every regular language constructed with $n$ or fewer
  operators has a finite number of different derivatives. If \R{} is a
  language constructed with $n+1$ operators, there are three cases.
  \begin{description}
  \item[\itemlabel{Case 1:}] $\R{} = \R{1}\union\R{2}$. We have $\D{s}{\R{}}
    =\D{s}{\R{1}}\union\D{s}{\R{2}}$. Thus $d_R \le d_{\R{1}}d_{\R{2}}$. 
  \item[\itemlabel{Case 2:}] $\R{} = \R{1}\R{2}$. Let $s=a_1a_2...a_k$. Then
    \begin{eqnarray*}
      \D{a_1}{\R{}} &=& \D{a_1}{(\R{1})}\R{2}\union\n{\R{1}}\D{a_1}{\R{2}}.\\
      \D{a_1a_2}{\R{}} &=& \D{a_1a_2}{(\R{1})}\R{2} \union
      \n{\D{a_1}{\R{1}}}\D{a_2}{\R{2}} \union
      \n{\R{1}}\D{a_1a_2}{\R{2}}.      
    \end{eqnarray*}
    In general the derivative with respect to a string $s$,
    $\length{s} = k$ will have the form
    \begin{eqnarray}
      && \D{a_1a_2...a_k}{\R{}} = \D{a_1a_2...a_k}{(\R{1})}\R{2} \union
      \n{\D{a_1a_2...a_{k-1}}{\R{1}}}\D{a_k}{\R{2}} \union \cdots \nonumber\\
      && \qquad \union \n{\D{a_1}{\R{1}}}\D{a_2...a_k}{\R{2}} \union
      \n{\R{1}}\D{a_1a_2...a_k}{\R{2}}\label{eqn-regular-finite-derivative-concat}
    \end{eqnarray}
    Thus \D{s}{\R{}} is the sum of $\D{s}{(\R{1})}\R{2}$ and at most
    $k$ derivatives of \R{2}. Therefore, if there are $d_{\R{1}}$
    derivatives of \R{1} and $d_{\R{2}}$ derivatives of \R{2}, there
    can be at most $d_{\R{}} \le d_{\R{1}}2^{d_{\R{2}}}$ different
    derivatives of \R{}.
  \item[\itemlabel{Case 3:}] $\R{} = \R[*]{0}$. Let us study the derivatives of \R[*]{0}.
    \begin{eqnarray*}
      \D{a_1}{\R[*]{0}} &=& \D{a_1}{(\R{0})}\R[*]{0} \\
      \D{a_1a_2}{\R[*]{0}} &=& \D{a_1a_2}{(\R{0})}\R[*]{0} \union
        \n{\D{a_1}{\R{0}}}\D{a_2}{\R[*]{0}} \\
      &=& \D{a_1a_2}{(\R{0})}\R[*]{0} \union
        \n{\D{a_1}{\R{0}}}\D{a_2}{(\R{0})}\R[*]{0} \\
      &\vdots&
    \end{eqnarray*}
    It is clear that \D{s}{\R{}} will be a sum of terms on the form
    $\D{t}{(\R{0})}\R[*]{0}$. If \R{0} has $d_{\R{0}}$ different
    derivatives, then \R{} will have at most $d_{\R{}} \le
    2^{d_{\R{0}}}$ different derivatives.\qedhere
  \end{description}
\end{proof}

\begin{theorem}\label{thm-cd-eqn-regular}
  Any language \L{} that has a finite number $d_{\L{}}$ of different
  derivatives is regular.
\end{theorem}

Before we give the proof we need to show two other results for
languages with a finite number of different derivatives.

\begin{theorem}\label{thm-char-derivative-length}
  Let \L{} be a language with a finite number $d_{\L{}}$ of different
  derivatives. Then each of the different derivatives occurs at least
  once\autocite[Theorem 4.3 b)]{brz} among the derivatives with
  respect to strings $s\in\Union_{k=0}^{d_{\L{}}-1}\A{k}$, i.e
  $s\in\A{*}$ such that $\length{s} \le d_{\L{}}-1$. These are called
  the \define{characteristic derivatives} of \L{}.
\end{theorem}

\begin{proof}
  Let $a_1, a_2,\cdots,a_n$ be some enumeration of the symbols in
  \A{}. We can now arrange the strings of \A{*} according to their
  length and content, starting with $\emptystring, a_1,
  a_2,\cdots,a_n,a_1a_1, a_1a_2,\cdots,a_1a_n,\cdots$.

  The derivatives are now found by taking the derivatives of \L{} with
  respect to the strings in the above order,
  i.e. \D{\emptystring}{\L{}}, \D{a_1}{\L{}}, \D{a_2}{\L{}},
  $\cdots$. If no new derivatives are found for strings $s$ with
  $\length{s} = k$, the process terminates. For, if no new derivative
  is found for $\length{s} = k$, then every \D{s}{\L{}} is equal to
  another derivative \D{t}{\L{}} with $\length{t} < k$. Consider
  $\D{sa}{\L{}} = \D{a}{(\D{s}{\L{}})} = \D{a}{(\D{t}{\L{}})} =
  \D{ta}{\L{}}$, where $a\in\A{1}$ and $\length{ta} < k+1$. Hence every
  derivative with respect to a string $sa$, $\length{sa} = k+1$ will
  be equal to some derivative with respect to a string $ta$,
  $\length{ta} < k+1$. Therefore, if no new derivatives are found for
  $\length{s} = k$, no new derivatives will be found for $\length{s} =
  k+1$, and the process can be terminated.
\end{proof}

\begin{theorem}\label{thm-cd-eqn}
  Let \L{} be a language with a finite number $d_{\L{}}$ of different
  derivatives. The relationship between the $d_{\L{}}$ characteristic derivatives
  of \L{} can be represented by a unique set of $d_{\L{}}$ equations
  of the form
  \begin{equation}\label{eqn-cd-eqn}
    \D{s}{\L{}} = \n{\D{s}{\L{}}} \union
    \Union_{a\in\A{1}}\set{a}\D{t_a}{\L{}},
  \end{equation}
  where \D{s}{\L{}} is a characteristic derivative and \D{t_a}{\L{}}
  is a characteristic derivative equal to \D{sa}{\L{}}. Such equations
  will be called the \define{characteristic equations} of \L{}.
\end{theorem}

\begin{proof}
  The theorem follows directly from
  \refthm{thm-derivative-expansion} and
  \refthm{thm-char-derivative-length}.
\end{proof}

We are now ready to give the proof of
\refthm{thm-cd-eqn-regular}\autocite[Theorem 4.7]{brz}.

\begin{proof}[Proof of \refthm{thm-cd-eqn-regular}]
  Since \L{} has a finite number of derivatives, \L{} has a set of
  characteristic equations. Assume that \D{s}{\L{}} is the last
  characteristic derivative, then
  \begin{eqnarray}\label{eqn-last-cd-eqn}
    \D{s}{\L{}}
    &=& \n{\D{s}{\L{}}} \union \Union_{a\in\A{1}}\set{a}\D{t_a}{\L{}} \nonumber\\
    &=& \underbrace{\n{\D{s}{\L{}}}
      \union \Union_{a\in\A{1} \setminus S}\set{a}\D{t_a}{\L{}}}_{\L{2}}
    \union \underbrace{\Union_{a\in S}\set{a}}_{\L{1}}\D{s}{\L{}} \nonumber\\
    &=& \L{1}\D{s}{\L{}} \union \L{2}
  \end{eqnarray}
  where $S = \{a : \D{s}{\L{}} = \set{a}\D{s}{\L{}}, a\in\A{1}\}$, and $t_a$
  are previous characteristic derivatives, i.e. we have split the
  terms in a recursive part, and a part consisting of previous
  derivatives. Obviously $\n{\L{1}} = \emptyset$, and we can apply
  Arden's rule (\refthm{thm-ardens-rule}) to equation
  \refeqn{eqn-last-cd-eqn}, we get
  \begin{equation}\label{eqn-last-cd-solution}
    \D{s}{\L{}} = \L[*]{1}\L{2} = \left(\Union_{a\in
      S}\set{a}\right)^* \left( \n{\D{s}{\L{}}} \union
    \Union_{a\in\A{1} \setminus S}\set{a}\D{t_a}{\L{}} \right)
  \end{equation}

  Thus, we can express the last characteristic derivative in terms of
  previous ones. The solution for this last derivative can then be
  substituted for in the first $d_{\L{}}-1$ equations, reducing the
  number of equations by 1. This process of elimination can then be
  repeated until the set of equations is solved for
  $\D{\emptystring}{\L{}} = \L{}$. This elimination also shows that
  \L{} is regular. All the coefficients in
  \refeqn{eqn-last-cd-solution} are regular, therefore the solution
  for \L{} will be an expression formed entirely of regular
  operations.
\end{proof}

\begin{corollary}\label{cor-cd-eqn-uniqueness}
  The set of characteristic equations can be solved for uniquely.
\end{corollary}

\begin{theorem}[Myhill-Nerode]\label{thm-derivative-myhill-nerode}
  A language \R{} is regular if, and only if, it has a finite number
  of different derivatives.
\end{theorem}

\begin{proof}
  Immediate from \refthm{thm-regular-finite-derivative} and
  \refthm{thm-cd-eqn-regular}
\end{proof}

As we will see in section \refsec{sec-non-regular-languages}
\refthm{thm-derivative-myhill-nerode} is the Myhill-Nerode theorem
formulated in terms of derivatives.

Why are the characteristic equations called characteristic? Because
they give information about the structure of the strings that are part
of the language. They characterize what strings are part of the
language, what substrings of those strings are repeatable, and what a
valid prefix and/or postfix look like. If $s\in\A{*}$ then
$\n{\D{s}{\R{}}}$ determines if $s\in\R{}$, and the other terms
describe what can follow $s$, expressed in terms of the characteristic
derivatives.

\begin{example}
  Suppose that $t\in\R{}$, then $\n{\D{t}{\R{}}} = \set{\emptystring}$ by
  \refcor{cor-nL-eq-inL}. It follows from
  \refthm{thm-regular-finite-derivative} that there exist a
  characteristic derivative \D{s}{\R{}} such that $\D{t}{\R{}} =
  \D{s}{\R{}}$. But then $\n{\D{s}{\R{}}} = \set{\emptystring}$ in
  \refeqn{eqn-cd-eqn} for that particular derivative. Thus, if
  $\n{\D{s}{\R{}}} = \set{\emptystring}$ for some characteristic
  derivative, then $\set{t:\D{t}{\R{}}=\D{s}{\R{}}}\subseteq\R{}$.
\end{example}

\begin{example}
  Study equation \refeqn{eqn-last-cd-solution}. \D{s}{\R{}} is the set
  of strings in \R{} with prefix $s$. Suppose $\n{\D{s}{\R{}}} =
  \set{\emptystring}$, then $s\in\R{}$ by
  \refthm{thm-derivative-decide}. Now, let $t\in\L[*]{1}$, then
  \begin{eqnarray}
    \D{st}{\R{}} &=& \D{t}{\D{s}{\R{}}} = \D{t}{(\L[*]{1}\L{2})} =
    \D{t}{(\L[*]{1}\L[*]{1}\L{2})} \nonumber\\
    &=& \D{t}{(\L[*]{1})}\L[*]{1}\L{2}\union rest =
    \D{t}{(\L[*]{1})}\D{s}{\R{}}\union rest
  \end{eqnarray}
  But then
  \begin{equation}
    \n{\D{st}{\R{}}} = \n{\D{t}{(\L[*]{1})}\D{s}{\R{}}\union rest} =
    \n{\D{t}{\L[*]{1}}}\n{\D{s}{\R{}}} \union \n{rest}
  \end{equation}
  Now, $\n{\D{t}{\L[*]{1}}} = \set{\emptystring}$ since $t\in\L[*]{1}$
  by choice. Also, $\n{\D{s}{\R{}}} = \set{\emptystring}$, and
  $\n{rest}$ is either \set{\emptystring} or $\emptyset$. Thus,
  $\n{\D{st}{\R{}}} = \set{\emptystring} \Leftrightarrow
  st\in\R{}$. So, in this case, since $s\in\R{}$, every string $st$
  with $t\in\L[*]{1}$ will also be in \R{}.
\end{example}

\section{Extending the operator set}

We are now ready to prove that regular languages are, in fact, closed
under all Boolean operations.

\begin{theorem}[Complement]\label{thm-regular-complement}
  If \R{} is a regular language, then so is its complement \inv{\R{}},
  and $d_{\inv{\R{}}} = d_{\R{}}$.
\end{theorem}

\begin{proof}
  By \refthm{thm-derivative-myhill-nerode} \R{} has $d_{\R{}}$
  characteristic equations
  \begin{equation}
    \D{s}{\R{}} = \n{\D{s}{\R{}}} \union
    \Union_{a\in\A{1}}\set{a}\D{t_a}{\R{}}.
  \end{equation}
  Take
  \begin{equation}\label{eqn-pre-invR-cd-eqn}
    \D{s}{\R{}} = \overline{\n{\D{s}{\R{}}}} \union
    \Union_{a\in\A{1}}\set{a}\D{t_a}{\R{}},
  \end{equation}
  where
  \begin{equation}
    \overline{\n{\D{s}{\R{}}}} = 
    \left\{ \begin{array}{cl}
      \{\emptystring\} & \mbox{if } \n{\D{s}{\R{}}} = \emptyset \\
      \emptyset & \mbox{if } \n{\D{s}{\R{}}} = \{\emptystring\}
    \end{array}\right. 
  \end{equation}
  to be the set of characteristic equations for a regular language
  $\bar{\R{}}$. That is, with $\n{\D{s}{\bar{\R{}}}} =
  \overline{\n{\D{s}{\R{}}}}$, and the rest of the coefficients the
  same,
  \begin{equation}\label{eqn-barR-cd-eqn}
    \D{s}{\bar{\R{}}} = \n{\D{s}{\bar{\R{}}}} \union
    \Union_{a\in\A{1}}\set{a}\D{t_a}{\bar{\R{}}}.
  \end{equation}
  By \refcor{cor-cd-eqn-uniqueness} equations
  \refeqn{eqn-barR-cd-eqn} can be solved for $\bar{\R{}}$
  uniquely.

  Let $t\in\R{}$, then by \refthm{thm-regular-finite-derivative} and
  \refthm{thm-char-derivative-length} there exists a characteristic
  derivative \D{s}{\R{}} such that $\D{t}{\R{}} = \D{s}{\R{}}$. Assume
  that
  \begin{equation}\label{eqn-R-barR-eq}
    \D{t}{\R{}} = \D{s}{\R{}} \Leftrightarrow
    \D{t}{\bar{\R{}}} = \D{s}{\bar{\R{}}}.
  \end{equation}
  Then we have 
  \begin{eqnarray}
    t\in\R{} &\Leftrightarrow& \n{\D{t}{\R{}}} = \set{\emptystring}
    \Leftrightarrow \n{\D{s}{\R{}}} = \set{\emptystring} \nonumber\\
    &\Leftrightarrow& \overline{\n{\D{s}{\R{}}}} = \emptyset
    \Leftrightarrow \n{\D{s}{\bar{\R{}}}} = \emptyset \nonumber\\
    &\Leftrightarrow& \n{\D{t}{\bar{\R{}}}} = \emptyset
    \Leftrightarrow t\not\in\bar{\R{}}
  \end{eqnarray}
  Thus $\bar{\R{}} = \inv{\R{}}$. Also, it is clear by construction 
  that $\bar{\R{}}$ and therefore \inv{\R{}} has $d_{\R{}}$ number of
  characteristic equations.

  Equation \refeqn{eqn-R-barR-eq} is proved by induction over the
  length of string $t$. It is trivially true for
  $t=\emptystring$. Assume that it is true for all strings $t$ with
  $\length{t} = k$, then it is also true for all strings $tb$,
  $\length{tb} = k+1$, $b\in\A{1}$, since for \R{}
  \begin{equation}
    \D{tb}{\R{}} = \D{b}{\D{t}{\R{}}} = \D{b}{\D{s}{\R{}}} =
    \D{b}{(\n{\D{s}{\R{}}})} \union \D{b}{\Union_{a\in\A{1}}\set{a}\D{t_a}{\R{}}}
    = \D{t_b}{\R{}},
  \end{equation}
  and for $\bar{\R{}}$
  \begin{equation}
    \D{tb}{\bar{\R{}}} = \D{b}{\D{t}{\bar{\R{}}}} = \D{b}{\D{s}{\bar{\R{}}}} =
    \D{b}{(\n{\D{s}{\bar{\R{}}}})} \union
    \D{b}{\Union_{a\in\A{1}}\set{a}\D{t_a}{\bar{\R{}}}} =
    \D{t_b}{\bar{\R{}}}.
  \end{equation}
  This concludes the proof.
\end{proof}

\begin{corollary}[Boolean operations]
  Since union and complement form a complete set of Boolean
  connectives, regular languages are closed under all such operations.
\end{corollary}

\section{Non-regular languages}\label{sec-non-regular-languages}

There are plenty of languages that are not regular. However, it is not
always easy to determine if a language is regular or not. In this
section we will investigate two conditions that can help. The
following theorem, built on \refdef{def-indistinguishable-strings},
gives a necessary and sufficient condition for a language to be
regular.

\begin{theorem}[Myhill-Nerode]
  A language $\L{}\subseteq\A{*}$ is regular if and only if the
  relation \eq{\L{}} has a finite number of equivalence classes.
\end{theorem}

\begin{proof}
  Suppose \L{} is regular. By \refthm{thm-derivative-myhill-nerode}
  \L{} has a finite set of different derivatives, it follows from
  \refeqn{eqn-eqclass-eq-derivative} that \eq{\L{}} has a finite set
  of equivalence classes.

  Conversely, suppose \eq{\L{}} has a finite set of equivalence classes
  for a language \L{}. Again, by \refeqn{eqn-eqclass-eq-derivative} it
  follows that \L{} has a finite set of derivatives, and thus by
  \refthm{thm-derivative-myhill-nerode} \L{} is regular.
\end{proof}

\begin{example}\label{xmp-ab-non-regular}
  Let \L{} be the language
  \begin{eqnarray}
    \L{} = \set{a^nb^n : n\in\N}
  \end{eqnarray}
  over the alphabet \set{a, b}. As we will see in the next example this
  language models the set of balanced parenthesized algebraic
  expressions. It can be shown to be non-regular using the
  Myhill-Nerode theorem as follows.

  Consider the set of derivatives with respect to $a^k$.
  \begin{equation}
    \D{a^k}{\L{}} = \set{a^{n-k}b^n : n,k\in\N,\; k \le n}
  \end{equation}
  Thus, the derivatives with respect to $a^k$ are all different. Hence
  \L{} does not have a finite set of different derivatives and is
  therefore not regular.
\end{example}

\begin{example}
Let \L{} be the set of algebraic expressions involving identifiers $a$
and $b$, operations $+$ and $*$ and left and right parentheses. Some
examples of algebraic expressions are $a$, $(a*b)$, $((a+b)*a)$ and
$(((a*b)+a)+(b*b))$. \L{} can be defined recursively as follows:
\begin{enumerate}
\item $a$ and $b$ are in \L{}.
\item If $s$ and $t$ are in \L{}, then $(s+t)$ and $(s*t)$ are in
  \L{}.
\item Nothing is in \L{} unless it is obtained from the above two
  clauses.
\end{enumerate}

Consider the set of strings $S_1=\set{s_1^k : s_1^k = (^ka,\;
  k\in\N}$, i.e the set of strings consisting of one or more left
parentheses followed by identifier $a$. Further, let $S_2=\set{s_2^k :
  s_2^k = [+b)]^k,\; k\in\N}$. Then $S = \set{s^k_1s^k_2:
  k\in\N}\subseteq\L{}$. It follows from \refxmp{xmp-ab-non-regular}
that $S$, and therefore \L{} has an infinite number of different
derivatives. Hence, \L{} is not regular.
\end{example}

\begin{example}
  Let \A{} be an alphabet, and
  \begin{equation}
    \L{} = \set{sts : s,t\in\A{*}}
  \end{equation}
  This language models that an identifier must be defined before it is
  used. It can be proven to be non-regular. (In fact it is not even
  context free.)

  Study the derivatives with respect to $st$.
  \begin{equation}
    \D{st}{\L{}} = \set{s}
  \end{equation}
  There are an infinite number of different strings $s$, and therefore
  an infinite number of different derivatives. Thus, \L{} is not
  regular.
\end{example}

The following theorem known as the Pumping lemma gives another
necessary, but not sufficient condition for a language to be regular.

\begin{theorem}[Pumping lemma]
  Let \R{} be a regular language, then there exists an integer $p \ge
  1$, called the \define{pumping length}, depending only on \R{} such
  that every string $r\in\R{}$ with $\length{r} \ge p$ can be written
  as $r=stu$ satisfying the conditions:
  \begin{enumerate}
  \item $\length{t} \ge 1$,
  \item $\length{st} \le p$, and
  \item $st^nu \in \R{}$ for all $n\in\N$.
  \end{enumerate}
\end{theorem}

\begin{proof}
  Since \R{} is regular, by \refthm{thm-derivative-myhill-nerode} it
  has a finite number $d_{\R{}}$ of different derivatives. Let $p =
  d_{\R{}}$ and let $r\in\R{}$ such that $\length{r} \ge p$. Further,
  let $s$ and $st$ be prefixes of $r$ such that\footnote{The following
    simply means that some characteristic derivative must be repeated
    if all derivatives with respect to prefixes of $r$ are
    calculated. It is guaranteed by
    \refthm{thm-char-derivative-length}.}
  \begin{equation}\label{eqn-pm-eqn1}
    \D{st}{\R{}} = \D{s}{\R{}},
  \end{equation}
    where $s$ is a characteristic derivative of \R{}, $\length{t} \ge
    1$, and $\length{st} \le p$. Such prefixes must exist since
    $\length{r} > d_{\R{}}-1$. Otherwise, we could use the prefixes of
    $r$ to get more than $d_{\R{}}$ different derivatives, which is
    impossible by \refthm{thm-char-derivative-length}. Thus we can
    write $r = stu$, and we have
  \begin{eqnarray}
    r\in\R{} &\Leftrightarrow& \emptystring \in \D{r}{\R{}}
    \Leftrightarrow \emptystring \in \D{r}{\R{}} \nonumber\\
    &\Leftrightarrow& \emptystring \in \D{stu}{\R{}}
    \Leftrightarrow \emptystring \in \D{st^nu}{\R{}}
    \Leftrightarrow  st^nu \in \R{} \quad n\in\N,
  \end{eqnarray}
  where the last two equivalences follow from
  \refeqn{eqn-pm-eqn1}.
\end{proof}

In simple words the lemma says that for any regular language \R{}, any
sufficiently long string $r\in\R{}$ can be split into three parts
$r=stu$, such that all strings $st^nu\in\R{}$ for $n\in\N$. That is,
there exists a substring of $r$ that can be repeated (pumped) any number of
times. Further, this substring must occur within the p first
symbols of $r$.

\begin{example}
  Let us prove once more that 
  \begin{eqnarray}
    \L{} = \{a^nb^n : n\in\N\}
  \end{eqnarray}
  is non-regular, this time by using the pumping lemma. Suppose that
  \L{} is regular. Then, pick $s = a^pb^p$, where $p$ is the pumping
  length. The $p$ first symbols of $s$ are all $a$'s. Therefore,
  $a^k$ for some $k<p$ will play the role of $t$ in the lemma. But
  then $a^pa^kb^p = a^{p+k}b^p\in\L{}$ which is not true. Therefore
  our assumption that \L{} is regular must be false.
\end{example}

\chapter{Regular expressions}\label{chp-regular-expressions}

We now introduce a notational device called regular expressions in
order to simplify our discussion of regular languages. The idea is to
let symbols and sequences of symbols of the alphabet denote the
corresponding languages. This allows us to drop the somewhat
cumbersome curly braces.

\begin{definition}[Regular expression]\label{def-regular-expression}
  A \define{regular expression} over an alphabet \A{}, where $a_1,
  a_2,\dots, a_n$ is an enumeration of the symbols of \A{}, is defined
  by the EBNF grammer:
  
  \begin{center}
    \setlength{\tabcolsep}{0.3em}
    \renewcommand{\arraystretch}{1.25}
    \begin{tabular}{rcl}
      \rx{r} &=& \nt{disj} \\
      \nt{disj} &=& \nt{disj}, "\union", \nt{conj} \\
      &|& \nt{conj}; \\
      \nt{conj} &=& \nt{conj}, "$\intersection$", \nt{concat} \\
      &|& \nt{concat}; \\
      \nt{concat} &=& \nt{concat}, \nt{clos} \\
      &|& \nt{clos}; \\
      \nt{clos} &=& \nt{atom}, "\textsuperscript{*}" \\
      &|& \nt{atom}; \\
      \nt{atom} &=& "\comp{}", \nt{atom} \\
      &|& "\rx{a_1}" | "\rx{a_2}" | $\cdots$ | "\rx{a_n}" \\
      &|& "$\emptyset$" \\
      &|& "" (* we usually write this as \emptystring\ *) \\
      &|& "(", \nt{disj}, ")";
    \end{tabular}
  \end{center}
  This grammar respects the usual precedence rules for the operators,
  as well as admits grouping by parentheses to override them. We also
  allow for intersection and complement since, by
  \refthm{thm-regular-complement}, regular languages are closed under
  these operations as well. Let $\rxlang{\cdot}$ denote the
  language associated with a regular expression. Then define 
  \begin{eqnarray*}
    \rxlang{\emptyset} &=& \emptyset \\
    \rxlang{\emptystring} &=& \set{\emptystring} \\
    \rxlang{a_i} &=& \{a_i\} \quad a_i\in\A{1} \\
    \rxlang{\comp{r}} &=& \inv{\rxlang{r}} \\
    \rxlang{r^*} &=& \rxlang{r}^* \\
    \rxlang{r_1r_2} &=& \rxlang{r_1}\rxlang{r_2} \\
    \rxlang{r_1 \intersection r_2} &=& \rxlang{r_1} \intersection \rxlang{r_2} \\
    \rxlang{r_1 \union r_2} &=& \rxlang{r_1} \union \rxlang{r_2}
  \end{eqnarray*}
  We also define, for some string $s\in\A{*}$, the notation $s\in\rx{r}$ to mean
  $s\in\rxlang{r}$. Further, we define
  \begin{displaymath}
    \rxlang{\D{a}{r}} = \D{a}{\rxlang{r}}
  \end{displaymath}
\end{definition}

With these definitions, everything that has been said about properties
of languages, Kleene closure, derivatives etc, are well-defined for
regular expressions as well.

\begin{definition}[Equality]
  Two regular expressions, \rx{r_1} and \rx{r_2}, that denote the same
  language are said to be \define{equal}, and we write $\rx{r_1} =
  \rx{r_2}$.
\end{definition}

\begin{example}
  A few examples of equal regular expressions over an alphabet
  $\A{} = \{a,b\}$.
  \begin{enumerate}
    \renewcommand{\labelenumi}{\alph{enumi})}
  \item $\rxlang{a \union b} = \set{a}\union\set{b} = \set{a,b} = \set{b,a} = \set{b}\union\set{a} = \rxlang{b \union a}$
  \item $\rxlang{(a \union b)(a \union b)} = \set{a,b}\set{a,b} = \set{aa, ab, ba, bb} = \rxlang{aa \union ab \union ba \union bb}$
  \item $\rxlang{(a \union b)^*} = \rxlang{(a^*b^*)^*}$
  \item $\rxlang{\D{a}{(\rx{abc})}} = \D{a}{\rxlang{abc}}
    = \D{a}{\set{abc}} = \set{bc} = \rxlang{bc}$
  \item $\rxlang{\D{a}{(\rx{r}^*)}} = \D{a}{\rxlang{r^*}}
    = \D{a}{(\rxlang{r})}\rxlang{r^*} = \rxlang{\D{a}{r}}\rxlang{r^*}
    = \rxlang{\D{a}{(r)}r^*}$
  \end{enumerate}
\end{example}

\begin{example}
  For a regular language, defined by a regular expression, equation
  \refeqn{eqn-derivative-decide} becomes
  \begin{equation}
      s\in\rx{r} \Leftrightarrow
      \emptystring\in\D{s}{\rx{r}}\label{eqn-rx-decide}
  \end{equation}
\end{example}

\begin{example}[Characteristic equations]
  Another illustrative example is to write the characteristic
  equations for a regular language in terms of one of its regular
  expressions. Let $\R{} = \rx{r}$, then the characteristic equations
  take the form
  \begin{equation}
    \D{s}{\rx{r}} = \n{\D{s}{\rx{r}}}\union\Union_{a\in\A{}}\rx{a}\D{t_a}{\rx{r}}
  \end{equation}
\end{example}

\begin{theorem}
  The characteristic equations expressed as regular expressions can be
  solved for uniquely up to regular expression equality.
\end{theorem}

\begin{proof}
  This follows from the fact that there can be several expressions
  that denote the same language. The final expression for the language
  will depend on the order in which the derivatives are eliminated,
  and what expressions where used to denote the derivatives.
\end{proof}

\section{Regular expression matching}\label{sec-re-matching}

\begin{definition}[Match]
  A regular expression $\rx{r}$ is said to \define{match} a string $s$
  when $s$ is in the language denoted by \rx{r}, i.e. \rx{r} matches $s$
  when $s\in\rx{r}$. Let $\rx{R} = \set{\rx{r}: \rx{r} \; \mbox{is a regular
      expression}}$. We define the operator (matches)
  \begin{equation}
    \sim : \rx{R} \times
    \A{*} \rightarrow \set{true, false}
  \end{equation}
  by
  \begin{equation}
    \rx{r}\sim s = \left\{ \begin{array}{cl}
      true & \mbox{if } s\in\rx{r} \\
      false & \mbox{if } s\notin\rx{r} 
    \end{array}\right.
  \end{equation}
\end{definition}

\refthm{thm-rec-derivative} together with
\refthm{thm-derivative-decide} and \refcor{cor-nL-eq-inL} give us an
algorithm to decide if a regular expression matches a string or not,
see \refalg{alg-lazy-match}. It iteratively calculates the derivative
of a regular expression with respect to a string, one symbol at a
time, and then checks if the result is nullable.

\begin{algorithm}[t]
  \caption{Regular expression matching using derivatives}\label{alg-lazy-match}
  \begin{algorithmic}[1]
    \Procedure{Match}{$\rx{r},s$}
      \While{$s \neq \emptystring $}
        \State $c \gets \Call{PopFront}{s}$
        \State $\rx{r} \gets \D{c}{\rx{r}}$
      \EndWhile
      \If{$\n{\rx{r}} = \emptystring $}
        \State matches $\gets true$
      \Else
        \State matches $\gets false$
      \EndIf
      \Return{matches}  
    \EndProcedure
  \end{algorithmic}
\end{algorithm}

\begin{example}\label{xmp-rx-match}
  Let us illustrate the algorithm on a simple example. Let $\rx{r} =
  \rx{ab^*}$, and $s = abb$, then
  \begin{eqnarray*}
    \rx{ab^*} \sim abb &\Leftrightarrow& \D{a}{\rx{ab^*}} \sim bb \\
    &\Leftrightarrow& \rx{b^*} \sim bb \\
    &\Leftrightarrow& \D{b}{\rx{b^*}} \sim b \\
    &\Leftrightarrow& \rx{b^*} \sim b \\
    &\Leftrightarrow& \D{b}{\rx{b^*}} \sim \emptystring \\
    &\Leftrightarrow& \rx{b^*} \sim \emptystring = true
  \end{eqnarray*}\qedhere
\end{example}

Since the time to calculate the derivative of a regular expression is
$O(m)$, where $m$ is the length of the expression, the above algorithm
runs is $O(mn)$ time for a string of length $n$. The space requirement
is $O(m)$ since we need to keep the expression in memory in order to
calculate each derivative.

While a time complexity of $O(mn)$ is not that bad, it is possible to
achieve better running times, but for that we need the theory of
finite state machines, which is where we turn to next.

\chapter{Finite automata}\label{chp-finite-automata}

In this chapter we will connect regular languages and their
derivatives with the theory of deterministic finite automata, a far
more common approach in the literature when discussing regular
expressions and their applications to string matching.

\section{Deterministic Finite Automaton}

A (deterministic) finite state machine, or deterministic finite
automaton (DFA) is an abstraction of a device that starts in a
specific state, receives some input, and ends up in a either the same
or a different state\footnote{Of course, for this to be useful there
  are in general some way to tell which state the device is in.}. More
formally we have the following definition.

\begin{definition}[DFA]
  A \define{deterministic finite automaton} or DFA, is a 5-tuple
  $(\A{}, Q, q_0,\d{}, F)$ where
  \begin{itemize}
  \item $\A{}$ is the input alphabet.
  \item $Q$ is a finite set, whose elements are called
    \define{states}.
  \item $q_0 \in Q$ is the \define{initial state}.
  \item $\d{}$ is the \define{transition function} $\d{}:Q \times
    \A{} \rightarrow Q$
  \item $F \subseteq Q$ is a set of \define{final} or
    \define{accepting states}.
  \end{itemize}
\end{definition}

\newpage

\begin{definition}[Recursive transition function]%
  \label{def-recursive-transition-function}
  The \define{recursive transition function}
  \begin{equation}
    \d[*]{}:Q \times \A{*} \rightarrow Q
  \end{equation}
  is defined by
  \begin{eqnarray}
    \d[*]{q,\emptystring} &=& q \quad q\in Q \\
    \d[*]{q, sa} &=& \d{\d[*]{q,s},a} \quad q\in Q, s\in\A{*}, a\in\A{}
  \end{eqnarray}
\end{definition}

\begin{example}\label{xmp-dfa-counter}
   Consider for example a one-figure digital counter with buttons for
   increase, decrease, and reset. The initial state is $0$, and the
   input is a series of increase, decrease and reset signals. The
   accepting states are those whose counter value are even, except for
   state $0$. The counter will be in one of 10 possible different
   states after some received input. Which state it is in will depend
   on the input, and the construction of the counter, whether it wraps
   around at $0$ and/or $9$ etc.
\end{example}

\begin{figure}
  \center
  \begin{minipage}[t]{0.55\textwidth}
    \centering
  \begin{tikzpicture}
    \footnotesize
    \node (0) at (0,0)     [state,initial]     {$0$};
    \node (1) at (160:3cm) [state]             {$1$};
    \node (2) at (120:3cm) [state,accepting]   {$2$};
    \node (3) at (80:3cm)  [state]             {$3$};
    \node (4) at (40:3cm)  [state,accepting]   {$4$};
    \node (5) at (0:3cm)   [state]             {$5$};
    \node (6) at (-40:3cm) [state,accepting]   {$6$};
    \node (7) at (-80:3cm) [state]             {$7$};
    \node (8) at (-120:3cm)[state,accepting]   {$8$};
    \node (9) at (-160:3cm)[state]             {$9$};

    \path
      (0) edge [loop left] node {$r$} ()
      (0) edge             node {$+$} (1)
      (1) edge [bend left] node [swap] {$-,r$} (0) 
      foreach \m/\n in {1/2, 2/3, 3/4, 4/5, 5/6, 6/7, 7/8, 8/9}
          {
            (\m) edge [bend left] node {$+$} (\n)
            (\n) edge [bend left] node [swap] {$-$} (\m)
          }
      (9) edge [bend right=10] node {$+,r$} (0)
      (0) edge [bend left] node {$-$} (9)
      foreach \m in {2,3,...,7}
          {
            (\m) edge node {$r$} (0)
          }
      (8) edge node [swap] {$r$} (0);
  \end{tikzpicture}
  \mbox{(a) Transition diagram.}
  \end{minipage}%
  \hfill%
  \begin{minipage}[t]{0.40\textwidth}
      \centering
      \setlength{\tabcolsep}{2ex}
      \begin{tikzpicture}
        \node [below right,text width=10cm,align=justify] at (4,3)
        {
          \begin{tabular}{c|c|c|c}
            \hline
            \hline
              & \multicolumn{3}{|c}{$\d{}$} \\
            \cline{2-4}
            \raisebox{1.5ex}[0mm]{\textsc{State}} & $+$ & $-$ & $r$ \\
            \hline
            0 & 1 & 9 & 0\\
            1 & 2 & 0 & 0\\
            \textbf{2} & 3 & 1 & 0\\
            3 & 4 & 2 & 0\\
            \textbf{4} & 5 & 3 & 0\\
            5 & 6 & 4 & 0\\
            \textbf{6} & 7 & 5 & 0\\
            7 & 8 & 6 & 0\\
            \textbf{8} & 9 & 7 & 0\\
            9 & 0 & 8 & 0\\
            \hline
          \end{tabular}
          \vspace{.5cm}
        };
      \end{tikzpicture}
      \mbox{(b) Transition table.}
  \end{minipage}

  \begin{minipage}{0.85\textwidth}
  \caption[\refxmp{xmp-dfa-counter} DFA representations]{The counter
      from \refxmp{xmp-dfa-counter} represented as (a) a transition
      diagram, and (b) a transition table. In the figure $+$ indicates
      increase, $-$ decrease, and $r$ is
      reset.}\label{fig-dfa-counter}
  \end{minipage}
\end{figure}

\subsection{Representations}

There are several ways to represent a DFA. Two common alternatives are
a directed graph, called a transition diagram, or a table called a
transition table.

The transition table is quite straight forward, with one row for each
state, and one column for each input. The entry in the table gives the
state the automaton moves to on respective input. Accepting states are
marked with \textbf{bold} state labels.

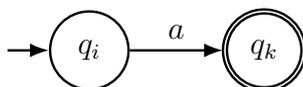
\begin{figure}[h]
  \center
  \begin{tikzpicture}
    \node[state,initial]   (q_i)                {$q_i$};
    \node[state,accepting] (q_k) [right=of q_i] {$q_k$};
    \path (q_i) edge node {$a$} (q_k);
  \end{tikzpicture}
  \caption{Transition diagram
    notation.}\label{fig-transition-diagram-notation}
\end{figure}

In a transition diagram the nodes of the graph, drawn as circles,
represents the states, and edges labeled by symbols from the input
alphabet represents the transition function. An edge from $q_i$ to
$q_k$ labeled $a$ corresponds to the transition $q_k =
\d{q_i,a}$. The initial state is indicated by an incoming arrow,
and the accepting states are drawn with double circles. This notation
is showed in \reffig{fig-transition-diagram-notation}. A more fully
worked out example is shown in \reffig{fig-dfa-counter}, which
displays both the transition diagram and the transition table for the
counter in \refxmp{xmp-dfa-counter}.

\section{DFA language recognizers}

Deterministic finite automatons, or finite state machines, have a wide
range of applications. In this section we will consider one of them,
that of recognizers of regular languages. We therefore make the
following definitions.

\begin{definition}[Accept, Reject]
  For a DFA $M=(\A{}, Q, q_0,\d{}, F)$, a state $q_i\in Q$ is said to
  \define{accept} a string $s\in\A{*}$ if, and only if, $q_k =
  \d[*]{q_i, s} \in F$. A string that is not accepted is said to be
  \define{rejected}. Moreover, $M$ is said to accept $s$ if, and only
  if, $q_0$ accepts $s$.
\end{definition}
Simply put, a state $q_i$ accepts a string if the DFA, starting
in $q_i$, is in an accepting state after processing the string.

\begin{definition}[Language recognition]\label{def-language-recognition}
  For a DFA $M=(\A{}, Q, q_0,\d{}, F)$, a state $q_i \in Q$ is said to
  \define{recognize} the language $\R{q_i} = \set{s: \d[*]{q_i,s} \in
    F}$. Further, $M$ is said to recognize the language \R{} = \R{q_0}
\end{definition}
Thus, the language recognized by a DFA consists of all the strings
accepted by its starting state. So, in the application as a language
recognizer, the input to the DFA is a string over some alphabet \A{},
and the states keep track of what strings are part of the language the
DFA recognizes. After a string has been processed, the DFA will be in
a state that tells if the string is contained in the language, an
accepting state, or not, in which case it is in a non-accepting state.

Note that the transition function \d{} is total, i.e. that \d{q_i, a_k}
is defined for all $q_i\in Q$ and all $a_k\in\A{}$. However, for our
purposes it is sometimes impractical to display all transitions, since
it would result in lots of transitions from every node to a single
node whose sole purpose is to capture an error or non-accepting
state. For clarity we will therefore sometimes delete such a state and
all transitions to it, and let their existence be implied.

\begin{definition}[Equivalent states]
  Let $M=(\A{}, Q, q_0,\d{}, F)$ be a DFA that recognizes a language
  \R{}. Two states $q_i$ and $q_k$ are said to be \define{equivalent},
  and we write $q_i \eq{M} q_k$ if, and only if, $\R{q_i} = \R{q_k}$,
  where \R{q_i} and \R{q_k} are the languages recognized by respective
  state.
\end{definition}

\begin{example}\label{xmp-rx-dfa-match}
  Let $\A{} = \set{a,b,c}$ and consider the regular expression in
  \refxmp{xmp-rx-match}, \rx{ab^*}. A transition diagram for a DFA
  that recognizes this language is shown in \reffig{fig-rx-dfa-match}.
  \begin{figure}
    \center
    \begin{minipage}[t]{0.55\textwidth}
      \centering
      \begin{tikzpicture}
        \node[state,initial]    (0)                    {$q_0$};
        \node[state]            (2) [below right=of 0] {$q_{\emptyset}$};
        \node[state, accepting] (1) [above right=of 2] {$q_1$};

        \path
        (0) edge              node        {$a$}   (1)
        (0) edge              node [swap] {$b,c$} (2)
        
        (1) edge [loop above] node        {$b$}   ()
        (1) edge              node        {$a,c$} (2)

        (2) edge [loop below] node        {$a,b,c$} ()
        ;
      \end{tikzpicture}
      \mbox{(a) Transition diagram.}
    \end{minipage}%
    \hfill%
    \begin{minipage}[t]{0.40\textwidth}
      \centering
      \setlength{\tabcolsep}{2ex}
      \begin{tikzpicture}
        \node [below right,text width=10cm,align=justify] at (0,0)
        {
          \begin{tabular}{c|c|c|c}
            \hline
            \hline
              & \multicolumn{3}{|c}{$\d{}$} \\
            \cline{2-4}
            \raisebox{1.5ex}[0mm]{\textsc{State}} & $a$ & $b$ & $c$ \\
            \hline
            $q_0$ & $q_1$ & $q_{\emptyset}$ & $q_{\emptyset}$ \\
            $\mathbf{q_1}$ & $q_{\emptyset}$ & $q_1$ & $q_{\emptyset}$ \\
            $q_{\emptyset}$ & $q_{\emptyset}$ & $q_{\emptyset}$ & $q_{\emptyset}$ \\
            \hline
          \end{tabular}
          \vspace{1.25cm}
        };
      \end{tikzpicture}
      \mbox{(b) Transition table.}
    \end{minipage}
    \begin{minipage}{0.85\textwidth}
      \caption[\refxmp{xmp-rx-dfa-match} Expression
        construction]{The DFA in \refxmp{xmp-rx-dfa-match} represented
        as (a) a transition diagram, and (b) a transition
        table.}\label{fig-rx-dfa-match}
    \end{minipage}
  \end{figure}  
  $q_{\emptyset}$ is the so called error state. It is needed if we require
  $\d{}$ t be a total function.
\end{example}

\subsection{Regular expression matching using DFAs}

We are now ready to give an algorithm for matching strings with the
aid of DFAs.
\begin{algorithm}
  \caption{Regular expression matching using DFAs}\label{alg-dfa-match}
  \begin{algorithmic}[1]
    \Procedure{Match}{$(\protect\A{}, Q, q_0, \d{}, F), s$}
      \State $q_c \gets q_0$ 
      \While{$s \neq \emptystring $}
        \State $c \gets \Call{PopFront}{s}$
        \State $q_c \gets \d{q_c,c}$
      \EndWhile
      \If{$q_c \in F$}
        \State matches $\gets true$
      \Else
        \State matches $\gets false$
      \EndIf
      \Return{matches}  
    \EndProcedure
  \end{algorithmic}
\end{algorithm}
By choosing a proper implementation $\d{q_c,c}$ can be done in $O(1)$
time, and \refalg{alg-dfa-match} will therefore run in $O(n)$ time for
a string of length $n$. This is clearly an improvement in time
complexity. Our former algorithm, \refalg{alg-lazy-match}, had a worst
case running time of $O(mn)$ where $m$ is the size of the
expression. However, a problem with DFAs are their worst case space
requirements. DFAs that recognizes certain languages grows
exponentially in the number of states with respect to the length of a
regular expression defining the language. A classic example is the
language defined by
\begin{equation}
  \rx{r} = \rx{(a+b)^*a(a+b)^{m-1}}
         = \rx{(a+b)^*a\underbrace{\rx{(a+b)(a+b)\dots(a+b)}}_{\mathnormal{m-1}}}
\end{equation}
It is the language consisting of strings of $a$ and $b$ in any order,
where the $m$th symbol from the right end is an $a$. Any DFA
recognizing the language will have at least $O(2^m)$ states.

\section{DFA construction}

We now turn to the problem of constructing a DFA given a regular
expression. The following theorem, due to Kleene, is central to the
solution.

\begin{theorem}[Kleene's theorem]\label{thm-kleenes-theorem}
  A language is regular if, and only if, it is recognized by a DFA.
\end{theorem}

\begin{proof}
  We break this proof into two separate theorems, the result will then
  be immediate from \refthm{thm-dfa-is-regular} and
  \refthm{thm-regular-is-dfa}.
\end{proof}

To begin with, we need some connection between a DFA and
derivatives. This turns out to be easy to find by using
\refdef{def-language-recognition}.

\begin{theorem}\label{thm-state-eq-derivative}
  Let $M=(\A{}, Q, q_0,\d{}, F)$ be a DFA, and let \R{q_s} be the
  language recognized by a state $q_s = \d[*]{q_0,s} \in Q$. Then
  \begin{eqnarray}
    \R{q_s} &=& \D{s}{\R{}} \\
    q_s \in F &\Leftrightarrow& \emptystring \in \D{s}{\R{}}
  \end{eqnarray}
\end{theorem}

\begin{proof}
  \begin{equation}
    \R{q_s} =
    \set{t:\d[*]{q_s,t} \in F} =
    \set{t:\d[*]{q_0,st} \in F} =
    \set{t: st \in \R{}} =
    \D{s}{\R{}}
  \end{equation}
  By which it is also obvious that $q_s\in F \Leftrightarrow
  \emptystring \in \R{q_s}$.
\end{proof}

\begin{corollary}
  \begin{equation}\label{eqn-state-string-der-eq}
    q_s \eq{M} q_t \Leftrightarrow
    \D{s}{\R{}} = \D{t}{\R{}} \Leftrightarrow
    s \eq{\R{}} t
  \end{equation}
  The last equivalence is repeated from
  \refeqn{eqn-indistinguishable-strings-eq-derivatives}.
\end{corollary}

\begin{theorem}\label{thm-dfa-is-regular}
  Let $M=(\A{}, Q, q_0,\d{}, F)$ be a DFA that recognizes a language
  \R{}, then \R{} is regular, and \R{} can be constructed from $M$.
\end{theorem}

\begin{proof}
  Let $s\in\R{}$, then \d[*]{q_0,s} is a state that recognizes
  \D{s}{\R{}}. Since there are a finite number of states in $Q$, \R{}
  has a finite number of derivatives. By
  \refthm{thm-derivative-myhill-nerode} \R{} must be regular.
  
  In order to construct \R{}, the idea is to derive a set of
  characteristic equations from $M$, these can then be solved for
  \R{}.

  First, for each $q_i\in Q$, let $s_i$ be the shortest string such
  that $q_i = \d{q_0, s_i}$. By \refthm{thm-state-eq-derivative} we
  have that each $q_i$ recognizes the language \D{s_i}{\R{}}. Since
  $s_i$ by choice are the shortest strings, each \D{s_i}{\R{}} must be
  a characteristic derivative of \R{}. Next, pick $q_m =
  \d[*]{q_0,s_m}\in Q$. Then $q_m$ recognizes the language
  \D{s_m}{\R{}}. According to \refthm{thm-derivative-expansion} it can
  be expanded to

  \begin{equation}\label{eqn-characteristic-equation-step1}
    \D{s_m}{\R{}} = \n{\D{s_m}{\R{}}} \union  \Union_{a\in\A{1}}\set{a}\D{s_ma}{\R{}}
  \end{equation}
  where $\n{\D{s_m}{\R{}}} = \set{\emptystring}$ if and only if $q_m$ is an
  accepting state, since $s_m\in\R{} \Leftrightarrow
  \emptystring\in\D{s_m}{\R{}}$ by \refthm{thm-derivative-decide}.

  Now, let $q_{n_a}$ be the state $M$ transitions to when it is in $q_m$
  and processes the symbol $a$, i.e.

  \begin{equation}
    q_{n_a} = \d{q_m, a} = \d{\d[*]{q_0,s_m}, a} = \d[*]{q_0,s_ma}
  \end{equation}
  Thus, $q_{n_a}$ recognizes the language \D{s_ma}{\R{}}. However, $q_{n_a}$
  also recognizes the language $\D{s_{n_a}}{\R{}}$, since $q_{n_a} =
  \d[*]{q_0,s_{n_a}}$ by definition. Therefore we have
  
  \begin{equation}
    \D{s_ma}{\R{}} = \D{s_{n_a}}{\R{}}  
  \end{equation}  
  and \refeqn{eqn-characteristic-equation-step1} can be written

  \begin{equation}\label{eqn-characteristic-equation-step2}
    \D{s_m}{\R{}} = \n{\D{s_m}{\R{}}} \union  \Union_{a\in\A{1}}\set{a}\D{s_{n_a}}{\R{}}
  \end{equation}
  But this is precisely a set of characteristic equations for \R{}.
\end{proof}

\begin{example}\label{xmp-dfa-lang-construction}
  Suppose that $\A{} = \set{a,b,c}$ and that we have the DFA shown in
  \reffig{fig-dfa-lang-construction}. Let us derive a regular
  expression \rx{r} for the language the DFA recognizes.
  \begin{figure}
    \vspace{-2cm}
    \center
    \begin{minipage}[t]{0.55\textwidth}
      \centering
      \begin{tikzpicture}
        \node[state,initial, accepting]   (0) {$q_0$};
        \node[state] (1) [above right=of 2] {$q_1$};
        \node[state] (2) [below right=of 0] {$q_2$};
        \path
        (0) edge [loop above] node {$a$} ()
        (0) edge [bend left] node  {$b$} (1)
        (0) edge [bend right] node [swap] {$c$} (2)
        
        (1) edge [bend left] node [swap] {$a$} (0)
        (1) edge [loop above] node {$b$} ()
        (1) edge [bend left] node  {$c$} (2)

        (2) edge [loop below] node {$a,b,c$} ()
        ;
      \end{tikzpicture}
      \mbox{(a) Transition diagram.}
    \end{minipage}%
    \hfill%
    \begin{minipage}[t]{0.40\textwidth}
      \centering
      \setlength{\tabcolsep}{2ex}
      \begin{tikzpicture}
        \node [below right,text width=10cm,align=justify] at (0,0)
        {
          \begin{tabular}{c|c|c|c}
            \hline
            \hline
              & \multicolumn{3}{|c}{$\d{}$} \\
            \cline{2-4}
            \raisebox{1.5ex}[0mm]{\textsc{State}} & $a$ & $b$ & $c$ \\
            \hline
            $\mathbf{q_0}$ & $q_0$ & $q_1$ & $q_2$\\
            $q_1$ & $q_0$ & $q_1$ & $q_2$\\
            $q_2$ & $q_2$ & $q_2$ & $q_2$\\
            \hline
          \end{tabular}
          \vspace{1.25cm}
        };
      \end{tikzpicture}
      \mbox{(b) Transition table.}
    \end{minipage}
    \begin{minipage}{0.85\textwidth}
      \caption[\refxmp{xmp-dfa-lang-construction} Expression
        construction]{The DFA in \refxmp{xmp-dfa-lang-construction}
        represented as (a) a transition diagram, and (b) a transition
        table.}\label{fig-dfa-lang-construction}
    \end{minipage}
  \end{figure}
  The shortest strings to reach each node, and thus the derivatives of
  \rx{r} that each state represents are as follows

  \begin{center}
    \begin{tabular}{c|c|c}
      \hline
      \hline
      \textsc{State} & \textsc{String} & \textsc{Derivative} \\
      \hline
      $q_0$ & \emptystring & \D{\emptystring}{\rx{r}}\\
      $q_1$ & $b$ & \D{b}{\rx{r}} \\
      $q_2$ & $c$ & \D{c}{\rx{r}} \\
      \hline
    \end{tabular}
  \end{center}
  
  By utilizing the information we have from the transition function we
  can write the characteristic equations
  \begin{eqnarray}
    \D{\emptystring}{\rx{r}} &=&
      \rx{\emptystring} \union
      \rx{a}\D{\emptystring}{\rx{r}} \union
      \rx{b}\D{b}{\rx{r}} \union
      \rx{c}\D{c}{\rx{r}} \label{eqn-dfa-lang-construction-1a} \\
    \D{b}{\rx{r}} &=&
      \rx{a}\D{\emptystring}{\rx{r}} \union
      \rx{b}\D{b}{\rx{r}} \union
      \rx{c}\D{c}{\rx{r}} \label{eqn-dfa-lang-construction-1b} \\
    \D{c}{\rx{r}} &=&
      \rx{a}\D{c}{\rx{r}} \union
      \rx{b}\D{c}{\rx{r}} \union
      \rx{c}\D{c}{\rx{r}} \label{eqn-dfa-lang-construction-1c}
  \end{eqnarray}
  Equation \refeqn{eqn-dfa-lang-construction-1c} can be rewritten as
  \begin{eqnarray}
    \D{c}{\rx{r}} &=&
      \rx{a}\D{c}{\rx{r}} \union
      \rx{b}\D{c}{\rx{r}} \union
      \rx{c}\D{c}{\rx{r}} \\
    &=& \rx{(a \union b \union c)}\D{c}{\rx{r}} \union \emptyset \\
    &=& \rx{(a \union b \union c)}^*\emptyset \label{eqn-dfa-lang-construction-2c} \\
    &=& \emptyset
  \end{eqnarray}
  where we have used Arden's rule (\refthm{thm-ardens-rule}) to reach
  \refeqn{eqn-dfa-lang-construction-2c}. Substituting $\emptyset$ for
  \D{c}{\rx{r}} in the other two equations, we are left with
  \begin{eqnarray}
    \D{\emptystring}{\rx{r}} &=&
      \rx{\emptystring} \union
      \rx{a}\D{\emptystring}{\rx{r}} \union
      \rx{b}\D{b}{\rx{r}}  \label{eqn-dfa-lang-construction-3a} \\
    \D{b}{\rx{r}} &=&
      \rx{a}\D{\emptystring}{\rx{r}} \union
      \rx{b}\D{b}{\rx{r}} \label{eqn-dfa-lang-construction-3b}
  \end{eqnarray}
  Again, by Arden's rule
  \begin{equation}
    \D{b}{\rx{r}}
    = \rx{a}\D{\emptystring}{\rx{r}} \union
      \rx{b}\D{b}{\rx{r}}
    = \rx{b^*}\rx{a}\D{\emptystring}{\rx{r}}
  \end{equation}
  Substituting this back into equation
  \refeqn{eqn-dfa-lang-construction-3a}, and yet another use of
  Arden's rule, we get
  
  \begin{eqnarray}
    \D{\emptystring}{\rx{r}}
    &=& \rx{\emptystring} \union
      \rx{a}\D{\emptystring}{\rx{r}} \union
      \rx{b}\rx{b^*}\rx{a}\D{\emptystring}{\rx{r}} \\
    &=& \rx{(a \union bb^*a)}\D{\emptystring}{\rx{r}} \union \emptystring \\
    &=& \rx{(a \union bb^*a)}\D{\emptystring}{\rx{r}} \union \emptystring \\
    &=& \rx{(a \union bb^*a)}^*
  \end{eqnarray}
  Finally, $\rx{r} = \D{\emptystring}{\rx{r}}$ by definition, hence
  \begin{equation}
    \rx{r} = \rx{(a \union bb^*a)}^*
  \end{equation}
\end{example}

\begin{theorem}\label{thm-regular-is-dfa}
  Let \R{} be a regular language over \A{}, then there is a DFA
  recognizing \R{}.
\end{theorem}

\begin{proof}
  By \refthm{thm-cd-eqn} and the Myhill-Nerode theorem
  (\refthm{thm-derivative-myhill-nerode}) there exist a set of
  characteristic equations for \R{},
  \begin{equation}
    \D{s}{\R{}} = \n{\D{s}{\R{}}} \union \Union_{a\in\A{1}}\set{a}\D{t_a}{\R{}}
  \end{equation}
  Define a DFA $M=(\A{}, Q, q_\emptystring,\d{}, F)$ by
  \begin{itemize}
  \item $Q = \set{q_s} = \set{\D{s}{\R{}}:\D{s}{\R{}}\mbox{ is a
      characteristic derivative of \R{}}}$
  \item $q_{\emptystring} = \D{\emptystring}{\R{}}$ 
  \item \d{}: $q_{t_a} = \d{q_s, a}$ for each term
    \set{a}\D{t_a}{\R{}} in the characteristic equation for
    \D{s}{\R{}}.
  \item $F = \set{\D{s}{\R{}}: \n{\D{s}{\R{}}}=\set{\emptystring}}$
  \end{itemize}
  $M$ recognizes \R{}, for assume that $u\in\A{*}$, then
  \begin{equation}\label{eqn-regular-is-dfa-1}
    \d[*]{q_\emptystring,u}
    = \d{\d{\d{\dots\d{\d{q_\emptystring,a_0},a_1},\dots},a_{n-1}}, a_n}
    = q_s
    = \D{s}{\R{}}
  \end{equation}
  for some characteristic derivative \D{s}{R}. Now, study
  \begin{equation}
    \D{u}{\R{}}
    = \D{a_0a_1\dots a_{n-1}a_n}{\R{}}
    = \D{a_n}{\D{a_{n-1}}{\dots\D{a_1}{\D{a_0}{\R{}}}}}
    = \D{s}{\R{}}
  \end{equation}
  The last equality follows from the fact that \R{} has a finite set
  of characteristic derivatives. We calculate \D{u}{R} recursively,
  one symbol at a time. In each step we can replace the result with a
  characteristic derivative. Hence, in calculating \D{u}{R} we move
  through a series of characteristic derivatives. This is exactly the
  same series of characteristic derivatives that
  \refeqn{eqn-regular-is-dfa-1} moves through, and thus $\D{u}{\R{}} =
  \D{s}{R}$. We have
  \begin{equation}
    u\in\R{} \Leftrightarrow
    \emptystring\in\D{u}{R{}} \Leftrightarrow 
    \emptystring\in\D{s}{\R{}} \Leftrightarrow 
    q_s \in F
  \end{equation}
  Hence \d[*]{q_\emptystring, u} is an accepting state if and only if
  $u\in\R{}$.
\end{proof}

\noindent\refthm{thm-regular-is-dfa} gives us a way to construct a DFA
for a regular expression \rx{r}, see \refalg{alg-make-dfa}. It
basically performs a breadth first search of all possible following
states, for each state, while constructing \d{}.

\begin{example}\label{xmp-dfa-construction}
  Suppose that $\A{} = \set{a,b,c}$ and that $\rx{q_0} = \rx{ab^*}$. Let us
  construct a DFA that recognizes \rx{q_0}.
  \begin{eqnarray}
    \D{a}{\rx{q_0}} &=& \D{a}{(\rx{ab^*})} = \rx{b^*} = \rx{q_1} \\
    \D{b}{\rx{q_0}} &=& \D{b}{(\rx{ab^*})} = \emptyset = \rx{q_{\emptyset}} \\
    \D{c}{\rx{q_0}} &=& \D{c}{(\rx{ab^*})} = \emptyset = \rx{q_{\emptyset}} \\
    \D{a}{\rx{q_1}} &=& \D{a}{(\rx{b^*})} = \emptyset = \rx{q_{\emptyset}} \\
    \D{b}{\rx{q_1}} &=& \D{b}{(\rx{b^*})} = \D{b}{(\rx{b})}\rx{b^*} = \rx{b^*} = \rx{q_1} \\
    \D{c}{\rx{q_1}} &=& \D{b}{(\rx{b^*})} = \emptyset = \rx{q_{\emptyset}}
  \end{eqnarray}
  The constructed DFA can be seen in \reffig{fig-rx-dfa-match}.
\end{example}

\begin{algorithm}[t]
  \caption{DFA construction using derivatives}\label{alg-make-dfa}
  \begin{algorithmic}[1]
    \Procedure{MakeDfa}{\rx{r},\protect\A{}}
      \State $q_0 \gets \rx{r}$
      \State $\d{} \gets \emptyset$
      \State $F \gets \emptyset$
      \State $Q \gets \set{\rx{r}}$
      \State $S \gets \set{\rx{r}}$
      \While{$S \neq \emptyset$}
        \State $\rx{r} \gets \Call{Pop}{S}$
        \If{\n{\rx{r}} = \emptystring }
          \State \Call{Push}{\rx{r}, $F$} 
        \EndIf
        \For{$a \in \A{}$}
          \State $\rx{d} \gets \D{a}{\rx{r}}$
          \If{$\exists \rx{\bar{d}} \in Q$ such that
            $\rx{\bar{d}} = \rx{d}$}\label{row-make-dfa-check-eq}
            \State \Call{Push}{$(\rx{r}, a) \mapsto \rx{\bar{d}}$, \d{}}
          \Else
            \State \Call{Push}{\rx{d}, $Q$}
            \State \Call{Push}{\rx{d}, $S$}
            \State \Call{Push}{$(\rx{r}, a) \mapsto \rx{d}$, \d{}}
          \EndIf
        \EndFor  
      \EndWhile
      \Return{$(\A{}, Q, q_0, \d{}, F)$}
    \EndProcedure
  \end{algorithmic}
\end{algorithm}

We finish this section with a theorem that characterizes the size of
DFAs constructed using this method.

\begin{definition}[DFA equivalence]
  Two DFAs that recognizes the same language are said to be
  \define{equivalent}.
\end{definition}

\begin{definition}[Minimal DFA]
  A DFA is said to be minimal if no DFA with fewer states recognizes
  the same language.
\end{definition}

\begin{theorem}
  For any DFA, there exists a unique (up to the label of the states)
  equivalent minimal DFA.
\end{theorem}

\begin{proof}
  From the proofs of \refthm{thm-dfa-is-regular},
  \refthm{thm-regular-is-dfa} it is clear that a DFA that recognizes
  \R{} corresponds to the set of $d_{\R{}}$ characteristic equations
  for \R{}. According to \refthm{thm-cd-eqn} this set of equations is
  unique. Hence, any DFA recognizing \R{} will correspond to the same
  set of equations. Further, if a DFA has more than $d_{\R{}}$ states,
  at least two of them must be equivalent, since they will correspond
  to the same characteristic equation, and therefore recognize the
  same language. Therefore, any minimal DFA for \R{} will have
  precisely one state corresponding to one equation, likewise \d{} and
  $F$ are determined by the equations.
\end{proof}

\begin{corollary}
  Constructing a DFA that recognizes \R{} by using the characteristic
  equations of \R{} results in a minimal DFA.
\end{corollary}

\section{Practical considerations}

Although \refalg{alg-make-dfa} works in theory it is problematic in
practice. On line \ref{row-make-dfa-check-eq} we check for equality
between two regular expressions. This task is of non-elementary
complexity\footnote{Owens et al.\autocite{owens} quotes Aho et
  al.\autocite{aho1974design}}. The algorithm also loops over all
symbols in \A{}, and calculates the derivative with respect to each
one, for every state. This is time consuming for large sets of
symbols. Both of these issues will be considered in the following
sections.

\subsection{Similar expressions}

We noted above that determining equality of regular expressions is
non-trivial. However, the following theorem \autocite[Theorem
  5.2]{brz} shows that it is possible to relax this requirement, at
the expense of possibly constructing a non-minimal DFA.

\begin{definition}[Similarity]\label{def-basic-similarity}
  Two regular expressions, \rx{r_1} and \rx{r_2}, are
  \define{similar}, written $\rx{r_1} \simeq \rx{r_2}$, if one
  can be transformed to the other by using only the identities:
  \begin{eqnarray}
    \rx{r \union r} &=& \rx{r} \label{eqn-similarity-absorb}\\
    \rx{p \union q} &=& \rx{q \union p} \label{eqn-similarity-commute}\\
    \rx{(p \union q) \union r} &=& \rx{p \union (q \union r)} \label{eqn-similarity-associate}
  \end{eqnarray}
  Expressions that are not similar are said to be \define{dissimilar}.
\end{definition}

\begin{theorem}
  Every regular expression has only a finite number of dissimilar
  derivatives.
\end{theorem}

\begin{proof}
  To prove this theorem we must show that the process of constructing
  derivatives will terminate in a finite number of steps, even if only
  similarity is recognized. However, it is enough to prove it for
  regular expressions \rx{r} containing only the basic operators
  $\union$, $\cdot$ and $*$, since $\union$ and $\comp{}$ form a
  complete set of Boolean connectives, and it follows from
  \refthm{thm-regular-complement} that any complement can be written
  as an expression containing only the basic operators.

  For such \rx{r} the proof is implicit in the proof of
  \refthm{thm-regular-finite-derivative}, but let us make it more
  explicit.
  
  The theorem is obviously true for \rx{r} with $n = 0$ number of
  operators. Now, suppose $\rx{r} = \rx{r_1 \union r_2}$. This
  corresponds to Case 1 in \refthm{thm-regular-finite-derivative}. We
  have $\D{s}{\rx{r}} = \D{s}{\rx{r_1}} \union
  \D{s}{\rx{r_2}}$. Further, as $s$ takes on all possible values,
  compare \D{s}{\rx{r}} with all the previously found
  derivatives. Since \D{s}{\rx{r_1}} and \D{s}{\rx{r_2}} appears only
  a finite number of times in $\D{s}{\rx{r_1}} \union
  \D{s}{\rx{r_2}}$, and $\rx{r_1}$ and $\rx{r_2}$ has a finite number
  of dissimilar derivatives by the induction step, the process will
  clearly terminate.

  Case 2, with $\rx{r} = \rx{r_1r_2}$ is not as straight
  forward. However, we can write 
  \begin{eqnarray}
    && \D{a_1a_2...a_k}{\rx{r}} = \D{a_1a_2...a_k}{(\rx{r_1})}\rx{r_2} \union
    \n{\D{a_1a_2...a_{k-1}}{\rx{r_1}}}\D{a_k}{\rx{r_2}} \union \cdots \nonumber\\
    && \qquad \union \n{\D{a_1}{\rx{r_1}}}\D{a_2...a_k}{\rx{r_2}} \union
    \n{\rx{r_1}}\D{a_1a_2...a_k}{\rx{r_2}}
  \end{eqnarray}
  which is simply \refeqn{eqn-regular-finite-derivative-concat} put in
  terms of regular expressions. Notice that the number of terms
  increases with the length of $s$, but also notice that the bound,
  $d_{\rx{r}} \le d_{\rx{r_1}}2^{d_{\rx{r_2}}}$, resulting from the
  above equation is independent of $s$. Property
  \refeqn{eqn-similarity-associate} have been used to remove
  parentheses, and \refeqn{eqn-similarity-commute} identify
  derivatives where the terms are in different order. However, it is
  property \refeqn{eqn-similarity-absorb} that let us reduce the
  number of terms in new derivatives, and by that terminate the
  process. Thus, each new derivative must first be simplified, and
  then compared with the already found derivatives.

  The same argument applies to Case 3, when $\rx{r} = \rx{r_0^*}$.  
\end{proof}

Brzozowski found that although it is possible to use this concept of
similarity to construct DFAs, it generally results in automata with
large number of states, far from the minimal. On the other hand, Owens
et al. have shown that by expanding the number properties in the
definition of similirarity to those stated in \refdef{def-similarity},
the resulting DFAs are, in practice, often minimal or close to
minimal.

\begin{definition}[Expanded similarity]\label{def-similarity}
  Two regular expressions, \rx{r_1} and \rx{r_2}, are
  \define{similar}, written $\rx{r_1} \simeq \rx{r_2}$, if one
  can be transformed to the other by using only the identities:
  \vspace{-1em}
  
  \noindent\begin{minipage}[t]{0.50\textwidth}
    \begin{eqnarray*}
      \rx{r \intersection r} &=& \rx{r} \\
      \rx{p \intersection q} &=& \rx{q \intersection p} \\
      \rx{(p \intersection q) \intersection r} &=& \rx{p \intersection (q \intersection r)} \\
      \emptyset \intersection \rx{r} &=& \emptyset \\
      \comp\emptyset \intersection \rx{r} &=& \rx{r}
    \end{eqnarray*}
    \begin{eqnarray*}
      \rx{(p \cdot q) \cdot r} &=& \rx{p \cdot (q \cdot r)} \\
      \rx{\emptyset \cdot r} &=& \rx{\emptyset} \\
      \rx{r \cdot \emptyset} &=& \rx{\emptyset} \\
      \rx{\emptystring \cdot r} &=& \rx{r} \\
      \rx{r \cdot \emptystring} &=& \rx{r} \\
    \end{eqnarray*}
  \end{minipage}
  \begin{minipage}[t]{0.50\textwidth}
    \begin{eqnarray*}
      \rx{r \union r} &=& \rx{r} \\
      \rx{p \union q} &=& \rx{q \union p} \\
      \rx{(p \union q) \union r} &=& \rx{p \union (q \union r)} \\
      \emptyset \union \rx{r} &=& \rx{r} \\
      \comp\emptyset \union \rx{r} &=& \comp\emptyset
    \end{eqnarray*}
    \begin{eqnarray*}
      \rx{(r^*)^*} &=& \rx{r^*} \\
      \emptystring^* &=& \emptystring \\
      \emptyset^* &=& \emptystring \\
      \comp(\comp\rx{r}) &=& \rx{r}
    \end{eqnarray*}
  \end{minipage}
\end{definition}
From now on, when we talk about similarity, it is the properties in
\refdef{def-similarity} that we consider and make use of.

\begin{example}\label{xmp-suboptimal-dfa-construction}
  Suppose that $\A{} = \set{a,b,c}$ and that $\rx{r_0} = (a \union ab
  \union b)^*$. The difference between using similarity and equaltiy
  when constructing the corresponding DFA is shown in
  \reffig{fig-suboptimal-dfa-construction}. As can be seen, the DFA
  when using similarity is not minimal. It is readily seen why.
  \begin{eqnarray}
    \D{a}{\rx{r_0}} &=& \D{a}{\rx{(a \union ab \union b)^*}} \nonumber\\
    &=& \D{a}{\rx{(a \union ab \union b)}}\rx{(a \union ab \union b)^*} \nonumber\\
    &=& \rx{(\emptystring \union b)(a \union ab \union b)^*} \nonumber\\
    &=& \rx{r_0}
  \end{eqnarray}
  The last equality follows from the fact that
  \begin{eqnarray}
    \rx{(a \union ab \union b)^*}
    &\subseteq& \rx{(\emptystring \union b)(a \union ab \union b)^*} \nonumber\\
    &\subseteq& \rx{(\emptystring \union a \union ab \union b)(a \union ab \union b)^*} \nonumber\\
    &=& \rx{(a \union ab \union b)^*}
  \end{eqnarray}
  which is an equality missed by the similarity definition.
  \begin{figure}
    \center
    \begin{minipage}[t]{0.45\textwidth}
      \centering
      \begin{tikzpicture}
        \node[state,initial, accepting]   (0) {\rx{r_0}};
        \node[state, accepting] (1) [above right=of 2] {\rx{r_1}};
        \node[state] (2) [below right=of 0] {\rx{r_{\emptyset}}};
        \path
        (0) edge [loop above] node {$b$} ()
        (0) edge [bend left] node  {$a$} (1)
        (0) edge [bend right] node [swap] {$c$} (2)
        
        (1) edge [bend left] node [swap] {$b$} (0)
        (1) edge [loop above] node {$a$} ()
        (1) edge [bend left] node  {$c$} (2)

        (2) edge [loop below] node {$a,b,c$} ()
        ;
      \end{tikzpicture}
      \mbox{(a) Constructed using similarity.}
    \end{minipage}%
    \begin{minipage}[t]{0.45\textwidth}
      \centering
      \begin{tikzpicture}
        \node[state,initial, accepting]   (0) {\rx{r_0}};
        \node[state] (2) [below right=of 0] {\rx{r_{\emptyset}}};
        \path
        (0) edge [loop above] node {$a,b$} ()
        (0) edge              node {$c$} (2)
        
        (2) edge [loop below] node {$a,b,c$} ()
        ;
      \end{tikzpicture}
      \mbox{(b) Constructed using equality.}
    \end{minipage}
    \begin{minipage}{0.85\textwidth}
      \caption[\refxmp{xmp-suboptimal-dfa-construction} Expression
        construction]{Resulting DFA in
        \refxmp{xmp-suboptimal-dfa-construction} as transition
        diagrams.}\label{fig-suboptimal-dfa-construction}
    \end{minipage}
  \end{figure}  
\end{example}
\pagebreak

\subsection{Sets of symbols instead of single symbols}\label{sec-sets-of-symbols}

In the beginning of this section we noted that when calculating the
transitions from a state it is impractical to calculate the derivative
for each symbol for large alphabets like e.g. Unicode. This can
be remedied by using derivative classes from section
\refsec{sec-derivative-classes}. There we mention that if two symbols
belong to the same derivative class for some language \L{}, the
derivative of \L{} with respect to them are equal. Hence, if we know
the derivative classes, we only need to calculate as many derivatives
as there are derivative classes. The problem is of course to calculate
the derivative classes without calculating the
derivatives.\footnote{According to Owens et al. it is possible to
  determine derivative classes without calculating any derivatives,
  although it generally requires $O(\length{\A{}})$ work. However,
  they never mention how this is done.} Owens et al. gives a function
that computes an approximation for this.

The first step is to reformulate the syntax of regular expressions and
slightly change what ``atom'' means by instead of the previous
definition in \refdef{def-regular-expression} use\footnote{Note that
  $\rx{a_i} = \rx{[a_i]}$, so we could do without \rx{a_i}, but we
  keep it for convenience. We don't keep the notation for $\emptyset$
  however, since it is covered by the \rx{[\;]} case.}

\begin{center}
  \setlength{\tabcolsep}{0.3em}
  \renewcommand{\arraystretch}{1.25}
    \begin{tabular}{rcl}
      \nt{atom} &=& "\comp{}", \nt{atom} \\
      &|& "\rx{a_1}" | "\rx{a_2}" | $\cdots$ | "\rx{a_n}" \\
      &|& "[", \nt{list}, "]" (* we usually write this as \rx{A} *)\\
      &|& "" (* we usually write this as \emptystring\ *) \\
      &|& "(", \nt{disj}, ")"\\
      \nt{list} &=& "\rx{a_i}", \nt{list} \\
      &|& ""
    \end{tabular}
\end{center}
with the corresponding additional languages
\begin{eqnarray*}
  \rxlang{[\;]} &=& \emptyset \\
  \rxlang{A} = \rxlang{[a_i a_j \cdots a_k]} &=& \set{a_i, a_j, \cdots, a_k} = A
      \quad a_i, a_j, \cdots, a_k\in\A{1}
\end{eqnarray*}
i.e we allow sets of symbols $A\subseteq\A{}$ to be atoms, not just
single symbols\footnote{This corresponds to changing rule
  \ref{eqn-regular-base} in \refdef{def-regular-language} to ``$A
  \subseteq \A{}$ is a regular language''.}.

\begin{definition}[Derivative classes approximation]
  \ \\
  Let $\rx{R} = \set{\rx{r}: \rx{r} \; \mbox{is a regular
      expression}}$ and define
  \begin{equation}
    \dca: \rx{R} \rightarrow 2^{2^{\A{}}}
  \end{equation}
  by structural recursion as follows
  \begin{eqnarray*}
    \dca[\emptystring] &=& \set{\A{}} \\
    \dca[\rx{A}] &=& \set{A, \inv{A}} \\
    \dca[\rx{rs}] &=& \left\{ \begin{array}{ll}
        \dca[\rx{r}] \wedge \dca[\rx{s}] & \n{\rx{r}} = \emptystring\\
        \dca[\rx{r}] & otherwise
      \end{array}\right. \\
    \dca[\rx{r} \union \rx{s}] &=& \dca[\rx{r}] \wedge \dca[\rx{s}]\\
    \dca[\rx{r} \intersection \rx{s}] &=& \dca[\rx{r}] \wedge \dca[\rx{s}]\\
    \dca[\rx{r^*}] &=& \dca[\rx{r}] \\
    \dca[\rx{\comp{r}}] &=& \dca[\rx{r}]
  \end{eqnarray*}
  where
  \begin{equation}
    \dca[\rx{r}] \wedge \dca[\rx{s}] = \set{A_{\rx{r}} \intersection A_{\rx{s}} :
      A_{\rx{r}} \in \dca[\rx{r}], \;
      A_{\rx{s}} \in \dca[\rx{s}]}
  \end{equation}
\end{definition}

The next theorem shows that the function \dca[] either gives the exact
derivative classes, or possibly overpartions \A{1}, in which case we
will have to calculate more derivatives than had we known the exact
classes.

\begin{theorem}\label{thm-derivative-classes-approximation}
  Let \rx{r} be a regular expression over some alphabet \A{}. Then for
  all $A\in\dca[\rx{r}]$ and $a \in A$, we have $A\subseteq
  [a]_{\eq{\rx{r}}}$.
\end{theorem}

\begin{proof}
  Let $\Pi$ be the set of all derivative classes of \rx{r} with
  respect to $a \in \A{1}$, i.e. the derivative classes of \rx{r} with
  respect to a string of length 1.

  The proof is by induction over the number $n$ of regular operators
  used to form \rx{r}, using
  \refthm{thm-derivative-class-composition}.
  \begin{description}
  \item[\itemlabel{Base Case $n = 0$:}] \rx{r} can have two different forms,
    \begin{eqnarray}
      \rx{r} = \emptystring &\Rightarrow& \D{a}{\rx{r}} =
        \D{a}{\rx{\emptystring}} = \emptyset \quad a\in\A{} \nonumber\\
      &\Rightarrow& [a]_{\eq{\emptystring}} = \A{} \Rightarrow \Pi =
        \set{[a]_{\eq{\emptystring}}} = \set{\A{}} = \dca[\emptystring] \\
      \rx{r} = \rx{A} &\Rightarrow& \D{a}{\rx{r}} =
        \D{a}{\rx{A}} = \left\{ \begin{array}{cll}
          \set{\emptystring} & a \in A & \Rightarrow [a]_{\eq{\rx{A}}} = A \\
          \emptyset & a \in \inv{A} & \Rightarrow [a]_{\eq{\rx{A}}} = \inv{A}
        \end{array}\right. \nonumber\\
      &\Rightarrow& \Pi = \set{A, \inv{A}} = \dca[\rx{A}]
    \end{eqnarray}
    Hence \dca[] gives exact results for the base cases.
  \item[\itemlabel{Induction step:}] Assume that the theorem holds for
    $n=k$, then it also holds for $n=k+1$. Begin by study the case
    when \rx{r} is of the form $\rx{r} = \rx{st}$. If $\n{\rx{s}} \neq
    \emptystring$ then $\dca[\rx{r}] = \dca[\rx{s}]$ in which case it
    holds by the induction hypothesis.

    Assume that $\n{\rx{s}} = \emptystring$. Let $a,b \in A \in
    \dca[\rx{s}] \wedge \dca[\rx{t}]$. Then, for some $A_\rx{s} \in
    \dca[\rx{s}]$ and $A_\rx{t} \in \dca[\rx{t}]$
    \begin{equation}
      A = A_\rx{s} \intersection A_\rx{t} \subseteq
      \left\{ \begin{array}{cll}
          A_\rx{s} \subseteq [a]_{\eq{\rx{s}}} &\Rightarrow& a \eq{\rx{s}} b \\
          A_\rx{t} \subseteq [a]_{\eq{\rx{t}}} &\Rightarrow& a \eq{\rx{t}} b
        \end{array}\right.
    \end{equation}
    But then it follows from \refthm{thm-derivative-class-composition}
    that $a \eq{\rx{st}} b$ and we therefore have
    \begin{equation}
       A \subseteq [a]_{\eq{\rx{st}}} = [a]_{\eq{\rx{r}}}
    \end{equation}
    and the theorem holds. Cases for the other operators follows by
    similar arguments.\qedhere
  \end{description}
\end{proof}

\chapter{Anchors}\label{chp-anchors}

A feature found in many regular expression matcher implementations are
called \define{anchors}. It is the ability to only match a symbol if
the symbol is found in a particular context, for example at the
beginning/end of a line, or at the beginning/end of a word
etc. These sort of properties are implicit in the input
string\footnote{Of course what constitutes a word or line boundary
  depends on the symbol set. In the following examples we will use the
  regular notion from the ASCII set of symbols.}.

\begin{example}\label{xmp-implicit-anchors}
  Let a string
  \begin{displaymath}
    s = \mbox{Hello, world!}
  \end{displaymath}
  Then there are implicit context markers, anchors, at several
  positions. There is a \emph{``beginning of text''} and a
  \emph{``beginning of line''} before the ``H'', an \emph{``end of
    word''} after ``o'' but before the ``,'', a \emph{``beginning av
    word''} after the space but before the ``w'', an \emph{``end of
    word''} after ``d'' but before ``!'', and \emph{``end of line''}
  and \emph{``end of text''} after the ``!''.
\end{example}

One way to be able to express anchoring properties in regular
expressions is to introduce a set of symbols \B{} with one symbol for
each kind of anchor we want to be able to express. We can then write
regular expressions over $\A[\dagger]{} = \A{}\union\B{}$. This way we
make the anchors explicit in the expression.

\begin{definition}[Anchor symbols]
  \begin{equation}
    \B{} = \set{\bot, \bol, \bow, \eow, \eol, \eot}
  \end{equation}%
  \noindent\begin{minipage}[t]{0.50\textwidth}
  \vspace{-1.5\baselineskip}
  \begin{eqnarray*}
    \bot &=& \mbox{beginning of text} \\
    \bol &=& \mbox{beginning of line} \\
    \bow &=& \mbox{beginning of word} \\
  \end{eqnarray*}
  \end{minipage}
  \noindent\begin{minipage}[t]{0.50\textwidth}
  \vspace{-1.5\baselineskip}
  \begin{eqnarray*}    
    \eot &=& \mbox{end of text} \\
    \eol &=& \mbox{end of line} \\
    \eow &=& \mbox{end of word} \\
  \end{eqnarray*}
  \end{minipage}
\end{definition}

When Using the above idea in a regular expression matcher, we also
need to make the anchors explicit in the input string, i.e. we need to
pre-process the input string to include the anchor symbols.

\begin{example}\label{xmp-explicit-anchors}
  Continuing with the string from \refxmp{xmp-implicit-anchors},
  processing it to include anchors it will look like
  \begin{displaymath}
    s = \mbox{\bot\bol\bow{}Hello\eow, \bow{}world\eow!\eol\eot}
  \end{displaymath}
\end{example}

While regular expressions and strings over \A[\dagger]{} as defined
above work fine, the introduction of \B{} and making anchors explicit
in strings result in some inconvenient side effects.

Consider the string $s$ in \refxmp{xmp-explicit-anchors}. It is a string
over \A[\dagger]{}. However, the the regular expression
\begin{equation}
  \rx{r} = \rx{Hello, world!}
\end{equation}
over \A[\dagger]{} doesn't match the string any longer. It fails
already in the first position since the first input symbol is \bot,
but the expression starts with \rx{H}. In order to match $s$ we would
need to include all the anchor symbols in \rx{r} as well, even if we
are not interested in whether a particular symbol occurs in a specific
context or not. This is inconvenient and it would be nice if we could
make regular expressions ``ignore'' anchor symbols unless we explicitly
include them in the expression. Put in another way, can we make a
regular expression such as \rx{Hello, world!} denote not just the
language \set{``Hello, world!''} (a single string) but also all
variations of it where symbols from \B{} have been inserted between
symbols in the string, e.g. ``He\bol{}llo,
wor\eow{}ld!''\footnote{Most such strings would never occur in
  practice, since the pre-processing of strings $s\in\A{}$ would only
  emit strings following the semantic rules of the symbols. The point
  is to make the regular expression denote all such strings.}?

As it turns out it is indeed possible, by once again alter the
definition of the languages that the atom elements in the regular
expression grammar correspond to. Instead of letting $\rxlang{a_i} =
\set{a_i}$ for $a_i\in\A[\dagger]{1}$ etc. we use

\begin{definition}\label{def-rx-Adagger-language}
  \begin{eqnarray}
    \rxlang{a_i} &=& (\inv{\set{a_i}}\intersection\B{1})^*\set{a_i} \\
    \rxlang{A} = \rxlang{[a_ia_j\cdots a_k]} &=& (\inv{A}\intersection\B{1})^*A \qquad
      A = \set{a_i, a_j,\cdots, a_k} \subseteq\A[\dagger]{1}
  \end{eqnarray}
\end{definition}
With this definition our regular expression example denotes the language
\begin{eqnarray}
  && \rxlang{Hello, world!} = \nonumber\\
  && \qquad \B{*}\set{H}\B{*}\set{e}\B{*}\set{l}\B{*}\set{l}\B{*}\set{o}\B{*}
    \set{,}\B{*}\set{\ }\nonumber\\
  && \qquad \B{*}\set{w}\B{*}\set{o}\B{*}\set{r}\B{*}\set{l}\B{*}\set{d}\B{*}\set{!}
\end{eqnarray}
which is exactly what we want.

\section{Derivatives of anchor symbols}

\refdef{def-rx-Adagger-language} does not alter the definition of
derivative, \refdef{def-derivative}. What we have done is simply to
change what language each symbol in the regular expression
denotes. With these adjustments we see that
\begin{eqnarray}
  \rxlang{\D{b}{\rx{A}}} &=& \D{b}{((\inv{A}\intersection\B{})^*A)} \\
  &=& \D{b}{(\inv{A}\intersection\B{})}(\inv{A}\intersection\B{})^*A \union
    \n{(\inv{A}\intersection\B{})^*}\D{b}{A} \\
  &=& \D{b}{(\inv{A}\intersection\B{})}(\inv{A}\intersection\B{})^*A \union \D{b}{A} \\
  &=& \left\{
  \begin{array}{ll}
    \rxlang{\emptystring} & \quad b\in A \\
    (\inv{A}\intersection\B{})^*A = \rxlang{A} & \quad b\in\inv{A}\intersection\B{1} \\
    \rxlang{\emptyset{}} & \quad b\in\inv{A}\intersection\A{1} \\
  \end{array} \right.
\end{eqnarray}

\begin{example}
  Let $\A{} = \set{a,1}$, and $\B{} = \set{\bow,\eow}$ as defined
  above. Let \rx{r} be a regular expression\footnote{We use \rx{\A{*}}
    to mean \rx{[a_1a_2\cdots a_n]^*} with $a_i\in\A{}$.} over
  \A[\dagger]{} $\rx{r} = \rx{\A{}\A{*}\bow\A{*}}$. Suppose we match
  \rx{r} against four different strings using \refalg{alg-lazy-match}.
  \begin{itemize}
  \item[\itemlabel{a})] $s_1 = \mbox{aaa}$. After pre-processing $s_1$ we have
    \begin{eqnarray*}
      \rx{\A{}\A{*}\bow\A{*}} \sim \bow aaa \eow
      &\Leftrightarrow& \D{\bow}{\rx{(\A{})\A{*}\bow\A{*}}} \sim aaa \eow \\
      &\Leftrightarrow& {\rx{\A{}\A{*}\bow\A{*}}} \sim aaa \eow \\
      &\Leftrightarrow& \D{a}{\rx{(\A{})\A{*}\bow\A{*}}} \sim aa \eow \\
      &\Leftrightarrow& {\rx{\A{*}\bow\A{*}}} \sim aa \eow \\
      &\Leftrightarrow& \D{a}{\rx{\A{*}\bow\A{*}}} \sim a \eow \\
      &\Leftrightarrow& \D{a}{\rx{(\A{})\A{*}\bow\A{*}}}\union\D{a}{\rx{\bow\A{*}}} \sim a \eow \\
      &\Leftrightarrow& {\rx{\A{*}\bow\A{*}}} \sim a \eow \\
      &\Leftrightarrow& \D{a}{\rx{\A{*}\bow\A{*}}} \sim \eow \\
      &\Leftrightarrow& {\rx{\A{*}\bow\A{*}}} \sim \eow \\
      &\Leftrightarrow& \D{\eow}{\rx{\A{*}\bow\A{*}}} \sim \emptystring\\
      &\Leftrightarrow& {\rx{\A{*}\bow\A{*}}} \sim \emptystring = false
    \end{eqnarray*}
  \item[\itemlabel{b})] $s_2 = \mbox{111}$. After pre-processing $s_2$ we have
    \begin{eqnarray*}
      \rx{\A{}\A{*}\bow\A{*}} \sim 111
      &\Leftrightarrow& \D{1}{\rx{(\A{})\A{*}\bow\A{*}}} \sim 11 \\
      &\Leftrightarrow& {\rx{\A{*}\bow\A{*}}} \sim 11 \\
      &\Leftrightarrow& \D{1}{\rx{\A{*}\bow\A{*}}} \sim 1 \\
      &\Leftrightarrow& \D{1}{\rx{(\A{})\A{*}\bow\A{*}}}\union\D{1}{\rx{\bow\A{*}}} \sim 1 \\
      &\Leftrightarrow& {\rx{\A{*}\bow\A{*}}} \sim 1 \\
      &\Leftrightarrow& \D{1}{\rx{\A{*}\bow\A{*}}} \sim \emptystring \\
      &\Leftrightarrow& {\rx{\A{*}\bow\A{*}}} \sim \emptystring = false
    \end{eqnarray*}
  \item[\itemlabel{c})] $s_3 = \mbox{aa1}$. After pre-processing $s_3$ we have
    \begin{eqnarray*}
      \rx{\A{}\A{*}\bow\A{*}} \sim \bow aa\eow1 
      &\Leftrightarrow& \D{\bow}{\rx{(\A{})\A{*}\bow\A{*}}} \sim aa \eow 1 \\
      &\Leftrightarrow& {\rx{\A{}\A{*}\bow\A{*}}} \sim aa \eow 1 \\
      &\Leftrightarrow& \D{a}{\rx{(\A{})\A{*}\bow\A{*}}} \sim a\eow 1 \\
      &\Leftrightarrow& {\rx{\A{*}\bow\A{*}}} \sim a \eow 1 \\
      &\Leftrightarrow& \D{a}{\rx{\A{*}\bow\A{*}}} \sim \eow 1 \\
      &\Leftrightarrow& \D{a}{\rx{(\A{})\A{*}\bow\A{*}}}\union\D{a}{\rx{\bow\A{*}}} \sim \eow 1 \\
      &\Leftrightarrow& {\rx{\A{*}\bow\A{*}}} \sim \eow 1 \\
      &\Leftrightarrow& \D{\eow}{\rx{\A{*}\bow\A{*}}} \sim 1 \\
      &\Leftrightarrow& {\rx{\A{*}\bow\A{*}}} \sim 1 \\
      &\Leftrightarrow& \D{1}{\rx{\A{*}\bow\A{*}}} \sim \emptystring\\
      &\Leftrightarrow& {\rx{\A{*}\bow\A{*}}} \sim \emptystring = false
    \end{eqnarray*}
  \item[\itemlabel{d})] $s_4 = \mbox{1a}$. After pre-processing $s_4$ we have
    \begin{eqnarray*}
      \rx{\A{}\A{*}\bow\A{*}} \sim 1 \bow a \eow 
      &\Leftrightarrow& \D{1}{\rx{(\A{})\A{*}\bow\A{*}}} \sim \bow a\eow \\
      &\Leftrightarrow& {\rx{\A{*}\bow\A{*}}} \sim \bow a \eow \\
      &\Leftrightarrow& \D{\bow}{\rx{\A{*}\bow\A{*}}} \sim a \eow \\
      &\Leftrightarrow& \D{\bow}{\rx{(\A{})\A{*}\bow\A{*}}} \union \D{\bow}{\rx{\bow\A{*}}} \sim a \eow \\
      &\Leftrightarrow& {\rx{\A{*}\bow\A{*}}} \union {\rx{\A{*}}} \sim a \eow \\
      &\Leftrightarrow& \D{a}{\rx{\A{*}\bow\A{*}}} \union \D{a}{\rx{\A{*}}} \sim \eow \\
      &\Leftrightarrow& {\rx{\A{*}\bow\A{*}}} \union {\rx{\A{*}}} \sim \eow \\
      &\Leftrightarrow& \D{\eow}{\rx{\A{*}\bow\A{*}}} \union \D{\eow}{\rx{\A{*}}} \sim \emptystring \\
      &\Leftrightarrow& {\rx{\A{*}\bow\A{*}}} \union {\rx{\A{*}}} \sim \emptystring = true
    \end{eqnarray*}
  \end{itemize}
  As expected the expression matches strings with at least one \bow{}
  (``beginning-of-word'') anchor within the string. Case a) and c)
  above has a \bow{} at the beginning of the string, but not inside it,
  and case b) does not contain the anchor at all.
\end{example}

\begin{example}
  Imagine that we would like to match a string containing the two
  words ``hello'' and ``world'' irrespective of the case of the
  initial letter in each word, and irrespective of if they occur on
  the same line or not. Also we want to ignore any number of spaces,
  and if there is a comma or not in between the
  words\footnote{Programmers around the world, and in different
    languages, have not agreed upon if it should be ``Hello World'',
    ``hello, world'', ``Hello, world!'' etc. It's all a mess really
    ;-)}. A possible regular expression could be\footnote{We use
    \rx{\A{*}} to mean \rx{[a_1a_2\cdots a_n]^*}, with $a_i\in\A{}$,
    and $\bar{n}, \bar{t}, \bar{s}$ to mean newline, tab, and space.}
  \begin{equation}
    \rx{r} = \rx{\A{*}[Hh]ello[,\emptystring][\bar{n}\bar{t}\bar{s}]^*[Ww]orld\A{*}}
  \end{equation}
  This expression would get much more complicated if we needed to take
  the possible anchors into account as well. However, the anchors come
  handy if we, for example, want to make sure that the ``hello'' part begins on
  a new line, then we could use
  \begin{equation}
    \rx{r} = \rx{\A{*}\bol[Hh]ello[,\emptystring][\bar{n}\bar{t}\bar{s}]^*[Ww]orld\A{*}}
  \end{equation}
\end{example}

\section{Single symbols, word boundaries and more}

In \refdef{def-rx-Adagger-language} we changed the languages denoted
by regular expressions. This change leads us to the question whether
we can still express the language $\set{a_i}\subseteq\A[\dagger]{*}$,
i.e. the language consisting of a string of length one without a
prefix of possible anchors, with a regular expression or not.

For this purpose, let $A\subseteq\A[\dagger]{1}$, and study the
expression
\begin{eqnarray}
  \D{b}{\rx{r}}
  &=& \D{b}{(\rx{\comp{(A\A[\dagger]{*})}\intersection s})} \nonumber\\
  &=& \D{b}{(\rx{\comp{(A\A[\dagger]{*})}})} \intersection \D{b}{\rx{s}} \nonumber\\
  &=& \comp{(\D{b}{\rx{(A\A[\dagger]{*})}})} \intersection \D{b}{\rx{s}} \nonumber\\
  &=& \comp{(\D{b}{(\rx{A})} \rx{\A[\dagger]{*}})} \intersection \D{b}{\rx{s}} \nonumber\\
  &=& \left\{
  \begin{array}{ll}
    \comp{(\rx{A\A[\dagger]{*}})} \intersection \D{b}{\rx{s}} &\quad b \in\inv{A}\intersection\B{1}\\
    \comp{(\rx{\emptyset})} \intersection \D{b}{\rx{s}} = \D{b}{\rx{s}}&
      \quad b \in \inv{A}\intersection\A{1}\\
    \comp{(\rx{\A[\dagger]{*}})} \intersection \D{b}{\rx{s}} = \rx{\emptyset}& \quad b \in A \\
    \emptyset & \quad \D{b}{\rx{s}} = \emptyset
  \end{array} \right. \nonumber\\
  &=& \left\{
  \begin{array}{ll}
    \comp{(\rx{A\A[\dagger]{*}})} \intersection \D{b}{\rx{s}} &\quad b \in\inv{A}\intersection\B{1}\\
    \D{b}{\rx{s}}& \quad b \in \inv{A}\intersection\A{1}\\
    \emptyset & \quad otherwise
    \end{array} \right. \label{eqn-boundery-conditions}
\end{eqnarray}

Let us investigate \refeqn{eqn-boundery-conditions} for a few different cases.
\begin{itemize}
\item[\itemlabel{a})] Let $A=\inv{\set{a_i}}\subseteq\A[\dagger]{1}$, and $\rx{s} = \rx{a_i}$, then
  \begin{eqnarray}
    \D{b}{\rx{r}} &=& \D{b}{(\rx{\comp{([a_1a_2\cdots a_{i-1}a_{i+1}\cdots a_{n}]\A[\dagger]{*})}
        \intersection a_i})} =
      \D{b}{(\rx{\comp{(\inv{[a_i]}\A[\dagger]{*})} \intersection a_i})} \nonumber \\
    &=& \left\{
    \begin{array}{ll}
      \rx{\comp{(\inv{[a_i]}\A[\dagger]{*})}}
        \intersection \D{b}{\rx{a_i}} = \emptystring & \quad b = a_i \in\B{1} \\
      \D{b}{\rx{a_i}} = \emptystring & \quad b = a_i \in\A{1} \\
      \emptyset & \quad otherwise
    \end{array} \right. \nonumber \\
    &=& \left\{
    \begin{array}{ll}
      \emptystring & \quad b = a_i \\
      \emptyset & \quad otherwise
    \end{array} \right.
  \end{eqnarray}
  Thus, by \refthm{thm-derivative-decide}, $\rx{r} =
  (\rx{\comp{([a_1a_2\cdots a_{i-1}a_{i+1}\cdots
        a_{n}]\A[\dagger]{*})} \intersection a_i})$ is a regular
  expression over \A[\dagger]{} that denotes the language \set{a_i}.
\item[\itemlabel{b})] Let $A=\B{1}$, then $\inv{A}\intersection\B{1} =
  \emptyset$ and we get
  \begin{equation}
    \D{b}{\rx{r}}
    = \D{b}{(\rx{\comp{(\B{}\A[\dagger]{*})}\intersection s})}
    = \left\{
    \begin{array}{ll}
      \D{b}{\rx{s}}& \quad b \in \A{1}\\
      \emptyset & \quad otherwise
    \end{array} \right. \label{eqn-no-anchor-beginning}
  \end{equation}
  The expression \rx{\comp{(\B{}\A[\dagger]{*})}\intersection s}
  will therefore match strings matching \rx{s} as long as they don't
  begin with an anchor (symbol in \B{}).
\item[\itemlabel{c)}] Let $A=[\bow\eow]$, then
  $\inv{A}\intersection\B{1} = [\bot\bol\eol\eot]$ and we have
  \begin{eqnarray}
    \D{b}{\rx{r}}
    &=& \D{b}{(\rx{\comp{([\bow\eow]\A[\dagger]{*})}\intersection s})} \nonumber\\
    &=& \left\{
    \begin{array}{ll}
      \comp{(\rx{[\bow\eow]\A[\dagger]{*}})} \intersection \D{b}{\rx{s}}
        &\quad b \in [\bot\bol\eol\eot] \\
      \D{b}{\rx{s}}& \quad b \in \A{1}\\
      \emptyset & \quad otherwise
    \end{array} \right. \label{eqn-no-word-anchor}
  \end{eqnarray}
  Thus, \rx{r} will match any string matching \rx{s}, as long as it does not
  start with either \bow{} or \eow.
\end{itemize}

\begin{example}
  Equation \refeqn{eqn-no-word-anchor} is useful if we want an
  expression that forbids a word boundary at a specific
  position. Suppose we want to match \rx{s} followed by \rx{t}, but
  only when \rx{t} is not at a word boundary. One such expression is
  \begin{equation}
    \rx{r} = \rx{s(\comp{([\bow\eow]\A[\dagger]{*})} \intersection t)}
  \end{equation}
\end{example}

\begin{example}
  If we instead want an expression that requires a word boundary
  between the two parts \rx{s} and \rx{t} we can use
  \begin{equation}
    \rx{r} = \rx{s[\bow\eow]t}
  \end{equation}
\end{example}

\chapter{Submatching}\label{chp-submatching}

One of the most useful features found in many regular expression
matcher implementations is the ability to tell which part of the input
string matched a certain part of the expression. This is often called
\define{submatching}.

\begin{example}\label{xmp-intro-memory-ambiguity}
  Let $\A{} = \set{a,b,c}$ and suppose we have the expression $\rx{r}
  = \rx{[ab]^*[bc]^*[ac]^*}$. The algorithms we have discussed so far
  can tell us that a string, e.g. $s = abbcc$, is part of the language
  denoted by \rx{r}, i.e. that \rx{r} matches $s$. However, they
  cannot tell us if the first part of \rx{r}, \rx{[ab]^*}, matched $a$,
  $ab$, or $abb$. There are, in fact, a number of ways that the
  different parts of \rx{r} can match the input $s$, we give two
  examples among others
  \begin{eqnarray}
    \underbrace{abb}_{\rx{[ab]^*}}\underbrace{cc}_{\rx{[bc]^*}}\underbrace{\emptystring}_{\rx{[ac]^*}} \\
    \underbrace{a}_{\rx{[ab]^*}}\underbrace{bbc}_{\rx{[bc]^*}}\underbrace{c}_{\rx{[ac]^*}}
  \end{eqnarray}
\end{example}

In this chapter we will discuss how submatching can be implemented
using derivatives.

\section{Tags, banks, and slots}

Regular expressions denote regular
languages. \refthm{thm-derivative-decide} can help us decide if a
string belongs to a language or not. As we have seen, derivatives (and
their corresponding states in a DFA) denotes what is \emph{left to
  match} after partially processing an input string, i.e. they do not
contain information about what has been seen before. In order track
which part of a particular expression \rx{r} that matches a certain
part of a string $s$ we need some sort of memory. Inspired by Ville
Laurikari and his work on tagged transitions\autocite{laurikari-2}, we
therefore introduce four formal symbols into the regular expression
syntax. We call them formal because they don't occur as parts of the
language the expression denotes, and they don't affect derivation, but
we manipulate them in a similar manner to other symbols. The idea is
to let these formal symbols encode memory state, and memory operations
to take place at certain points in the matching process.

First, in order to be able to track when a certain point in the
expression is reached, we introduce \define{tags}. Each tag represents
a position, and a pair of tags will work as submatch delimiters. Tags
come in two flavors, \define{early} and \define{late}. The benefit of
having two kinds of tags will become clear in section
\refsec{sec-matching-modes}. Moreover, tags are enumerated, so each
tag has a unique number.

Second, we note that each term in an expression denotes an alternative
sequence to match. Thus, an expression of the form $\rx{r} = \rx{s
  \union t}$ has two different sequences, or paths, that can
match\footnote{Note that these paths need not to be exclusive, a
  particular input string may actually match both paths.}. In order to
keep track of which path/paths that match, we augment each term in an
expression with a \define{memory bank}. Each memory bank contains a
set of \define{memory slots}, the number of slots in each bank equals
the number of tags in the expression. A slot in a bank remembers a
certain position in the input string. Initially each slot contains a
value which indicates that the point in the expression has not been
reached, e.g. $-1$.

Third, we introduce the notion of \define{memory slot update}. A slot
update represents an operation to update the slot in a particular bank
with a position value.

The basic idea is to let derivation of tags, issue memory slot updates
towards memory banks.

\begin{definition}[Tags, bank, slot]
  \begin{equation}
    \T{} = \set{\etag{i}, \ltag{j}, \bank{k}, \slot{l}{p}} \qquad i,j,k,l,p\in\N
  \end{equation}
  \begin{eqnarray*}
    \etag{i} &=& \mbox{early tag, numbered $i$.} \\
    \ltag{j} &=& \mbox{late tag, numbered $j$.}\\
    \bank{k} &=& \mbox{memory bank $k$} \\
    \slot{l}{p} &=& \mbox{update memory slot $l$ with value $p$}
  \end{eqnarray*}
  We also define $\A[+]{} = \A{}\union\T{}$, and let \slotvalue{m}{n}
  mean the value of slot $m$ in bank $n$.
\end{definition}

Having introduced the symbols in \T{} we also extend the regular
expression grammar for the non-terminal ``atom'' to account for these
new terminals.

\begin{center}
  \setlength{\tabcolsep}{0.3em}
  \renewcommand{\arraystretch}{1.25}
  \begin{tabular}{rcl}
    \nt{atom} &=& "\comp{}", \nt{atom} \\
    &|& "\rx{a_1}" | "\rx{a_2}" | $\cdots$ | "\rx{a_n}" \quad $a_j\subseteq\A{1}$\\ 
    &|& "[", \nt{list}, "]" (* we usually write this as \rx{A} *)\\
    &|& "" (* we usually write this as \emptystring\ *) \\
    &|& "(", \nt{disj}, ")"\\
    &|& "\etag{2i}", \nt{disj}, "\ltag{2i+1}"
     |  "\etag{2i}", \nt{disj}, "\etag{2i+1}"  \quad $i\in\N$\\
    &|& "\bank{k}" | "\slot{l}{p}" \quad $k, l, p\in\N$\\
    \nt{list} &=& "\rx{a_i}", \nt{list} \quad $a_i\subseteq\A{1}$\\
    &|& ""
  \end{tabular}
\end{center}

Study the productions we have added for tags. Tags always occurs in
pairs either an early and a late, or two early ones. The semantics is
the same as for parentheses, i.e. a grouping that may change the
precedence. However, in addition to that, they will also record
positions. The numbering of tags will be discussed further in
\refsec{sec-bank-order}.

As mentioned earlier, we do not want these symbols to affect the
language denoted by an expression. Formally, we therefore let all the
symbols correspond to the \set{\emptystring} language\footnote{This
  will preserve the denoted language except for pathological edge cases
  where symbols $\gamma\in\T{}$ occur on their own, such as
  \rx{(\emptyset\union\etag{i})} or
  \rx{(r\intersection\ltag{i}}). While a theoretical possibility, such
  expressions are rather pointless to use in practice since the
  submatch will not capture anything.}, i.e.
\begin{equation}
  \rxlang{\;\etag{i}\;}, \rxlang{\;\ltag{j}\;}, \rxlang{\bank{k}}, \rxlang{\slot{l}{p}}
    = \set{\emptystring}
\end{equation}

Now, let \rx{r} be a regular expression over \A[+]{}. The expression
$\bank{k}\rx{r}$ is to be interpreted as the expression \rx{r}
augmented with memory bank \bank{k}. \bank{k} keeps a history record
of how \rx{r} was formed. Banks such as \bank{k} will be discussed in
more detail later, when we discuss derivatives of symbols in \T{}.

We postulate the following ``copy-on-distribution'' property for
memory banks. That is, when we use the distributive law together with
a memory bank, each term gets an independent copy of the bank. As we
will see later, it is sometimes useful to keep track of which bank
made which copy in intermediate expressions. We use the
$\hat{\bank{k}}$ as a notation for that, meaning that $\hat{\bank{k}}$
is a copy of \bank{k}.
\begin{equation}
  \bank{k}\rx{(r \union s)}
  = \bank{k}\rx{r} \union \hat{\bank{k}}\rx{s}
  = \bank{k}\rx{r} \union \bank{m}\rx{s}
\end{equation}
Bank \bank{m} has initially the same content as \bank{k}, but slots in
each bank can be updated independently, and hence the memory banks
will evolve independent of each other.

Also, since the formal language of \bank{} is \set{\emptystring}, it
is reasonable (and useful, as we will see later on) to define the
equalities
\begin{equation}
  \emptystring \union \bank{k} = \bank{k}\label{eqn-emptystring-bank-absorbtion}
\end{equation}

Further, we define an expression of the form
\begin{equation}
  \bank{k} = \bank{k}\slot{i}{p}
\end{equation}
to mean that the value of slot number $i$ in \bank{k} is to be set
to $p$. Moreover, slot updates have the following properties
\begin{eqnarray}
  (\rx{r}\intersection\slot{i}{p}\rx{s})
  &=& (\slot{i}{p}\rx{r}\intersection\rx{s})
    = \slot{i}{p}(\rx{r}\intersection\rx{s})\\
  \comp{(\slot{i}{p}\rx{r})} &=& \slot{i}{p}\comp{(\rx{r})}\\
  \emptystring \union \slot{i}{p} &=& \slot{i}{p}
\end{eqnarray}
We also set\footnote{The case with concatenations of banks
  does not occur in practice. However, the definition simplifies the
  proof of \refthm{thm-teval-idempotence}}
\begin{equation}
  \bank{k}\bank{k} = \bank{k}
\end{equation}

\begin{definition}[Tag, bank, slot derivative and nullify function]
  Let \rx{r} be a regular expression over \A[+]{}, $s\in\A{*}$ be a
  string, and let $p$ be the current position\footnote{The current
    position is a value defined by the environment. In a typical
    matching scenario, using an algorithm like \refalg{alg-lazy-match},
    this will be the current position in the input string}. We make
  the following definitions for the nullify function

  \begin{eqnarray}
    \n{\etag{i}} &=& \slot{i}{p}\\
    \n{\ltag{i}} &=& \slot{i}{p}\\
    \n{\slot{i}{p}} &=& \slot{i}{p}\\
    \n{\bank{k}} &=& \bank{k}
  \end{eqnarray}  

  Further, symbols in \T{} behave as \emptystring, and follows the
  ordinary rules of derivation, see \refthm{thm-derivative}. In
  particular we have for $\gamma\in\T{}$
  
  \begin{eqnarray}
    \D{s}{\gamma} &=& \left\{
    \begin{array}{ll}
      \gamma & \quad s = \emptystring\\
      \emptyset & \quad otherwise
    \end{array}\right.\\
    \D{s}{(\gamma\rx{r})} &=&
    \D{s}{(\gamma)}\rx{r}\union\n{\gamma}\D{s}{\rx{r}}\label{eqn-DaCat-tag}
  \end{eqnarray}
  For future reference we state rule \refeqn{eqn-DaCat-tag} explicitly
  for $s=a\in\A{1}$.
  \begin{eqnarray}
    \D{a}{(\etag{i}\rx{r})} &=& \D{a}{(\etag{i})}\rx{r}
      \union \n{\etag{i}}\D{a}{\rx{r}} = \slot{i}{p}\D{a}{\rx{r}}\\
    \D{a}{(\ltag{i}\rx{r})} &=& \D{a}{(\ltag{i})}\rx{r}
      \union \n{\ltag{i}}\D{a}{\rx{r}} = \slot{i}{p}\D{a}{\rx{r}}\\
    \D{a}{(\slot{i}{p}\rx{r})} &=& \D{a}{(\slot{i}{p})}\rx{r}
      \union \n{\slot{i}{p}}\D{a}{\rx{r}} = \slot{i}{p}\D{a}{\rx{r}}\\
    \D{a}{(\bank{k}\rx{r})} &=& \D{a}{(\bank{k})}\rx{r}
      \union \n{\bank{k}}\D{a}{\rx{r}} = \bank{k}\D{a}{\rx{r}}
  \end{eqnarray}
\end{definition}
Thus, an informal way of stating the above is to say that a derivation
of a tag gives a memory update operation towards the corresponding
slot, while memory banks and memory updates are transparent to
derivations.

With these definitions, equation \refeqn{eqn-rx-decide} becomes
\begin{equation}
  s\in\rx{r} \Leftrightarrow \bank{k}\in\D{s}{\rx{r}}
\end{equation}
for some memory bank \bank{k}. The slots of \bank{k} will hold the
position values in $s$ for the tags passed in \rx{r}, and thus be a
record of which parts of \rx{r} matched certain parts of $s$.

Now, let $s\in\A{*}$ be a string, and let \rx{r} and \rx{t} be regular
expressions over \A[+]{}. Suppose that
\begin{equation}
  \D{s}{(\bank{k}\rx{t})} = \bank{k}\slot{i}{p}\slot{j}{q}\rx{r} = \bank{k}\rx{r}
\end{equation}
Such an expression means that \rx{r} was formed through derivation of
\rx{t}, and during that derivation, tag $i$ was passed at position $p$
in $s$, while tag $j$ was passed at position $q$.

\begin{example}
  Let $\rx{r} = \rx{\bank{1}\etag{0}a\ltag{1}b}$, and match \rx{r} against
  the string $s=ab$ using \refalg{alg-lazy-match}.
  \begin{eqnarray*}
    \rx{\bank{1}\etag{0}a\ltag{1}b} \sim ab
    &\Leftrightarrow& \D{a}{(\rx{\bank{1}\etag{0}a\ltag{1}b})} \sim b \\
    &\Leftrightarrow& \bank{1}\D{a}{(\rx{\etag{0}a\ltag{1}b})} \sim b \\
    &\Leftrightarrow& \bank{1}\slot{0}{0}\D{a}{(\rx{a\ltag{1}b})} \sim b \\
    &\Leftrightarrow& \bank{1}\slot{0}{0}\rx{\ltag{1}b} \sim b \\
    &\Leftrightarrow& \bank{1}\slot{0}{0}\D{b}{(\rx{\ltag{1}b})} \sim \emptystring \\
    &\Leftrightarrow& \bank{1}\slot{0}{0}\slot{1}{1}\D{b}{\rx{b}} \sim \emptystring \\
    &\Leftrightarrow& \bank{1}\slot{0}{0}\slot{1}{1} \sim \emptystring\\
    &\Leftrightarrow& \bank{1} \sim \emptystring = true
  \end{eqnarray*}
  The example is trivial, but shows the principle. The expression
  \bank{1}\slot{0}{0}\slot{1}{1} means that slot 0 and 1 in \bank{1}
  will be updated with the values 0 and 1 respectively. Hence, when
  the matching is finished the slots of \bank{1} will hold the
  positions when \etag{0} and \ltag{1} where passed. With the aid of
  the information in \bank{1} we can point out what section of the
  input string that matched between \etag{0} and \ltag{1} in the
  expression.
\end{example}

\begin{example}\label{xmp-end-tag-problem}
  Let $\rx{r} = \rx{\bank{1}\etag{0}a\ltag{1}}$, and match \rx{r} against
  the string $s=a$ using \refalg{alg-lazy-match}.
  \begin{eqnarray*}
    \rx{\bank{1}\etag{0}a\ltag{1}} \sim a
    &\Leftrightarrow& \D{a}{(\rx{\bank{1}\etag{0}a\ltag{1}})} \sim \emptystring\\
    &\Leftrightarrow& \bank{1}\D{a}{(\rx{\etag{0}a\ltag{1}})} \sim \emptystring\\
    &\Leftrightarrow& \bank{1}\slot{0}{0}\D{a}{(\rx{a\ltag{1}})} \sim \emptystring\\
    &\Leftrightarrow& \bank{1}\slot{0}{0}\ltag{1} \sim \emptystring\\
    &\Leftrightarrow& \bank{1}\ltag{1} \sim \emptystring = true
  \end{eqnarray*}
  This example exposes a problem. While \slot{0}{0} correctly updates
  the value of the first slot in \bank{1}, and thus records the
  position when we pass \etag{0}, the second tag \ltag{1} never gets
  transformed into a memory update operation. Therefore, \bank{1} will
  not hold the correct information about the end position. This is the
  problem we will turn to next.
\end{example}

\section{Tag evaluation operation}

In order to solve the ``dangling tag'' problem in
\refxmp{xmp-end-tag-problem} we introduce a new operation.

\begin{definition}[Tag evaluation]
  Let \rx{r} and \rx{s} be a regular expressions over \A[+]{}, and $p$
  the current position. Then we define \define{tag
    evaluation}, \teval{}, on expressions over \A[+]{} recursively
  as follows.
  \begin{eqnarray}
    \teval{(\emptyset)} &=& \emptyset \label{eqn-store-emptyset}\\
    \teval{(\emptystring)} &=& \emptystring \label{eqn-store-emptystring}\\
    \teval{(\rx{a})} &=& \rx{a} \quad \rx{a}\subseteq\A{1} \\
    \teval{(\bank{i})} &=& \bank{i} \label{eqn-store-bank}\\
    \teval{(\etag{i})} &=& \slot{i}{p} \label{eqn-store-btag}\\
    \teval{(\ltag{i})} &=& \slot{i}{p} \label{eqn-store-etag}\\
    \teval{(\slot{i}{p})} &=& \slot{i}{p} \label{eqn-store-slot}\\
    \teval{(\rx{r^*})} &=& (\emptystring\union\n{\teval{(\rx{r})}})\rx{r^*}\label{eqn-store-clos} \\
    \teval{(\comp{\rx{r}})} &=& \comp{(\teval{(\rx{r})})} \label{eqn-store-comp}\\
    \teval{(\rx{r}\union\rx{s})} &=& \teval{(\rx{r})}\union\teval{(\rx{s})} \label{eqn-store-union}\\
    \teval{(\rx{r}\intersection\rx{s})} &=& \teval{(\rx{r})}\intersection\teval{(\rx{s})} \label{eqn-store-intersection}\\
    \teval{(\rx{rs})} &=&
    \left\{ \begin{array}{ll}
      \teval{(\rx{r})}\teval{(\rx{s})} & \rx{r}=\gamma \in \T{}\label{eqn-store-concat}\\
      \teval{(\rx{r})}\rx{s} & \n{\rx{r}} = \emptyset \\
      (\emptystring\union\n{\teval{(\rx{s})}})\teval{(\rx{r})}\rx{s} & otherwise
    \end{array}\right.
  \end{eqnarray}
\end{definition}

\begin{theorem}[Tag evaluation idempotence]\label{thm-teval-idempotence}
  Let \rx{r} be a regual expression over \A[+]{}. Then
  \begin{equation}
    \teval{(\teval{(\rx{r})})} = \teval{(\rx{r})} 
  \end{equation}
\end{theorem}

\begin{proof}
  The proof is by induction over the number of operations, $n$, used
  to form \rx{r}. It directly follows from equations
  \refeqn{eqn-store-emptyset} through \refeqn{eqn-store-slot} that the
  theorem holds for $n = 0$. Assume that it holds for $n = k$. Let
  \rx{r} be an expression with $n=k+1$ We can easily conclude that the
  theorem holds by the induction hypothesis if \rx{r} is of one of the
  forms in equations \refeqn{eqn-store-comp} through
  \refeqn{eqn-store-intersection}. Left to prove are the cases for
  equations \refeqn{eqn-store-clos} and \refeqn{eqn-store-concat}.

  We first note that for any $\rx{r}$ we have that
  $\teval{(\n{\rx{r}})} = \n{\rx{r}}$ since $\n{\rx{r}} \in
  \set{\emptyset, \emptystring, \bank{k}}$ for some $k$, and
  $(\emptystring \union \n{\rx{r}})(\emptystring \union \n{\rx{r}}) =
  (\emptystring \union \n{\rx{r}} \union \n{\rx{r}}\n{\rx{r}}) =
  (\emptystring \union \n{\rx{r}})$.
  
  Let $\rx{r} = \rx{s^*}$. Then
  \begin{eqnarray}
    \teval{(\teval{(\rx{s^*})})}
    &=& \teval{(\emptystring\union\n{\teval{(\rx{s})}})}\teval{(\rx{s^*})}\nonumber\\
    &=& (\emptystring\union\n{\teval{(\rx{s})}})
        (\emptystring\union\n{\teval{(\rx{s})}})\rx{s^*}\nonumber\\
    &=& (\emptystring\union\n{\teval{(\rx{s})}})\rx{s^*}\nonumber\\
    &=& \teval{(\rx{s^*})}
  \end{eqnarray}
  and thus the theorem holds for \refeqn{eqn-store-clos}.

  Let instead $\rx{r} = \rx{st}$. Then we have three cases
  \begin{description}
  \item[\itemlabel{Case 1:}] $\rx{s} = \gamma \in \Gamma \Rightarrow
    \teval{(\rx{s})} \in \Gamma$ and we have
    \begin{equation}
      \teval{(\teval{(\rx{st})})}
      = \teval{( \teval{(\rx{s})}\teval{(\rx{t})} )}
      = \teval{(\teval{(\rx{s})})}\teval{(\teval{(\rx{t})})}
      = \teval{(\rx{s})}\teval{(\rx{t})}
      = \teval{(\rx{st})}
    \end{equation}
    Where the third equality follows from the induction hypothesis.
  \item[\itemlabel{Case 2:}] $\n{\rx{s}} = \emptyset \Rightarrow
    \n{\teval{(\rx{s})}} = \emptyset$ since, by
    \refthm{thm-store-same-language}, \rx{s} and \teval{(\rx{s})}
    denotes the same language. Thus,
    \begin{equation}
      \teval{(\teval{(\rx{st})})}
      = \teval{( \teval{(\rx{s})}\rx{t} )}
      = \teval{(\teval{(\rx{s})})}\rx{t}
      = \teval{(\rx{s})}\rx{t}
      = \teval{(\rx{st})}
    \end{equation}
    Where, again, the third equality follows from the induction
    hypothesis.
  \item[\itemlabel{Case 3:}] In general we have 
    \begin{eqnarray}
      \teval{(\teval{(\rx{st})})}
      &=& \teval{( (\emptystring\union\n{\teval{(\rx{t})}})
            \teval{(\rx{s})}\rx{t} )}\nonumber\\
      &=& \teval{ (\emptystring\union\n{\teval{(\rx{t})}}) }
          \teval{( \teval{(\rx{s})}\rx{t} )}\nonumber\\
      &=& (\emptystring\union\n{\teval{(\rx{t})}})
          (\emptystring\union\n{\teval{(\rx{t})}})\teval{( \teval{(\rx{s})} )}\rx{t}\nonumber\\
      &=& (\emptystring\union\n{\teval{(\rx{t})}})\teval{( \teval{(\rx{s})} )}\rx{t}\nonumber\\
      &=& (\emptystring\union\n{\teval{(\rx{t})}})\teval{(\rx{s})}\rx{t}\nonumber\\
      &=& \teval{(\rx{st})}
    \end{eqnarray}
    The fifth equaltiy follows from the induction hypothesis.
  \end{description}
  Hence, the theorem also holds for \refeqn{eqn-store-concat}, which
  concludes the proof \footnote{Although this result may seem of
    limited theoretical value, it can simplify some parts of an
    implementation.}.
\end{proof}

\begin{theorem}\label{thm-store-same-language}
  Let \rx{r} be a regular expression over \A[+]{}, then \rx{r} and
  \teval{(\rx{r})} denotes the same language.
\end{theorem}

\begin{proof}
  First we note that the structure of \rx{r} is preserved by
  \teval{}. This obvious from the definition in most cases, the only
  question is with regard to equations \refeqn{eqn-store-clos}
  and \refeqn{eqn-store-concat}. At first it seems that this may
  introduce new terms into the expression. However,
  $\n{\rx{s}}\in\set{\emptyset, \emptystring, \bank{k}}$ for some
  $k$. Therefore, by property
  \refeqn{eqn-emptystring-bank-absorbtion},
  $(\emptystring\union\n{\rx{s}})$ will collapse into a single term,
  whose corresponding language is \set{\emptystring}.

  Second, the corresponding language of all symbols are preserved by
  \teval{}. The only symbols that are altered by \teval{} are \etag{i}
  and \ltag{i}. Since \etag{i}, \ltag{i} and \bank{i} all correspond
  to \set{\emptystring}, the denoted language is the same.
\end{proof}

\begin{example}\label{xmp-simple-tag-expression}
  Let us study the expression in \refxmp{xmp-end-tag-problem} again,
  but now with an added tag evaluation.
  \begin{eqnarray}
    \teval{(\D{a}{(\rx{\bank{1}\etag{0}a\ltag{1}})})}
    &=& \teval{(\bank{1}\D{a}{(\rx{\etag{0}a\ltag{1}})})} 
      = \teval{(\bank{1}\slot{0}{0}\D{a}{(\rx{a\ltag{1}})})}\nonumber\\
    &=& \teval{(\bank{1}\slot{0}{0}\ltag{1})} 
      = \bank{1}\slot{0}{0}\teval{(\ltag{1})}
      = \bank{1}\slot{0}{0}\slot{1}{1}
  \end{eqnarray}
  As expected \bank{1} gets updated with values for both slot 0 and
  slot 1, and will hold correct information about the submatch.
\end{example}

\refxmp{xmp-simple-tag-expression} is the simplest expression
possible. Let us take a look at a slightly more complex example, and
study how \teval{} behaves when repetition is involved.

\begin{example}
  Let $\rx{r_0} = \bank{1}\rx{\etag{0}a^*\ltag{1}}$.
  \begin{eqnarray}
    \rx{r_1} &=& \teval{(\D{a}{(\rx{r_0})})}
    = \teval{(\D{a}{(\rx{\bank{1}\etag{0}a^*\ltag{1}})})}\nonumber\\
    &=& \teval{(\bank{1}\D{a}{(\rx{\etag{0}a^*\ltag{1}})})} 
    = \teval{(\bank{1}\slot{0}{0}\D{a}{(\rx{a^*\ltag{1}})})}\nonumber\\
    &=& \teval{(\bank{1}\slot{0}{0}(\D{a}{(\rx{a^*)}\ltag{1}}
      \union \n{\rx{a^*}}\D{a}{(\ltag{1})} ))}\nonumber\\
    &=& \bank{1}\slot{0}{0} \teval{( \D{a}{(\rx{a^*)}\ltag{1}} )}
    = \bank{1}\slot{0}{0} \teval{( \D{a}{(\rx{a})}\rx{a^*}\ltag{1} )}\nonumber\\
    &=& \bank{1}\slot{0}{0} \teval{( \rx{a^*}\ltag{1} )}
    = \bank{1}\slot{0}{0} (\emptystring \union \n{\teval{(\ltag{1})}}) \teval{( \rx{a^*} )}\ltag{1}\nonumber\\
    &=& \bank{1}\slot{0}{0}\slot{1}{1} \teval{( \rx{a^*} )}\ltag{1}
    = \bank{1}\slot{0}{0}\slot{1}{1}\rx{a^*}\ltag{1}
  \end{eqnarray}
  \begin{eqnarray}
    \rx{r_2} &=& \teval{(\D{a}{(\rx{r_1})})}
    = \teval{(\D{a}{(\rx{\bank{1}\slot{0}{0}\slot{1}{1}a^*\ltag{1}})})}\nonumber\\
    &=& \bank{1}\slot{0}{0}\slot{1}{1} \teval{( \D{a}{(\rx{a^*)}\ltag{1}} )}
    = \bank{1}\slot{0}{0}\slot{1}{1} \teval{( \D{a}{(\rx{a})}\rx{a^*}\ltag{1} )}\nonumber\\
    &=& \bank{1}\slot{0}{0}\slot{1}{1} \teval{( \rx{a^*}\ltag{1} )}
    = \bank{1}\slot{0}{0}\slot{1}{1} (\emptystring \union \n{\teval{(\ltag{1})}})
      \teval{( \rx{a^*} )}\ltag{1}\nonumber\\
    &=& \bank{1}\slot{0}{0}\slot{1}{1}\slot{1}{2} \teval{( \rx{a^*} )}\ltag{1}
      = \bank{1}\slot{0}{0}\slot{1}{2}\rx{a^*}\ltag{1}
      = \rx{r_1}\label{eqn-tag-eval-repeat}
  \end{eqnarray}
  In \refeqn{eqn-tag-eval-repeat} we see that each derivative of
  \rx{r_1} results in the same expression, but with an update of slot
  1 with the current position. Hence, after each iteration the memory \bank{1} will
  keep information about what has been matched.\footnote{The current position is
    updated after each derivation.}
\end{example}

\begin{theorem}\label{thm-store-properties}
  Let \rx{r} be a regular expression over \A[+]{}. We note the
  following two properties for \teval{(\rx{r})}.
  \begin{enumerate}
  \item[\itemlabel{a)}]\label{eqn-store-prop-1} If $\rx{r} =
    \gamma\rx{s}$ for $\gamma\in\set{\etag{i}, \ltag{i}}$, then
    $\teval{(\rx{r})} = \slot{i}{p}\teval{(\rx{s})}$. That is, \teval{} transforms
    \etag{i} and \ltag{i} symbols in the beginning of an expression
    into memory update operations.
  \item[\itemlabel{b)}]If \rx{r} is nullable, \teval{(\rx{r})} will
    issue memory update operations for all tags that are part of a
    nullable subexpression of \rx{r}.
  \end{enumerate}
\end{theorem}

\begin{proof}
  \ 
  \begin{enumerate}
  \item[\itemlabel{a)}] follows directly from the definition.
  \item[\itemlabel{b)}] The proof is by induction over the number of
    operations, $n$, used to form \rx{r}. The theorem is true for $n =
    0$, since the only nullable expressions are given by equations
    \refeqn{eqn-store-emptystring}, \refeqn{eqn-store-bank},
    \refeqn{eqn-store-btag}, \refeqn{eqn-store-etag} and
    \refeqn{eqn-store-slot}.

    Assume that the theorem holds for $n=k$. Then, it also holds for
    $n=k+1$. By the induction hypothesis it is true for equations,
    \refeqn{eqn-store-union}, \refeqn{eqn-store-intersection}. Left to
    prove are \refeqn{eqn-store-comp}, \refeqn{eqn-store-clos} and
    \refeqn{eqn-store-concat}.

    Equation \refeqn{eqn-store-comp} is a bit special. If
    \rx{\comp{r}} is nullable, then \rx{r} is not. But then the
    theorem holds, since there are no tags that are part of a nullable
    subexpression.
    
    Let us continue with \refeqn{eqn-store-clos}. If \rx{r} is
    nullable, then, by the induction hypothesis, the theorem holds for
    \teval{(\rx{r})} and $(\emptystring\union\n{\teval{(\rx{r})}})$
    will collapse to that set of memory update operations. However,
    compared with \rx{r}, \rx{r^*} does not contain any new tags, and
    thus the theorem holds. If \rx{r} is not nullable, then, by
    \refthm{thm-store-same-language}, \teval{(\rx{r})} will not be
    nullable either. In this case,
    $(\emptystring\union\n{\teval{(\rx{r})}}) = \emptystring$, and the
    theorem still holds.

    Lastly, assume $\rx{r} = \rx{st}$ and study equation
    \refeqn{eqn-store-concat}. Since \rx{r} is nullable, so is both
    \rx{s} and \rx{t}. But then the second case in
    \refeqn{eqn-store-concat} is ruled out, and the theorem holds by
    the induction hypothesis.
  \end{enumerate}
\end{proof}

\begin{example}
  Let $\rx{r} = (\etag{0}(\etag{2}a\ltag{3})^{*}\ltag{1})^{*}$. Then
  \begin{eqnarray*}
    \rx{\teval{(r)}} &=& \rx{\teval{((\etag{0}(\etag{2}a\ltag{3})^{*}\ltag{1})^{*})}} \\
    &=& \rx{(\emptystring \union \nu(
        \underbrace{\teval{(\etag{0}(\etag{2}a\ltag{3})^{*}\ltag{1})}}_{\mathnormal q}
      ) )r}\\
    q &=&  \teval{(\etag{0}(\etag{2}a\ltag{3})^{*}\ltag{1} )}\\
    &=& \slot{0}{p}\slot{1}{p}\teval{((\etag{2}a\ltag{3})^{*})}\\
    &=& \slot{0}{p}\slot{1}{p}
    (\emptystring \union \n{\teval{(\etag{2}a\ltag{3})}}   )
    (\etag{2}a\ltag{3})^{*}\\
    &=& \slot{0}{p}\slot{1}{p} (\emptystring \union \emptyset) (\etag{2}a\ltag{3})^{*}\\
    &=& \slot{0}{p}\slot{1}{p}(\etag{2}a\ltag{3})^{*}
  \end{eqnarray*}
  Hence,
  \begin{equation}
    \rx{\teval{(r)}}
    = \rx{(\emptystring \union \n{  \slot{0}{p}\slot{1}{p}(\etag{2}a\ltag{3})^{*}  } )r}
    = \rx{(\emptystring \union \slot{0}{p}\slot{1}{p}\n{(\etag{2}a\ltag{3})^{*}}) r}
    = \slot{0}{p}\slot{1}{p}\rx{r}
  \end{equation}
  Which is the expected result. Tags \etag{0} and \ltag{1} wraps
  something nullable, whereas tags \etag{2} and \ltag{3} does not.
\end{example}

\section{Memory disambiguation}

In the previous sections we have presented a mechanism based on
derivation that records when a certain position in an expression is
matched. We are now ready take on the problem presented in the
beginning of this chapter.

\begin{example}\label{xmp-ambiguous-memory-1}
  Let $\rx{r_0} = \bank{1}\rx{\etag{0}a^*\ltag{1}\etag{2}a^*\ltag{3}a}$, and let
  input be $s=aa$.
  \begin{eqnarray}
    \rx{r_1} &=& \teval{( \D{a}{\rx{r_0}} )} 
    = \teval{( \D{a}{(\bank{1}\rx{\etag{0}a^*\ltag{1}\etag{2}a^*\ltag{3}a})} )}
    = \bank{1} \teval{( \D{a}{(\rx{\etag{0}a^*\ltag{1}\etag{2}a^*\ltag{3}a})} )}\nonumber\\
    &=& \bank{1} \teval{( \slot{0}{0}\D{a}{(\rx{a^*\ltag{1}\etag{2}a^*\ltag{3}a})} )}
    = \bank{1}\slot{0}{0} \teval{( \D{a}{(\rx{a^*\ltag{1}\etag{2}a^*\ltag{3}a})} )}\nonumber\\
    &=& \bank{1}\slot{0}{0} \teval{( \D{a}{(\rx{a^*})}\rx{\ltag{1}\etag{2}a^*\ltag{3}a}
      \union \D{a}{(\rx{\ltag{1}\etag{2}a^*\ltag{3}a})} )}\nonumber\\
    &=& \bank{1}\slot{0}{0} \teval{( \rx{a^*\ltag{1}\etag{2}a^*\ltag{3}a}
      \union \slot{1}{0}\slot{2}{0}\D{a}{\rx{(a^*\ltag{3}a})} )}\nonumber\\
    &=& \bank{1}\slot{0}{0} \teval{( \rx{a^*\ltag{1}\etag{2}a^*\ltag{3}a}
      \union \slot{1}{0}\slot{2}{0} (\D{a}{(\rx{a^*})}\rx{\ltag{3}a}
      \union \D{a}{(\rx{\ltag{3}a})} ))}\nonumber\\
    &=& \bank{1}\slot{0}{0} \teval{( \rx{a^*\ltag{1}\etag{2}a^*\ltag{3}a}
      \union \slot{1}{0}\slot{2}{0} (\rx{a^*}\rx{\ltag{3}a}
      \union \slot{3}{0}\D{a}{(\rx{a})} ))}\nonumber\\
    &=& \bank{1}\slot{0}{0} \teval{( \rx{a^*\ltag{1}\etag{2}a^*\ltag{3}a}
      \union \slot{1}{0}\slot{2}{0}\rx{a^*}\rx{\ltag{3}a}
      \union \slot{1}{0}\slot{2}{0}\slot{3}{0} )}\nonumber\\
    &=& \underbrace{\bank{1}\slot{0}{0}\rx{a^*\ltag{1}\etag{2}a^*\ltag{3}a}}_{1} \union
    \underbrace{\hat{\bank{1}}\slot{0}{0}\slot{1}{0}\slot{2}{0}\rx{a^*\ltag{3}a}}_2 \union
    \underbrace{\hat{\bank{1}}\slot{0}{0}\slot{1}{0}\slot{2}{0}\slot{3}{0}}_3\nonumber\\
    &=& \underbrace{\bank{1}\rx{a^*\ltag{1}\etag{2}a^*\ltag{3}a}}_{1} \union
    \underbrace{\bank{2}\rx{a^*\ltag{3}a}}_2 \union
    \underbrace{\bank{3}}_3 \label{eqn-ambiguous-memory-1}
  \end{eqnarray}
  For the last equality we have applied the slot updates to respective bank.
  We can interpret these terms in order. Term $1$ is what is left to
  match if the first \rx{a^*} is used to match the first input
  $a$. Term $2$ is what is left to match if the first \rx{a^*} matches
  \emptystring{} and the second \rx{a^*} matches the $a$ input
  symbol. Term $3$ is what it is left to match (i.e. nothing) if both
  \rx{a^*} matches \emptystring{} and the last \rx{a} matches the
  first input symbol. Since $\n{\rx{r_1}} =
  \hat{\bank{1}}\slot{0}{0}\slot{1}{0}\slot{2}{0}\slot{3}{0} = \bank{3}$, $a$
  is part of the language denoted by $\rx{r_0}$. Further, we see that
  when matching input $a$, all slot values are $0$ for the matching
  term 3.

  Continue to study
  \begin{eqnarray*}
    \rx{r_2} &=& \teval{(\D{a}{\rx{r_1}})} \\ 
    &=& \teval{(\D{a}{(\bank{1}\slot{0}{0}\rx{a^*\ltag{1}\etag{2}a^*\ltag{3}a})} )} \union
    \teval{(\D{a}{(\bank{2}\slot{0}{0}\slot{1}{0}\slot{2}{0} \rx{a^*\ltag{3}a})} )}\\
    && \union \teval{(\D{a}{(\bank{3}\slot{0}{0}\slot{1}{0}\slot{2}{0}\slot{3}{0})} )}\\
    &=& \underbrace{\bank{1}\slot{0}{0}\teval{(\D{a}{(\rx{a^*\ltag{1}\etag{2}a^*\ltag{3}a})} )}}_1 \union
    \underbrace{\bank{2}\slot{0}{0}\slot{1}{0}\slot{2}{0} \teval{(\D{a}{(\rx{a^*\ltag{3}a})} )}}_2\\
    &=& \underbrace{\underbrace{\bank{1}\slot{0}{0}\rx{a^*\ltag{1}\etag{2}a^*\ltag{3}a}}_{a} \union
    \underbrace{\hat{\bank{1}}\slot{0}{0}\slot{1}{1}\slot{2}{1} \rx{a^*\ltag{3}a}}_{b} \union
    \underbrace{\hat{\bank{1}}\slot{0}{0}\slot{1}{1}\slot{2}{1}\slot{3}{1}}_{c}}_1\\
    && \union \underbrace{
      \underbrace{\bank{2}\slot{0}{0}\slot{1}{0}\slot{2}{0} \rx{a^*\ltag{3}a}}_{d} \union
      \underbrace{\hat{\bank{2}}\slot{0}{0}\slot{1}{0}\slot{2}{0}\slot{3}{1}}_{e}}_2
  \end{eqnarray*}
  Obviously \rx{r_0} matches $aa$. There are two terms, $c$ and $e$,
  in the above expression that signals the match since they both
  contain \emptystring. However, they differ in their memory,
  i.e. they correspond to two different ways that \rx{r_0} matches
  $aa$. From the slot values we see that term $c$ corresponds to the
  case when the first \rx{a^*} has matched the first $a$, and term $e$
  represents the case when the second \rx{a^*} matched the first
  $a$. We can illustrate it in a similar manner to what we did in
  \refxmp{xmp-intro-memory-ambiguity} in the introduction to this
  chapter.
  \begin{eqnarray}
    \mbox{\small $term\;c$} \qquad
    \underbrace{a}_{\rx{a^*}}\underbrace{\emptystring}_{\rx{a^*}}\underbrace{a}_{\rx{a}}\\
    \mbox{\small $term\;e$}
    \qquad \underbrace{\emptystring}_{\rx{a^*}}\underbrace{a}_{\rx{a^*}}\underbrace{a}_{\rx{a}}
  \end{eqnarray}
  Likewise, there are two other terms that are equal and only differ
  in memory, $b$ and $d$.
  \begin{eqnarray}
    \mbox{\small $term\;b$} \qquad
    \underbrace{a}_{\rx{a^*}}\underbrace{a}_{\rx{a^*}}\underbrace{}_{\rx{a}}\\
    \mbox{\small $term\;d$} \qquad
    \underbrace{\emptystring}_{\rx{a^*}}\underbrace{aa}_{\rx{a^*}}\underbrace{}_{\rx{a}}
  \end{eqnarray}
  The term left, term $a$, captures the case when the first \rx{a^*}
  matched both $a$'s in the input.
  \begin{eqnarray}
    \mbox{\small $term\;a$} \qquad
    \underbrace{aa}_{\rx{a^*}}\underbrace{}_{\rx{a^*}}\underbrace{}_{\rx{a}}
  \end{eqnarray}
  In general, regular expression matcher implementations do not report
  all possible ways the expression matches a string. Instead, they
  employ some sort of disambiguation policy to select a particular
  match in an ambiguous case. The same policy is applied in order to
  choose between other ambiguous terms like $b$ and $d$. Before we
  continue this example, we discuss such a policy in the following
  section.
\end{example}

\subsection{\POSIX{} semantics}\label{sec-posix-semantics}

As the above example illustrates, there may be many different ways to
``distribute'' a match over a regular expressions constituents. A
popular disambiguation policy to make the matching predictable is
given in the \POSIX{} specification\autocite[ch 9]{posix}. There
submatches are delimited by parentheses\footnote{Hence, a submatch in
  \POSIX{} corresponds to the parenthesized production in the
  non-terminal \nt{atom} in the grammar given in
  \refdef{def-regular-expression}}. Glenn Fowler, formerly of Bell
Labs and AT\&T Labs Research, has given a succinct interpretation of
the matching rules\autocite{fowler} which we restate here with a
slight adaption to our terminology.

\begin{enumerate}
\item Determine the longest of the first-most matches for the complete
  expression. Let this be submatch \submatch{0}.
\item Consistent with the whole match being the longest of the
  first-most matches, each subpattern\footnote{Subpattern in the
    \POSIX{} terminology corresponds to the non-terminal \nt{clos} in
    \refdef{def-regular-expression}}, from left to right, shall match
  the longest possible string. For this purpose, an empty string shall
  be considered to be longer than no match at all.
\item Enumerate the submatches. Submatch \submatch{i} begins at the
  $i$th opening bracket, counting from $1$ to $n$.
\item For $1 \leq i \leq n$ determine the longest match for
  \submatch{i} consistent with the matches already determined for
  \submatch{j}, $0 \leq j < i$.
\end{enumerate}
\newpage

\begin{example}
  Let $\rx{r} = \rx{(a+\emptystring)((ab)+\emptystring)}$, and input
  $s = ab$ then
  \begin{center}
    \begin{tabular}{cll}
      \hline
      \hline
      \textsc{submatch} & \textsc{expression} & \textsc{match}\\
      \hline
      \submatch{0} & \rx{(a+\emptystring)((ab)+\emptystring)} & $ab$\\
      \submatch{1} & \rx{(a+\emptystring)} & \emptystring\\
      \submatch{2} & \rx{((ab)+\emptystring)} & $ab$\\
      \submatch{3} & \rx{(ab)} & $ab$\\
      \hline
    \end{tabular}
  \end{center}
\end{example}

\begin{example}
  Let $\A{} = \set{a,b,c}$, $\rx{r} = \rx{[ab]^*(([bc])^*)}$, and
  input $s = abbcc$. Then the \POSIX{} policy would give
  \begin{center}
    \begin{tabular}{cll}
      \hline
      \hline
      \textsc{submatch} & \textsc{expression} & \textsc{match}\\
      \hline
      \submatch{0} & \rx{[ab]^*(([bc])^*)} & $abbcc$\\
      \submatch{1} & \rx{(([bc])^*)} & $cc$\\
      \submatch{2} & \rx{([bc])} & $c$\\
      \hline
    \end{tabular}
  \end{center}
\end{example}

\begin{example}
  Let $\rx{r} = \rx{(a^*)(a^*)a}$, and input $s = aa$. This is the
  \POSIX{} equivalent of the situation in
  \refxmp{xmp-ambiguous-memory-1}. Applying the policy above we have
  \begin{center}
    \begin{tabular}{cll}
      \hline
      \hline
      \textsc{submatch} & \textsc{expression} & \textsc{match}\\
      \hline
      \submatch{0} & \rx{(a^*)(a^*)a} & $aa$\\
      \submatch{1} & \rx{(a^*)} & $a$\\
      \submatch{2} & \rx{(a^*)} & \emptystring\\
      \hline
    \end{tabular}
  \end{center}
\end{example}

\subsection{Bank order and disambiguation}\label{sec-bank-order}

\begin{definition}[Bank order]
  Let \bank{i} and \bank{j} be memory banks with $n$ slots each, and
  suppose $\slotvalue{l}{i} = \slotvalue{l}{j}$ for $0 \leq l < k <
  n$, i.e. the $k$ first slot values are equal. Bank \bank{i} is said
  to have \define{higher priority} than \bank{j}, written $\bank{i} >
  \bank{j}$, when either
  \begin{itemize}
  \item[\itemlabel{a)}] Slot $k$ is of type early and
    $\slotvalue{k}{i} < \slotvalue{k}{j}$, i.e the value of slot $k$ in
    \bank{i} is less than the value of slot $k$ in \bank{j}.
  \end{itemize}
  or
  \begin{itemize}
  \item[\itemlabel{b)}] Slot $k$ is of type late and $\slotvalue{k}{i}
    > \slotvalue{k}{j}$, i.e. the value of slot $k$ in \bank{i} is
    greater than the value of slot $k$ in \bank{j}.
  \end{itemize}
  If all slot values are equal, the banks are said to have
  \define{equal priority}, otherwise \bank{i} has \define{lower
    priority} than \bank{j}, written $\bank{i} < \bank{j}$.
\end{definition}

The definition of bank order gives us a policy that we can use to
disambiguate terms in a regular expression matching scenario.

\begin{example}\label{xmp-ambiguous-memory-2}
  Consider once more the expression for \rx{r_2} in
  \refxmp{xmp-ambiguous-memory-1}, and compare the bank order for
  terms $c$ and $e$. We have
  \begin{equation}
    \bank{5} = \underbrace{\hat{\bank{1}}\slot{0}{0}\slot{1}{1}\slot{2}{1}\slot{3}{1}}_{c} >
    \underbrace{\hat{\bank{2}}\slot{0}{0}\slot{1}{0}\slot{2}{0}\slot{3}{1}}_{e} = \bank{6}
  \end{equation}
  since after applying the slot updates $\slotvalue{0}{5} =
  \slotvalue{0}{6} = 0$, and $\slotvalue{1}{5} = 1$ but
  $\slotvalue{1}{6} = 0$ and tag 1 is of late type.

  Similarly, applying the slot updates and by comparing the bank order
  between terms $b$ and $d$ we have
  \begin{equation}
    \bank{4} = \hat{\bank{1}}\slot{0}{0}\slot{1}{1}\slot{2}{1} >
    \bank{2}\slot{0}{0}\slot{1}{0}\slot{2}{0} = \bank{2}
  \end{equation}
  Thus, the full expression for \rx{r_2} after applying this
  disambiguation is
  \begin{equation}\label{eqn-ambiguous-memory-2}
    \rx{r_2} = \bank{1}\rx{a^*\ltag{1}\etag{2}a^*\ltag{3}a}
    \union \bank{4}\rx{a^*\ltag{3}a}
    \union \bank{5}
  \end{equation}
  Since $\n{\rx{r_2}} = \bank{5}$, \rx{r_0} matches the string $aa$,
  and the reported submatches are
  \begin{center}
    \begin{tabular}{cll}
      \hline
      \hline
      \textsc{submatch} & \textsc{expression} & \textsc{match}\\
      \hline
      \rxstrut \submatch{0} & \rx{\etag{0}a^*\ltag{1}\etag{2}a^*\ltag{3}a} & $aa$\\
      \rxstrut \submatch{1} & \rx{\etag{0}a^*\ltag{1}} & $a$\\
      \rxstrut \submatch{2} & \rx{\etag{2}a^*\ltag{3}} & \emptystring\\
      \hline
    \end{tabular}
  \end{center}
  or, in a more graphical form in terms of the input string
  \begin{equation}
    \underbrace{a}_{\rx{a^*}}\underbrace{\emptystring}_{\rx{a^*}}\underbrace{a}_{\rx{a}}
  \end{equation}
\end{example}

\begin{theorem}
  Let \rx{r} be a regular expression over \A[+]{}. Whenever there is
  an ambiguity in matching \rx{r} against $s\in\A{*}$, choosing the
  term with the highest bank priority as a disambiguation policy will
  comply with \POSIX{} matching rules.
\end{theorem}

\begin{proof}
  Note that the order of the tag enumeration corresponds to the
  submatches in \POSIX{}. The submatch captured by tag \etag{2i} and
  \ltag{2i+1}, precisely correspond to submatch $i+1$ in \POSIX{}. Bank
  order will make sure that the position in the input string where
  \etag{2i} is passed, will be as early as possible, consistent with
  tags \etag{2j}, $j<i$. Likewise, tag \ltag{2i+1} will be passed as
  late as possible, consistent with tags \ltag{2j+1}. This gives the
  longest possible match for each submatch.

  The so called subpatterns in \POSIX{} poses a slight problem. These
  are parts of the pattern that may vary in length of what they match,
  but are not necessarily enclosed in parentheses,
  e.g. \rx{[abc]^*}. According to the \POSIX{} policy such subpatterns
  are considered of equal importance as submatches. However, this
  problem can be solved by letting the parser wrap subpatterns between
  tags as well.\footnote{This will effectively turn them into
    submatches. Such an implementation would need to keep track of
    tags inserted by the parser, and tags provided by the user, so
    that submatches can be mapped to their expected numbers.}
  
  We also note that \refthm{thm-store-properties} ensures that the
  second part of property two in \refsec{sec-posix-semantics},
  i.e. that a empty string shall be considered to be longer than no
  match at all, is satisfied.
\end{proof}

\subsection{Memory bank rearrangement}
Consider equation \refeqn{eqn-ambiguous-memory-2} in the previous
example, and compare it to equation \refeqn{eqn-ambiguous-memory-1},
we have
\begin{eqnarray}
  \rx{r_2} &=& \bank{1}\rx{a^*\ltag{1}\etag{2}a^*\ltag{3}a}
  \union \bank{4}\rx{a^*\ltag{3}a}
  \union \bank{5} = \left\{ \begin{array}{c}
    \bank{2} \gets \bank{4}\\
    \bank{3} \gets \bank{5}
  \end{array}\right\}\\
  &=& \bank{1}\rx{a^*\ltag{1}\etag{2}a^*\ltag{3}a}
  \union \bank{2}\rx{a^*\ltag{3}a}
  \union \bank{3} = \rx{r_1} 
\end{eqnarray}
That is, by copying the contents of \bank{4} and \bank{5} into
\bank{2} and \bank{3} respectively, we see that \rx{r_1} and \rx{r_2}
are really the same expression. This is important to notice in case we
are building a DFA. The machine would otherwise contain equivalent
states, and thus be non-minimal. Such rearrangements of memory banks
are always possible by using at most $m+1$ banks if there are $m$
banks in the expression.

\subsection{Greedy and lazy operators}\label{sec-matching-modes}
The concept of bank order, and early and late tags provides us with an
opportunity to implement two different matching strategies for the
Kleene closure, also called multiplicity operator (${}^*$). Historically
they have gone by different names, but the most popular seems to be
the terminology used by Perl, where they are called \define{greedy}
and \define{lazy}.

In the greedy case the closure matches as many symbols as possible of
the input string, while still being consistent with the overall
matching rules. In the lazy case, it matches as few symbols as
possible.

It is possible to realize lazy operators using derivatives by wrapping
the expression affected by the operator between two early tags.

\begin{example}
  As before, let $\rx{r} = \rx{(a^*)(a^*)a}$, and input $s = aa$.
  \begin{center}
    \begin{tabular}{clll}
      \hline
      \hline
        & & \multicolumn{2}{c}{\textsc{match}} \\
      \raisebox{1.5ex}[0mm]{\textsc{submatch}} &
      \raisebox{1.5ex}[0mm]{\textsc{expression}} &
      \textsc{greedy} &
      \textsc{lazy}\\
      \hline
      \submatch{0} & \rx{(a^*)(a^*)a} & $aa$ & $aa$ \\
      \submatch{1} & \rx{(a^*)} & $a$ & \emptystring \\
      \submatch{2} & \rx{(a^*)} & \emptystring & $a$ \\
      \hline
    \end{tabular}
  \end{center}
  We can achieve the lazy result by considering the expression
  \begin{equation}
    \rx{r_0} = \bank{1}\rx{\etag{0}a^*\etag{1}\etag{2}a^*\etag{3}a}
  \end{equation}
  Note that the derivatives will be the same as for the expression in
  \refxmp{xmp-ambiguous-memory-1}, the only difference will be the
  bank order comparisons. Thus, in this case, we
  have
  \begin{equation}
    \bank{5} = \hat{\bank{1}}\slot{0}{0}\slot{1}{1}\slot{2}{1}\slot{3}{1} <
    \hat{\bank{2}}\slot{0}{0}\slot{1}{0}\slot{2}{0}\slot{3}{1} = \bank{6}
  \end{equation}
  since all tags are of early type. Similarly, comparing the bank
  order between terms $b$ and $d$ we have
  \begin{equation}
    \bank{4} = \hat{\bank{1}}\slot{0}{0}\slot{1}{1}\slot{2}{1} <
    \bank{2}\slot{0}{0}\slot{1}{0}\slot{2}{0} = \bank{2}
  \end{equation}
  In this case the second derivative becomes
  \begin{equation}
    \rx{r_2} = \D{aa}{\rx{r_0}} = \bank{1}\rx{a^*\etag{1}\etag{2}a^*\etag{3}a}
    \union \bank{2}\rx{a^*\etag{3}a}
    \union \bank{6}
  \end{equation}
  Hence, the submatches reported are those in table above.
\end{example}

\subsection{More on disambiguation policies}\label{sec-disambiguation-policies}
Having both greedy and lazy operators, it is possible to define different
disambiguation policies.

\begin{definition}[First most longest policy]
  This is the \POSIX{} policy formulated in section
  \refsec{sec-posix-semantics}.
\end{definition}

Instead of searching for the overall longest string, i.e. compare two
matches and select the longest one, we can compare bank orders between
two different matches and select the one with highest priority.

\begin{definition}[First-most pre-order tag enumeration policy]
  Enumerate the tag pairs, the $i$th tag pair begins at the $i$th
  opening tag. Find the first-most matching string in the input. If
  there is more than one such string, select the one with highest
  bank priority.
\end{definition}

\begin{definition}[First-most post-order tag enumeration policy]
  Enumerate the tag pairs, the $i$th tag pair ends at the $i$th
  closing tag. Find the first-most matching string in the input. If
  there is more than one such string, select the one with highest
  bank priority.
\end{definition}

\begin{example}
  Let $\rx{r} =
  \rx{\etag{}\etag{}a^*\etag{}\etag{}a^*\ltag{}\etag{}a}$ and $s =
  aaa$. We have not enumerated the tags because different policies
  yield different enumerations. The table below shows the different
  cases and their corresponding submatches. As can be seen, for this
  particular \rx{r} the \POSIX{} and post-order policy give the same
  results, but for different reasons. \POSIX{} matches the whole
  string because of the first-most \emph{longest} property (overriding
  the lazy early-early tag-pair $0/1$), whereas in the post-order case
  it is the greedy early-late combination of tag-pair $2/3$ that
  allows the enclosed \rx{a^*} to match the two first $a$ symbols. The
  pre-order case only matches a single $a$ since the early-early
  tag-pair $0/1$ prevents both \rx{a^*} groups from growing.
  \begin{center}
    \begin{tabular}{llcccc}
      \hline
      \hline
      \textsc{policy} &
      \textsc{expression} &
      \submatch{0} &
      \submatch{1} &
      \submatch{2} &
      \submatch{3} \\
      \hline
      \rxstrut \POSIX{} & \rx{\etag{0}\etag{2}a^*\etag{3}\etag{4}a^*\ltag{5}\etag{1}a}
         & $aaa$ & \emptystring & $aa$ & $aa$ \\
      \rxstrut pre-order & \rx{\etag{0}\etag{2}a^*\etag{3}\etag{4}a^*\ltag{5}\etag{1}a}
         & $a$ & \emptystring & \emptystring & \emptystring \\
      \rxstrut post-order & \rx{\etag{4}\etag{0}a^*\etag{1}\etag{2}a^*\ltag{3}\etag{5}a}
         & $aaa$ & \emptystring & $aa$ & $aa$ \\
      \hline
    \end{tabular}
  \end{center}  
\end{example}

\chapter{Putting it all together}\label{chp-putting-it-together}

In this chapter we put all the discussed techniques together, and
modify algorithms \refalg{alg-lazy-match}, \refalg{alg-dfa-match}, and
\refalg{alg-make-dfa} to include sets of symbols, similarity,
anchoring and submatching.

We begin with the algorithm for lazy matching, i.e. where the
derivatives are calculated on the fly.

\begin{algorithm}
  \caption{Regular expression matching using derivatives}\label{alg-full-lazy-match}
  \begin{algorithmic}[1]
    \Procedure{Match}{$\rx{r},s$}
      \State $t \gets \Call{InjectAnchors}{s}$
      \State $p \gets 0$
      \While{$t \neq \emptystring $}
        \State $c \gets \Call{PopFront}{t}$
        \State $\rx{r} \gets \D{c}{\rx{r}}$
        \State $p \gets p+1$
        \State $\rx{r} \gets \Call{Disambiguate}{\teval{(\rx{r})}}$
      \EndWhile
      \If{$\Call{Disambiguate}{\n{\rx{r}}} = \bank{k}$ for some $k$}
        \State matches $\gets \bank{k}$
      \Else
        \State matches $\gets false$
      \EndIf
      \Return{matches}  
    \EndProcedure
  \end{algorithmic}
\end{algorithm}

As can be seen in \refalg{alg-full-lazy-match} the implementation for
lazy matching is straight forward, and fairly similar to
\refalg{alg-lazy-match}. The \textsc{Disambiguate} procedure realizes,
for example, one of the suggested policies in section
\refsec{sec-disambiguation-policies}.

If the lazy match algorithms are similar, this is even more true for a
matcher using a DFA. Compare algorithms \refalg{alg-dfa-match} and
\refalg{alg-full-dfa-match}. The only difference is the pre-processing
of the input string to insert anchors, and the return value to include
submatches. This is no surprise because most of the work goes into the
algorithm for DFA construction.

\begin{algorithm}
  \caption{Regular expression matching using DFAs}\label{alg-full-dfa-match}
  \begin{algorithmic}[1]
    \Procedure{Match}{$(\protect\A{}, Q, q_0, \d{}, F), s$}
      \State $t \gets \Call{InjectAnchors}{s}$
      \State $q_c \gets q_0$
      \State $p \gets 0$
      \While{$t \neq \emptystring $}
        \State $c \gets \Call{PopFront}{t}$
        \State $q_c \gets \d{q_c,c}$
        \State $p \gets p+1$
      \EndWhile
      \If{$q_c \in F$}
        \State matches $\gets \bank{q_c}$
      \Else
        \State matches $\gets false$
      \EndIf
      \Return{matches}  
    \EndProcedure
  \end{algorithmic}
\end{algorithm}

The algorithm for creating a DFA for a given regular expression
including submatches becomes a little bit more involved. In this case
each state is associated with a set of memory banks, and each
transition is associated with a set of memory slot updates. Hence, the
transition function does not only contain information about how states
are related to each other, but also what memory operations to perform
during transitions between them. This information of course needs to
be recorded during DFA construction.

\begin{algorithm}
  \caption{DFA construction using derivative classes and submatches}\label{alg-full-make-dfa}
  \begin{algorithmic}
    \Procedure{MakeDfa}{\rx{r},\protect\A{}}
      \State $p \gets 0$
      \State $(\rx{r}, I) \gets \Call{Disambiguate}{\teval{(\rx{r})}}$
      \State $q_0 \gets \rx{r}$
      \State $\d{} \gets \emptyset$
      \State $F \gets \emptyset$
      \State $Q \gets \set{\rx{r}}$
      \State $S \gets \set{(\rx{r}, p)}$
      \While{$S \neq \emptyset$}
        \State $(\rx{r}, p) \gets \Call{Pop}{S}$
        \State $(\rx{q}, U) \gets \Call{Disambiguate}{\n{\rx{r}}}$
        \If{\rx{q} = \bank{k} for some $k$}
          \State \Call{Push}{(\rx{r}, \bank{k}), $F$} 
        \EndIf
        \For{$P \in \dca{(\rx{r})}$}
          \State $a \gets \Call{PickSomeElement}{P}$
          \State $(\rx{d}, U_{\D{}{}}) \gets \D{a}{\rx{r}}$
          \State $p \gets p+1$
          \State $(\rx{d}, U_{\rx{d}}) \gets \Call{Disambiguate}{\teval{(\rx{d})}}$
          \State $U \gets U_{\D{}{}} \union U_{\rx{d}}$
          \If{$\exists \rx{\bar{d}} \in Q$ such that
            $\rx{\bar{d}} \simeq \rx{d}$}\label{row-full-make-dfa-check-eq}
            \State $U \gets \Call{RearrangeMemory}{\rx{d},\rx{\bar{d}}, U}$
            \State \Call{Push}{$(\rx{r}, P, U) \mapsto \rx{\bar{d}}$, \d{}}
          \Else
            \State \Call{Push}{\rx{d}, $Q$}
            \State \Call{Push}{$(\rx{d}, p)$, $S$}
            \State \Call{Push}{$(\rx{r}, P, U) \mapsto \rx{d}$, \d{}}
          \EndIf
        \EndFor  
      \EndWhile
      \Return{$(\A{}, Q, q_0, \d{}, F, I)$}
    \EndProcedure
  \end{algorithmic}
\end{algorithm}

Disambiguation is done by keeping track of $p$, the number of
derivatives calculated to reach the current expression, counted from
the initial expression.\footnote{Assume the initial expression is
  \rx{r_0}, and the current expression is \rx{r_q}. Assume further
  that $s$ is the shortest string such that $\rx{r_q} =
  \D{s}{\rx{r_0}}$, then $p = \length{s}$.} The \textsc{Disambiguate}
procedure in \refalg{alg-full-make-dfa} returns two values, a
disambiguated expression, and a set of memory operations to
perform. Note also the set $I$ in the returned machine, it is a set of
initial memory operations to perform before processing any input
string. Further, the \textsc{RearrangeMemory} procedure calculates the
set of memory bank operations that needs to be performed in order to
transform \rx{d} to \rx{\bar{d}}. (Providing $U$, the set of memory
operations used to form \rx{d}, as additional input allows for some
optimizations.)

\begin{example}\label{xmp-dfa-submatch}
  Let $\A{} = \set{a,b}$
  Construct the DFA for $\rx{r} =
  \rx{\bank{1}\etag{0}a^*\ltag{1}\etag{2}a^*\ltag{3}a}$ using algorithm
  \refalg{alg-full-make-dfa}, and employing the first-most longest
  disambiguation policy.
  \begin{eqnarray*}
    \rx{r_0} &=& \teval{(\rx{r})}
      = \teval{( \bank{1}\rx{\etag{0}a^*\ltag{1}\etag{2}a^*\ltag{3}a} )} \\
    &=& \bank{1}\slot{0}{p} \teval{( \rx{a^*\ltag{1}\etag{2}a^*\ltag{3}a} )}
      = \bank{1}\slot{0}{p} \rx{a^*\ltag{1}\etag{2}a^*\ltag{3}a}
      = \bank{1}\rx{a^*\ltag{1}\etag{2}a^*\ltag{3}a}\\
    \rx{r_1} &=& \teval{( \D{a}{\rx{r_0}} )}\\
    &=& \bank{1}\rx{a^*\ltag{1}\etag{2}a^*\ltag{3}a} \union
    \hat{\bank{1}}\slot{1}{p}\slot{2}{p} \rx{a^*\ltag{3}a} \union
    \hat{\bank{1}}\slot{1}{p}\slot{2}{p}\slot{3}{p}\\
    &=& \bank{1}\rx{a^*\ltag{1}\etag{2}a^*\ltag{3}a} \union
        \bank{2}\slot{1}{p}\slot{2}{p} \rx{a^*\ltag{3}a} \union
        \bank{3}\slot{1}{p}\slot{2}{p}\slot{3}{p}\\
    &=& \bank{1}\rx{a^*\ltag{1}\etag{2}a^*\ltag{3}a} \union
        \bank{2}\rx{a^*\ltag{3}a} \union
        \bank{3}\\
    \rx{r_2} &=& \teval{( \D{a}{\rx{r_1}} )}\\
    &=& \bank{1}\rx{a^*\ltag{1}\etag{2}a^*\ltag{3}a} \union
    \hat{\bank{1}}\slot{1}{p}\slot{2}{p} \rx{a^*\ltag{3}a} \union
    \hat{\bank{1}}\slot{1}{p}\slot{2}{p}\slot{3}{p}\\
    &=& \bank{1}\rx{a^*\ltag{1}\etag{2}a^*\ltag{3}a} \union
        \bank{2} \rx{a^*\ltag{3}a} \union
        \bank{3} = \rx{r_1}\\
    \rx{r_3} &=& \teval{( \D{b}{\rx{r_0}} )} = \emptyset = \rx{r_{\emptyset}}\\
    \rx{r_4} &=& \teval{( \D{b}{\rx{r_1}} )} = \emptyset = \rx{r_{\emptyset}}
  \end{eqnarray*}
  The calculations and disambiguation for \rx{r_1} and \rx{r_2} have
  previously been done in \refxmp{xmp-ambiguous-memory-1} and
  \refxmp{xmp-ambiguous-memory-2} and is therefore not repeated here.
\end{example}

\begin{figure}
  \center
  \begin{tikzpicture}
    \node[]                 (3) [left=of 0]        {};
    \node[state,initial]    (0)                    {$r_0$};
    \node[state]            (2) [below right=of 0] {$r_{\emptyset}$};
    \node[state, accepting] (1) [above right=of 2] {$r_1$};
    
    \path
    (3) edge              node        {\scriptsize\bank{1}\slot{0}{p}}   (0)
    (0) edge              node        {{\scriptsize \arraycolsep=1.4pt $\begin{array}{rcl}
            \bank{2} &=& \hat{\bank{1}}\slot{1}{p}\slot{2}{p}\\
            \bank{3} &=& \hat{\bank{1}}\slot{1}{p}\slot{2}{p}\slot{3}{p}\\
            &a&
          \end{array}$}}   (1)
    (0) edge              node [swap] {\scriptsize $b$}   (2)
    
    (1) edge [loop above] node        {{\scriptsize \arraycolsep=1.4pt $\begin{array}{rcl}
            \bank{2} &=& \hat{\bank{1}}\slot{1}{p}\slot{2}{p}\\
            \bank{3} &=& \hat{\bank{1}}\slot{1}{p}\slot{2}{p}\slot{3}{p}\\
            &a&
          \end{array}$}}   ()
    (1) edge              node        {\scriptsize $b$}   (2)
    
    (2) edge [loop below] node        {\scriptsize $a,b$} ()
    ;
  \end{tikzpicture}
  \caption{The DFA in \refxmp{xmp-dfa-submatch} represented
    as a transition diagram.}\label{fig-dfa-submatch}
\end{figure}

\section{Related work}
Regular languages and regular expressions have been known and used at
least since 1951 when Kleene gave the proof of equivalence between
DFAs and regular languages\autocite*{kleene-1951}. In 1968 Thompson
published his algorithm for constructing a non-deterministic finite
automaton (NFA) from a regular
expressions\autocite*{thompson-1968}. This type of construction has
gained a fair amount of attention. Russ Cox has written a series of
articles covering this approach to regular expression
matching\autocites*{cox-2007}{cox-2009}{cox-2010}.

Thompson's construction and NFAs are also the starting point for Ville
Laurikari and his tagged transitions approach to solving the
submatching problem, using either NFAs or
DFAs\autocite*{laurikari-2}. Tagged DFAs have been further
investigated and adapted to comply with \POSIX{} semantics by Kuklewicz
in a Haskell implementation\autocite*{kuklewicz-2007}, and by
Trofimovich in a lexer generator\autocite*{trofimovich-17}. They make
use of so called minimize and maximize tags, similar to our early and
late tags.

Submatching together with Brzozowski derivatives have to our knowledge
only been discussed by Sulzmann et al.\autocite*{sulzmann-14}. They
employ an interesting and novel approach by using parse trees.

\section{Future work}
We are currently implementing a C++ regular expression library based
on the ideas presented in this report. We hope to present results,
benchmarks, and comparisons with other implementations in a future
article.

\clearpage
\printbibliography[title={\vspace{-2cm}References}]
\addcontentsline{toc}{chapter}{Index}
\end{document}